\documentclass[11pt,prd,aps,showpacs,nofootinbib]{revtex4}

\usepackage{graphicx}
\usepackage{amssymb}
\usepackage{epstopdf}
\usepackage{amsmath}

\newcommand{\be}{\begin{equation}}
\newcommand{\ee}{\end{equation}}
\newcommand{\bea}{\begin{eqnarray}}
\newcommand{\eea}{\end{eqnarray}}

\newcommand{\sla}[1]%
        {\kern .25em\raise.18ex\hbox{$/$}\kern-.6em #1}

\newcommand{\gdll}{\raisebox{-0.4\totalheight}{\includegraphics[scale=.3]{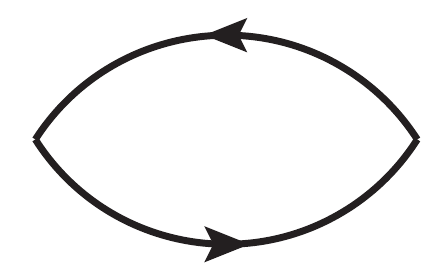}}}
\newcommand{\gdsi}{\raisebox{-0.4\totalheight}{\includegraphics[scale=.3]{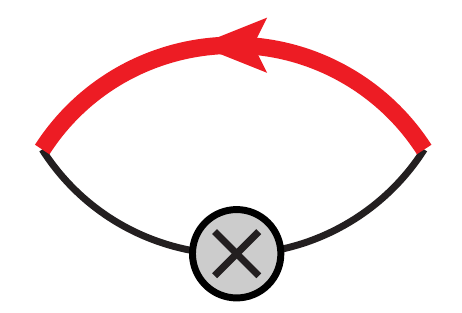}}}
\newcommand{\gdsip}{\raisebox{-0.4\totalheight}{\includegraphics[scale=.3]{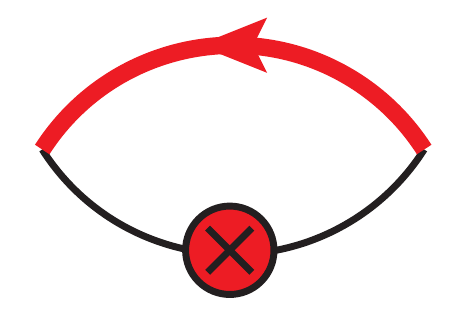}}}
\newcommand{\gdsil}{\raisebox{-0.4\totalheight}{\includegraphics[scale=.3]{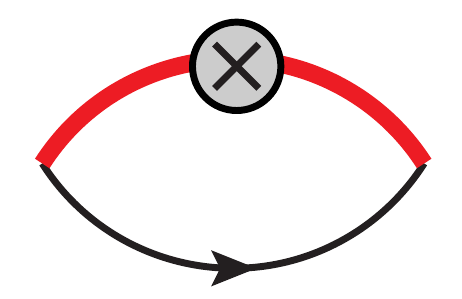}}}
\newcommand{\gdsipl}{\raisebox{-0.4\totalheight}{\includegraphics[scale=.3]{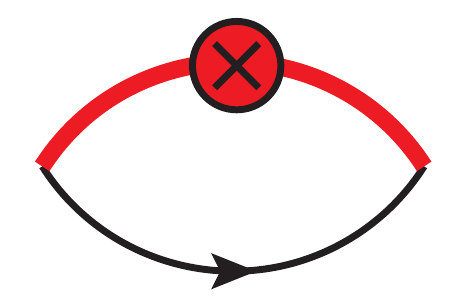}}}

\newcommand{\gdli}{\raisebox{-0.4\totalheight}{\includegraphics[scale=0.3]{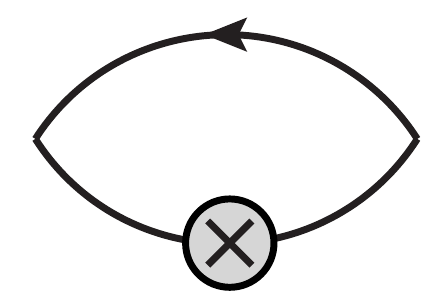}}}
\newcommand{\gdlip}{\raisebox{-0.4\totalheight}{\includegraphics[scale=0.3]{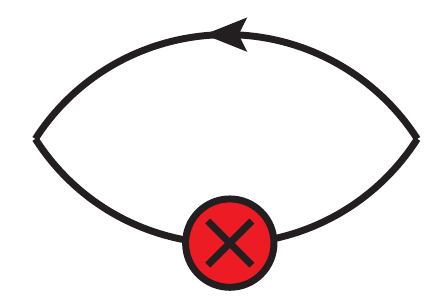}}}

\newcommand{\gdsl}{\raisebox{-0.4\totalheight}{\includegraphics[scale=.3]{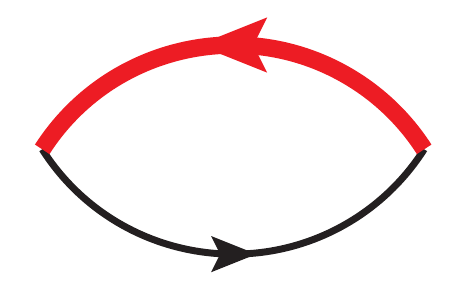}}}

\newcommand{\gdslselfs}{\raisebox{-0.22\totalheight}{\includegraphics[scale=.3]{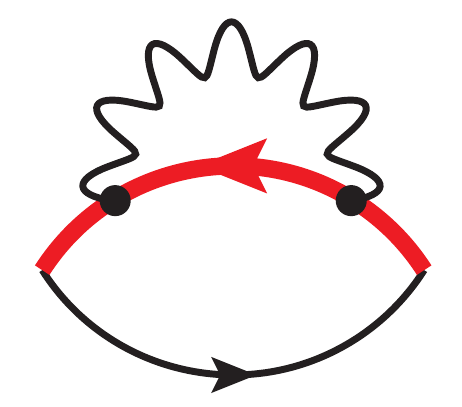}}}
\newcommand{\gdslselfl}{\raisebox{-0.6\totalheight}{\includegraphics[scale=.3]{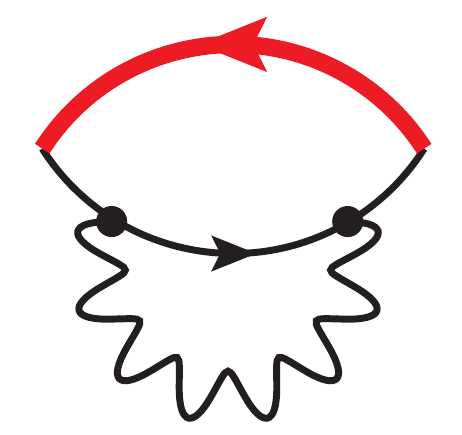}}}
\newcommand{\gdslexch}{\raisebox{-0.4\totalheight}{\includegraphics[scale=.3]{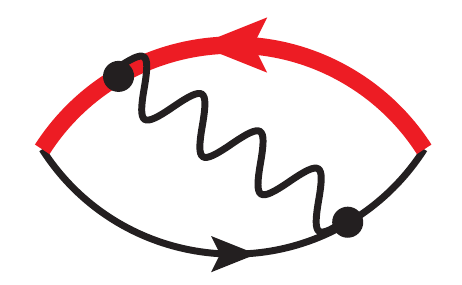}}}

\newcommand{\gdllself}{\raisebox{-0.4\totalheight}{\includegraphics[scale=.3]{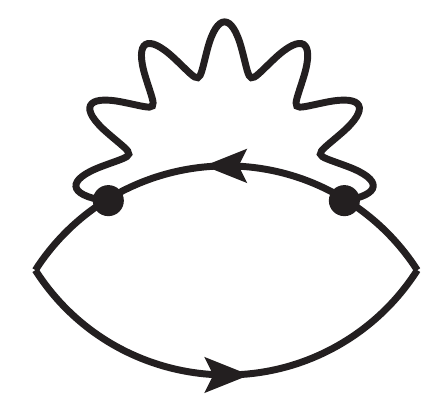}}}
\newcommand{\gdllexch}{\raisebox{-0.4\totalheight}{\includegraphics[scale=.3]{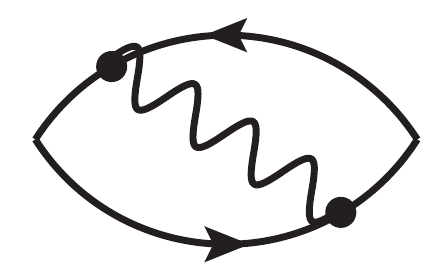}}}
\newcommand{\discgdllexch}{\raisebox{-0.2\totalheight}{\includegraphics[scale=0.4]{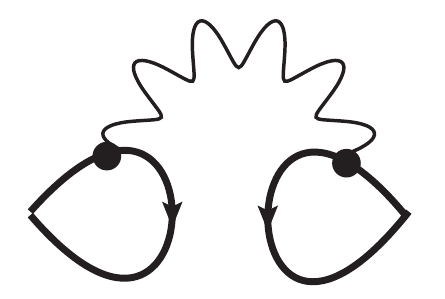}}}

\newcommand{\gdllphtad}{\raisebox{-0.3\totalheight}{\includegraphics[scale=.3]{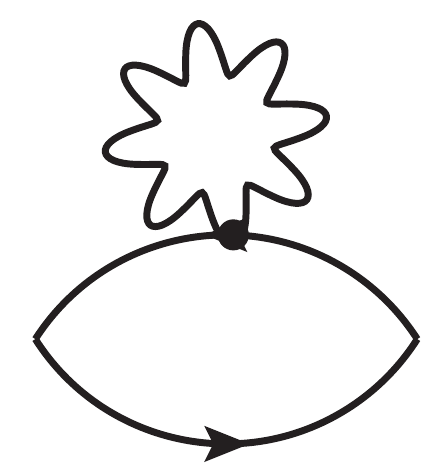}}}

\newcommand{\gdslphtads}{\raisebox{-0.2\totalheight}{\includegraphics[scale=.3]{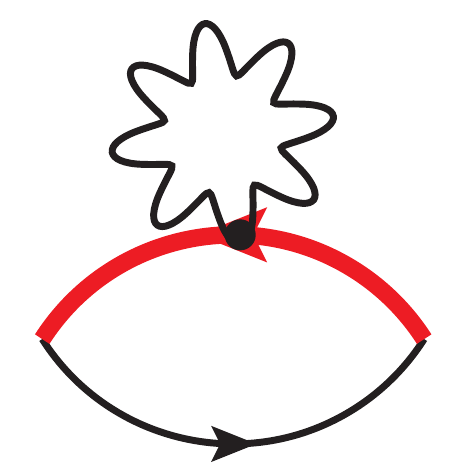}}}
\newcommand{\gdslphtadl}{\raisebox{-0.6\totalheight}{\includegraphics[scale=.3]{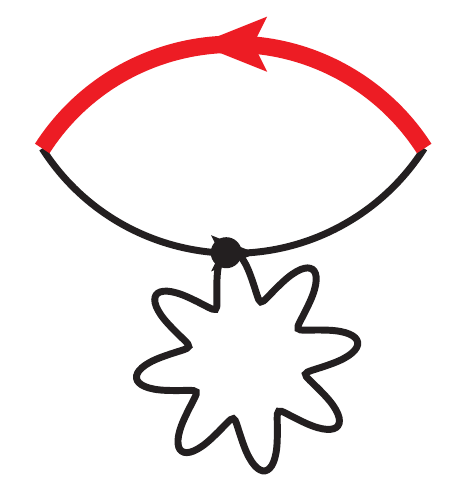}}}

\newcommand{\gdclphtadl}{\raisebox{-0.6\totalheight}{\includegraphics[scale=.25]{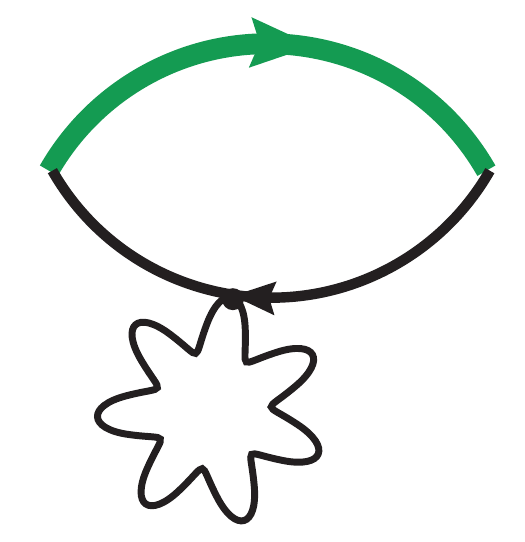}}}
\newcommand{\gdclselfl}{\raisebox{-0.6\totalheight}{\includegraphics[scale=.25]{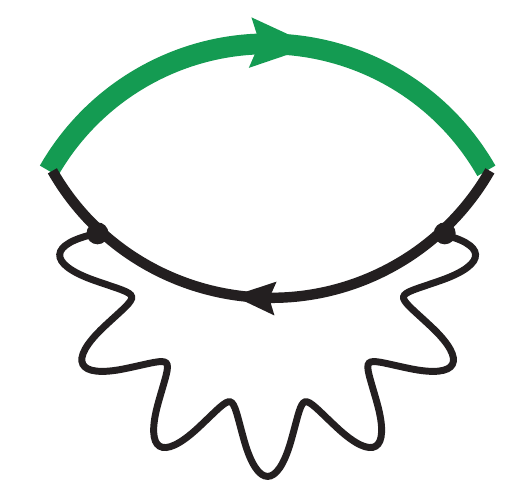}}}
\newcommand{\gdclexch}{\raisebox{-0.4\totalheight}{\includegraphics[scale=.25]{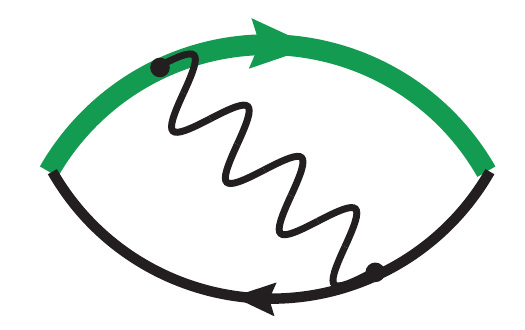}}}
\newcommand{\gdcl}{\raisebox{-0.4\totalheight}{\includegraphics[scale=.25]{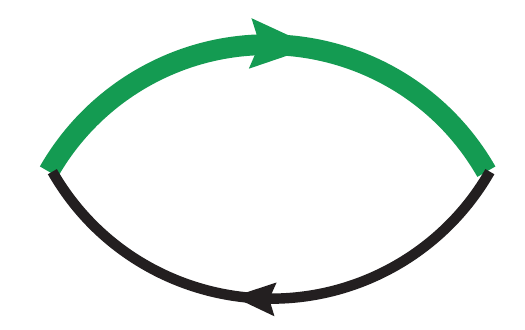}}}
\newcommand{\gdlic}{\raisebox{-0.4\totalheight}{\includegraphics[scale=.25]{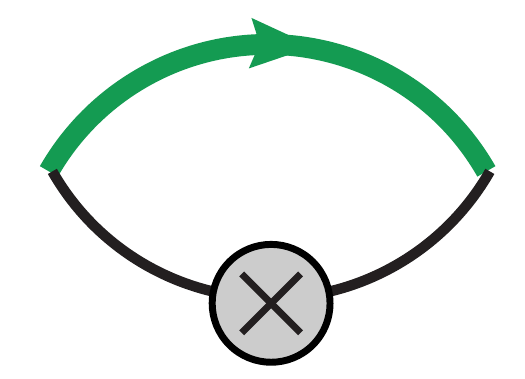}}}
\newcommand{\gdlipc}{\raisebox{-0.4\totalheight}{\includegraphics[scale=.25]{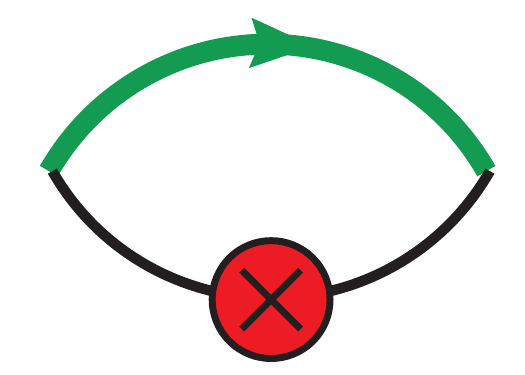}}}
\newcommand{\gdci}{\raisebox{-0.4\totalheight}{\includegraphics[scale=.25]{pics/g2lic.pdf}}}
\newcommand{\gdclphtadc}{\raisebox{-0.6\totalheight}{\includegraphics[scale=.25]{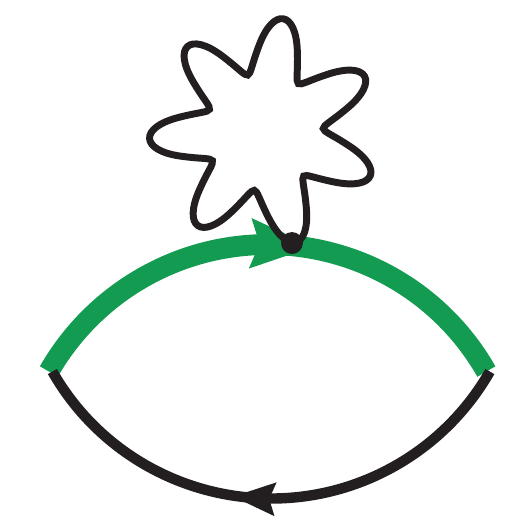}}}
\newcommand{\gdclselfc}{\raisebox{-0.6\totalheight}{\includegraphics[scale=.25]{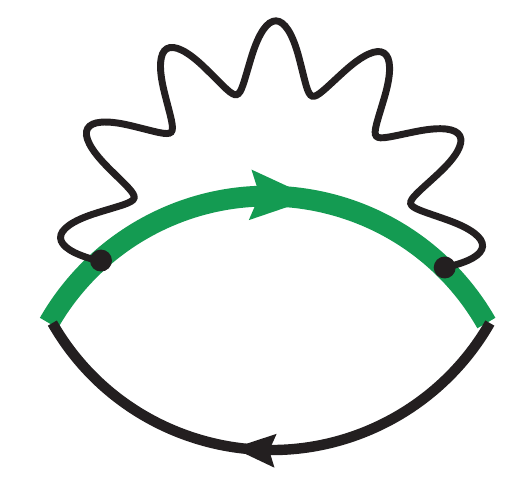}}}
\newcommand{\gdcipl}{\raisebox{-0.4\totalheight}{\includegraphics[scale=.25]{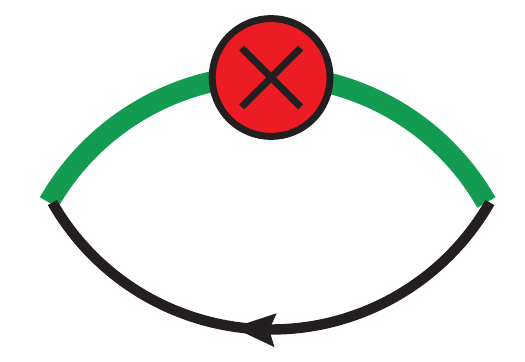}}}
\newcommand{\gdcil}{\raisebox{-0.4\totalheight}{\includegraphics[scale=.25]{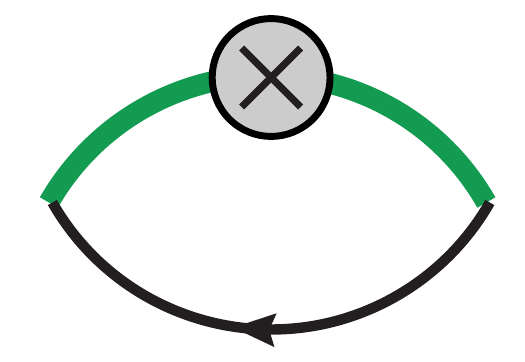}}}
\newcommand{\gdcsexch}{\raisebox{-0.4\totalheight}{\includegraphics[scale=.25]{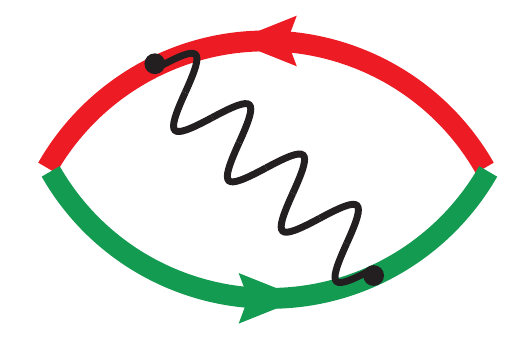}}}
\newcommand{\gdcs}{\raisebox{-0.4\totalheight}{\includegraphics[scale=.25]{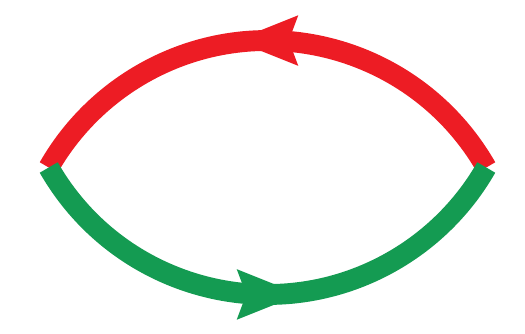}}}
\newcommand{\gdcsphtads}{\raisebox{-0.6\totalheight}{\includegraphics[scale=.25]{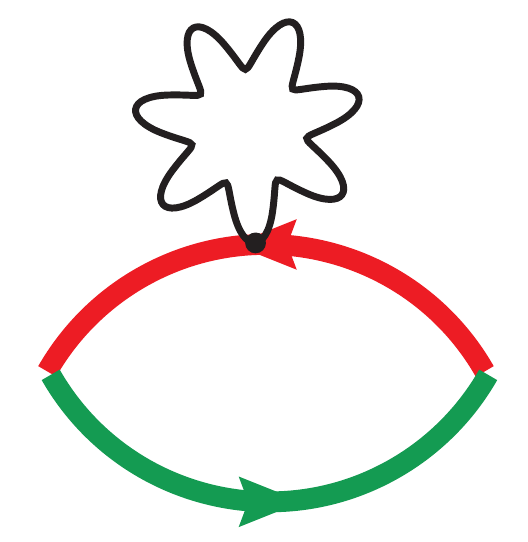}}}
\newcommand{\gdcsselfs}{\raisebox{-0.6\totalheight}{\includegraphics[scale=.25]{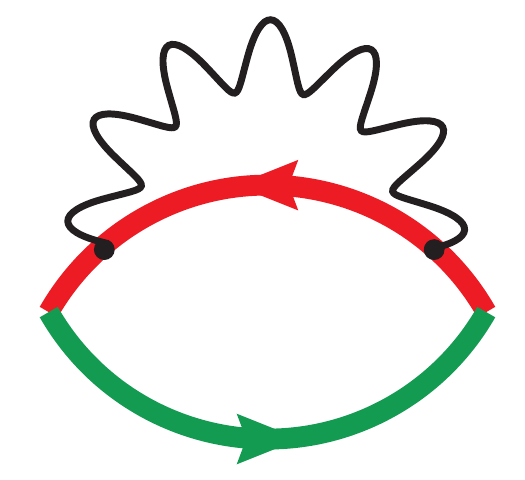}}}
\newcommand{\gdcsphtadc}{\raisebox{-0.6\totalheight}{\includegraphics[scale=.25]{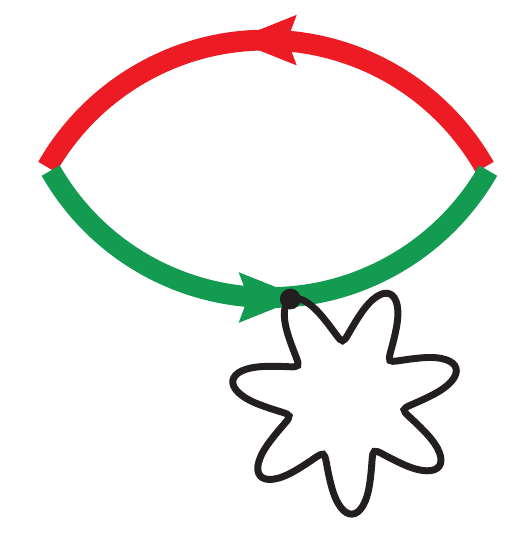}}}
\newcommand{\gdcsselfc}{\raisebox{-0.6\totalheight}{\includegraphics[scale=.25]{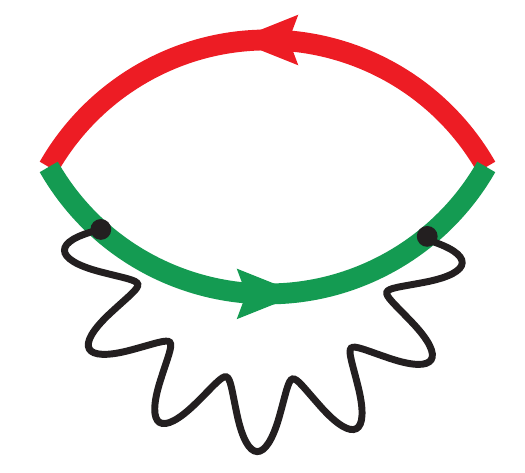}}}
\newcommand{\gdsipc}{\raisebox{-0.4\totalheight}{\includegraphics[scale=.25]{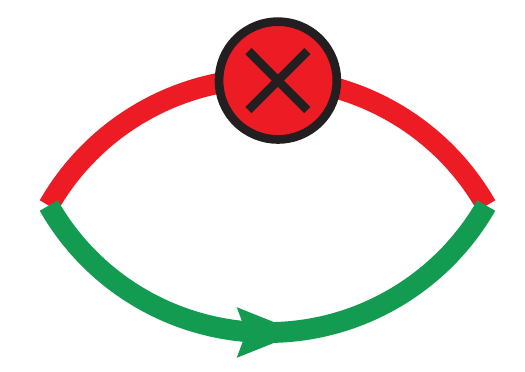}}}
\newcommand{\gdcips}{\raisebox{-0.4\totalheight}{\includegraphics[scale=.25]{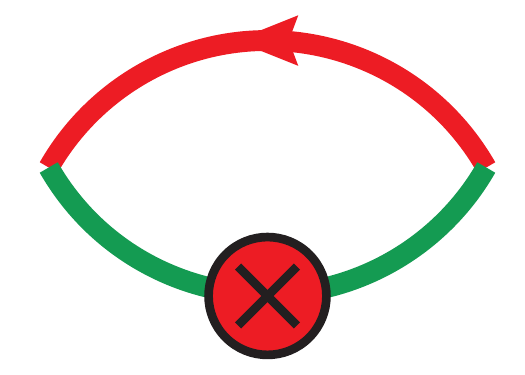}}}
\newcommand{\gdsic}{\raisebox{-0.4\totalheight}{\includegraphics[scale=.25]{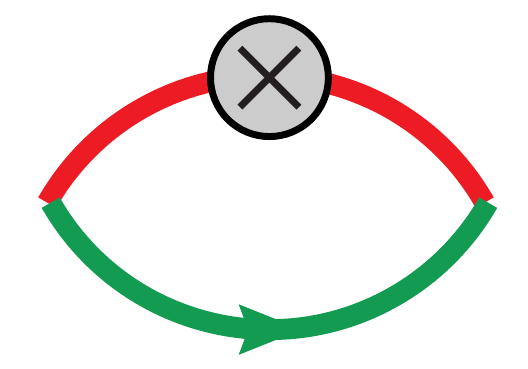}}}
\newcommand{\gdcis}{\raisebox{-0.4\totalheight}{\includegraphics[scale=.25]{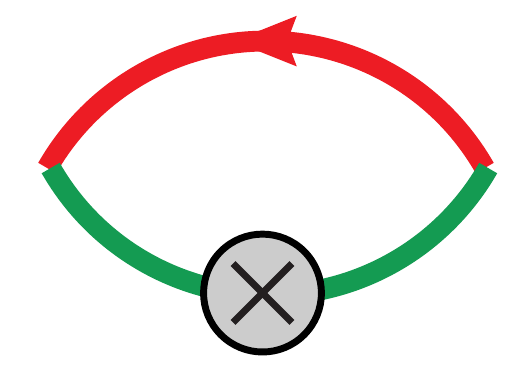}}}

\newcommand{\bear}[1]{\begin{equation}\begin{array}{#1}}
\newcommand{\eear}{ \end{array}\end{equation}}
\newcommand{\barr}[1]{\begin{array}{#1}}
\newcommand{\earr}{\end{array}}

\newcommand{\Romatre}{Dipartimento di Matematica e Fisica and INFN, Universit\`a Roma Tre,\\ Via della Vasca Navale 84, I-00146 Rome, Italy}
\newcommand{\ToV}{Dipartimento di Fisica and INFN, Universit\`a di Roma ``Tor Vergata'',\\ Via della Ricerca Scientifica 1, I-00133 Rome, Italy}
\newcommand{\LaSapienza}{Dipartimento di Fisica and INFN, Universit\`a Roma ``La Sapienza'',\\ Piazzale Aldo Moro 5, 00185 Roma, Italy}
\newcommand{\RomatreINFN}{Istituto Nazionale di Fisica Nucleare, Sezione di Roma Tre,\\ Via della Vasca Navale 84, I-00146 Rome, Italy}

\begin{document}


\title{Leading isospin-breaking corrections to pion, kaon\\[2mm] and charmed-meson masses with Twisted-Mass fermions}

\author{D.~Giusti} \affiliation{\Romatre}
\author{G.~Martinelli} \affiliation{\LaSapienza}
\author{V.~Lubicz} \affiliation{\Romatre}
\author{F.~Sanfilippo} \affiliation{\RomatreINFN}
\author{S.~Simula} \affiliation{\RomatreINFN}
\author{N.~Tantalo} \affiliation{\ToV}
\author{C.~Tarantino} \affiliation{\Romatre}

\author{(RM123 collaboration)} \noaffiliation


\begin{abstract}

We present a lattice computation of the isospin-breaking corrections to pseudoscalar meson masses using the gauge configurations produced by the European Twisted Mass collaboration with $N_f = 2 + 1 + 1$ dynamical quarks at three values of the lattice spacing ($a \simeq 0.062, 0.082$ and $0.089$ fm) with pion masses in the range $M_\pi \simeq 210 - 450$ MeV. 
The strange and charm quark masses are tuned at their physical values. 
We adopt the RM123 method based on the combined expansion of the path integral in powers of the $d$- and $u$-quark mass difference ($\widehat{m}_d - \widehat{m}_u$) and of the electromagnetic coupling $\alpha_{em}$.
Within the quenched QED approximation, which neglects the effects of the sea-quark charges, and after the extrapolations to the physical pion mass and to the continuum and infinite volume limits, we provide results for the pion, kaon and (for the first time) charmed-meson mass splittings, for the prescription-dependent parameters $\epsilon_{\pi^0}$, $\epsilon_\gamma(\overline{MS}, 2~\mbox{GeV})$, $\epsilon_{K^0}(\overline{MS}, 2~\mbox{GeV})$, related to the violations of the Dashen's theorem, and for the light quark mass difference $(\widehat{m}_d - \widehat{m}_u)(\overline{MS}, 2~\mbox{GeV})$.

\end{abstract}

\maketitle

\newpage

\section{Introduction}
\label{sec:intro}

In the last few years the determination of several observables in flavor physics by lattice QCD reached such a precision that both electromagnetic (e.m.) effects and strong isospin breaking (IB) corrections, generated by the light-quark mass difference $(\widehat{m}_d - \widehat{m}_u)$, cannot be neglected any more (see e.g.~Ref.~\cite{FLAG} and references therein).  
Typical examples are the calculations of the leptonic decay constants $f_K$ and $f_\pi$ relevant for $K_{\ell 2}$ and $\pi_{\ell 2}$ decays, and the determination of the vector form factor at zero four-momentum transfer $f_+(0)$ appearing in semileptonic $K_{\ell 3}$ decays. 
These quantities are used to extract the CKM entries $\vert V_{us} \vert$ and $\vert V_{us}\vert / \vert V_{ud} \vert$ from the experimental decay rates, and they have been computed on the lattice with a precision at the few per mille level~\cite{FLAG}.
Such a precision is of the same order of the uncertainties of the e.m.~and strong IB corrections to the leptonic and semileptonic decay rates~\cite{FlaviaNet}. 

The issue of how to include electromagnetic effects in the hadron spectrum and in the determination of quark masses from {\it ab-initio} lattice calculations was addressed for the first time in Ref.~\cite{Duncan:1996xy}. 
Using a variety of different methods to include QED effects in lattice QCD simulations, several collaborations have recently obtained remarkably accurate results for the hadron spectrum, such as the determination of the charged-neutral mass splittings of light pseudoscalar (PS) mesons and baryons~\cite{Blum:2010ym,deDivitiis:2011eh,Ishikawa:2012ix,Aoki:2012st,deDivitiis:2013xla,Borsanyi:2014jba,Endres:2015gda,Fodor:2015pna,Horsley:2015eaa,Fodor:2016bgu,Boyle:2016lbc} (see Ref.~\cite{Patella:2017fgk} for a recent review).

Till now the inclusion of QED effects in lattice QCD simulations has been carried out following mainly two methods: in the first one QED is added directly to the action and QED+QCD simulations are performed at few values of the electric charge (see, e.g., Ref.~\cite{Borsanyi:2014jba,Boyle:2016lbc}), while the second one, the RM123 approach of Ref.~\cite{deDivitiis:2013xla}, consists in an expansion of the lattice path-integral in powers of the two {\it small} parameters $(\widehat{m}_d - \widehat{m}_u)$ and $ \alpha_{em}$, namely $\alpha_{em} \approx (\widehat{m}_d - \widehat{m}_u) / \Lambda_{QCD} \approx 1 \%$.
Since it suffices to work at leading order in the perturbative expansion, the attractive feature of the RM123 method is that the small values of the two expansion parameters are factorized out, so that one can get relatively large numerical signals for the {\it slopes} of the corrections with respect to the two expansion parameters. 
Moreover the slopes can be determined using isospin symmetric QCD gauge configurations.
In this work we adopt the RM123 method.

Using the gauge ensembles generated by the European Twisted Mass Collaboration (ETMC) with $N_f = 2 + 1 + 1$ dynamical quarks \cite{Baron:2010bv,Baron:2011sf} and the quenched QED approximation, we have calculated the pion, kaon, charmed-meson mass splittings and various $\epsilon$ parameters describing the violations of the Dashen's theorem \cite{Dashen:1969eg} (see Ref.~\cite{FLAG}).
The precise definition of the latter ones depend on the separation between QED and QCD effects, which we implement using the prescription of Ref.~\cite{deDivitiis:2013xla} discussed in detail in Section \ref{sec:master}.

Within the quenched QED approximation, which neglects the effects of the sea-quark electric charges, our results\footnote{The quark mass ratio $m_u / m_d$ is renormalization group invariant in pure QCD only. 
In the presence of QED effects the running of the quark mass depends on its electric charge and, therefore, the ratio $\widehat{m}_u / \widehat{m}_d$ depends on the renormalization scheme and scale.} are:
 \bea
      \label{eq:pion}
      M_{\pi^+} - M_{\pi^0} & = & 4.21 ~ (26) ~ \mbox{MeV} \quad [4.5936 ~ (5) ~ \mbox{MeV}]_{exp} ~ , \\[2mm]
       \left[ M_{K^+} - M_{K^0} \right]^{QED}(\overline{MS}, 2~\mbox{GeV}) & = & 2.07 ~ (15) ~ \mbox{MeV} ~ , \\[2mm]
       \left[ M_{K^+} - M_{K^0} \right]^{QCD}(\overline{MS}, 2~\mbox{GeV}) & = & - 6.00 ~ (15) ~ \mbox{MeV} ~ , \\ \nonumber \\
       \label{eq:mdmu}
       (\widehat{m}_d - \widehat{m}_u)(\overline{MS}, 2~\mbox{GeV}) & = & 2.38 ~ (18) ~ \mbox{MeV} ~ , \\[2mm]
       \frac{\widehat{m}_u}{\widehat{m}_d}(\overline{MS}, 2~\mbox{GeV}) & = & 0.513 ~ (30) ~ , \\[2mm]
       \widehat{m}_u(\overline{MS}, 2~\mbox{GeV}) & = & 2.50 ~ (17) ~ \mbox{MeV} ~ , \\[2mm]
       \widehat{m}_d(\overline{MS}, 2~\mbox{GeV}) & = & 4.88 ~ (20) ~ \mbox{MeV} ~ , \\ \nonumber \\
       \epsilon_{\pi^0} & = & 0.03 ~ (4) ~ , \\[2mm]
        \epsilon_\gamma(\overline{MS}, 2~\mbox{GeV}) & = & 0.80 ~ (11) ~ , \\[2mm]
       \epsilon_{K^0}(\overline{MS}, 2~\mbox{GeV}) & = & 0.15 ~ (3) ~ , \\ \nonumber \\
       \label{eq:Dmeson_QED}
       \left[ M_{D^+} - M_{D^0} \right]^{QED}(\overline{MS}, 2~\mbox{GeV}) & = & 2.42 ~ (51) ~ \mbox{MeV} ~ , \\[2mm]
       \label{eq:Dmeson_QCD}
       \left[ M_{D^+} - M_{D^0} \right]^{QCD}(\overline{MS}, 2~\mbox{GeV}) & = & 3.06 ~ (27) ~ \mbox{MeV} ~ , \\[2mm]
       \label{eq:Dmeson}
       M_{D^+} - M_{D^0} & = & 5.47 ~ (53) ~ \mbox{MeV} \quad [4.75 ~ (8) ~ \mbox{MeV}]_{exp} ~ , \\[2mm]
       \delta M_{D^+} + \delta M_{D^0} & = & 8.2 ~ (9) ~ \mbox{MeV} ~ , \\[2mm]
       \label{eq:Dsplus}
       \delta M_{D_s^+} & = & 5.5 ~ (6) ~ \mbox{MeV} ~ , 
 \eea
where the errors include an estimate of the effects of the QED quenching, while by $\widehat{m}$ we indicate a quark mass renormalized in QCD+QED.
In Eqs.~(\ref{eq:pion}) and (\ref{eq:Dmeson}) the experimental values from PDG \cite{PDG} are given in squared brackets for comparison.
Instead the experimental value of the kaon mass splitting $M_{K^+} - M_{K^0} = -3.934 (20)$ MeV~\cite{PDG} is used as the input to determine the quark mass difference $(\widehat{m}_d - \widehat{m}_u)$ given in Eq.~(\ref{eq:mdmu}).
We point out that Eqs.~(\ref{eq:Dmeson_QED}-\ref{eq:Dsplus}) represent the first lattice determinations of e.m.~and strong IB corrections for charmed meson masses (within the quenched QED approximation).

Using the above results and the experimental values of the meson masses \cite{PDG}, we have estimated the pion, kaon, $D$- and $D_s$-meson masses in isospin-symmetric QCD:
 \bea
      \label{eq:pion_QCD}
       M_\pi^{QCD} & = & 134.9 ~ (2) ~ \mbox{MeV} \qquad [134.8 ~ (3) ~ \mbox{MeV}]_{FLAG} ~ , \\
      \label{eq:kaon_QCD}
       M_K^{QCD}  & = & 494.4 ~ (1) ~ \mbox{MeV} \qquad [494.2 ~ (3) ~ \mbox{MeV}]_{FLAG} ~ , \\
      \label{eq:Dmeson_QCD}
       M_D^{QCD} & = & 1863.1 ~ (6) ~ \mbox{MeV} ~ , \\
      \label{eq:Dsmeson_QCD}
       M_{D_s}^{QCD} & = &1963.5 ~ (1.5) ~ \mbox{MeV} ~ ,    
 \eea
where the current estimates from FLAG \cite{FLAG} are given in squared brackets for comparison.

The paper is organized as follows. 
In section~\ref{sec:simulations} we describe the lattice setup and give the simulation details. 
In section~\ref{sec:master} we present the calculations of the relevant correlators within the RM123 approach.
The results of our analysis for the pion mass splitting $M_{\pi^+} - M_{\pi^0}$ and for the $\epsilon_{\pi^0}$ parameter are given in sections~\ref{sec:pion} and~\ref{sec:pi0}, respectively.
In section \ref{sec:kaon} we determine the light quark mass difference $\widehat{m}_d - \widehat{m}_u$ using the experimental value of the kaon mass splitting $M_{K^+} - M_{K^0}$, while section \ref{sec:K0} is devoted to the evaluation of the $\epsilon_{K^0}$ parameter.
In section~\ref{sec:charm} we evaluate the IB corrections in the charmed $D^+$, $D^0$ and $D_s^+$ mesons. 
Using our result for $\widehat{m}_d - \widehat{m}_u$, we present the first lattice determination of the $D$-meson mass difference $M_{D^+} - M_{D^0}$.
 Finally, section~\ref{sec:conclusions} contains our conclusion and outlooks for future developments.

\section{Simulation details}
\label{sec:simulations}

The gauge ensembles used in this work are the ones generated by ETMC with $N_f = 2 + 1 + 1$ dynamical quarks, which include in the sea, besides two light mass-degenerate quarks, also the strange and charm quarks with masses close to their physical values \cite{Baron:2010bv,Baron:2011sf}. 

The lattice actions for sea and valence quarks are the same used in Ref.~\cite{Carrasco:2014cwa} to determine the up, down, strange and charm quark masses in isospin symmetric QCD.
They are the Iwasaki action for gluons and the Wilson Twisted Mass Action for sea quarks.
In the valence sector, in order to avoid the mixing of strange and charm quarks a non-unitary set up was adopted, in which the valence strange and charm quarks are regularized as Osterwalder-Seiler fermions, while the valence up and down quarks have the same action of the sea.
Working at maximal twist such a setup guarantees an automatic ${\cal{O}}(a)$-improvement.

We considered three values of the inverse bare lattice coupling $\beta$ and different lattice volumes, as shown in Table \ref{tab:simudetails}, where the number of configurations analyzed ($N_{cfg}$) corresponds to a separation of $20$ trajectories. 
At each lattice spacing, different values of the light sea quark masses have been considered. 
The light valence and sea quark masses are always taken to be degenerate. 
The bare mass of the strange valence quark $a \mu_s$ is obtained, at each $\beta$, using the physical strange mass and the mass renormalization constants determined in Ref.~\cite{Carrasco:2014cwa}. 

\begin{table}[htb!]
{\small
\begin{center}
\begin{tabular}{||c|c|c|c|c|c|c|c|c||}
\hline
ensemble & $\beta$ & $V / a^4$ &$a\mu_{sea} = a\mu_{val}$&$a\mu_\sigma$&$a\mu_\delta$&$N_{cfg}$& $a\mu_s$ & $a\mu_c$\\ \hline \hline
$A30.32$ & $1.90$ & $32^{3}\times 64$ &$0.0030$ &$0.15$ &$0.19$ &$150$& $0.02363$ & $0.27903$ \\
$A40.32$ & & & $0.0040$ & & & $100$ & & \\
$A50.32$ & & & $0.0050$ & & &  $150$ & & \\
\cline{1-1} \cline{3-4} \cline{7-7} 
$A40.24$ & & $24^{3}\times 48 $ & $0.0040$ & & & $150$ &  &\\
$A60.24$ & & & $0.0060$ & & &  $150$ &  &\\
$A80.24$ & & & $0.0080$ & & &  $150$ &  &\\
$A100.24$ &  & & $0.0100$ & & &  $150$ &  &\\
\cline{1-1} \cline{3-4} \cline{7-7}
$A40.20$ & & $20^{3}\times 48 $ & $0.0040$ & & & $150$ &  &\\
\hline \hline
$B25.32$ & $1.95$ & $32^{3}\times 64$ &$0.0025$&$0.135$ &$0.170$& $150$& $0.02094$ & $0.24725$ \\
$B35.32$ & & & $0.0035$ & & & $150$ & & \\
$B55.32$ & & & $0.0055$ & & & $150$ & & \\
$B75.32$ &  & & $0.0075$ & & & $~80$ & & \\
\cline{1-1} \cline{3-4} \cline{7-7}
$B85.24$ & & $24^{3}\times 48 $ & $0.0085$ & & & $150$ & & \\
\hline \hline
$D15.48$ & $2.10$ & $48^{3}\times 96$ &$0.0015$&$0.1200$ &$0.1385 $& $100$& $0.01612$ & $0.19037$ \\ 
$D20.48$ & & & $0.0020$  &  &  & $100$ & & \\
$D30.48$ & & & $0.0030$ & & & $100$ & & \\
 \hline   
\end{tabular}
\end{center}
}
\vspace{-0.25cm}
\caption{\it \small Values of the simulated sea and valence quark bare masses, of the pion ($M_\pi$) and kaon ($M_K$) masses for the $16$ ETMC gauge ensembles with $N_f = 2+1+1$ dynamical quarks generated within the isospin symmetric theory (see Ref.~\cite{Carrasco:2014cwa} for details). The values of the strange and charm quark bare masses $a \mu_s$ and $a \mu_c$ correspond to the physical strange and charm quark masses, respectively, determined in Ref.~\cite{Carrasco:2014cwa}.}
\label{tab:simudetails}
\end{table}

In Ref.~\cite{Carrasco:2014cwa} eight branches of the analysis were considered. 
They differ in: 
\begin{itemize}
\item the continuum extrapolation adopting for the scale parameter either the Sommer parameter $r_0$ or the mass of a fictitious PS meson made up of strange(charm)-like quarks; 
\item the chiral extrapolation performed with fitting functions chosen to be either a polynomial expansion or a Chiral Perturbation Theory (ChPT) Ansatz in the light-quark mass;
\item the choice between two methods, denoted as M1 and M2, which differ by $O(a^2)$ effects, used to determine in the RI$^\prime$-MOM scheme the mass renormalization constant (RC) $Z_m = 1 / Z_P$. 
\end{itemize}
In the present analysis we made use of the input parameters corresponding to each of the eight branches of Ref.~\cite{Carrasco:2014cwa}.
The central values and the errors of the input parameters, evaluated using bootstrap samplings with ${\cal{O}}(100)$ events, are collected in Table \ref{tab:8branches}.
Throughout this work all the results obtained within the above branches are averaged according to Eq.~(28) of Ref.~\cite{Carrasco:2014cwa}.

\begin{table}[htb!]
\begin{center} 
\begin{tabular}{||c|l|c|c|c|c|c||} 
\hline 
\multicolumn{1}{||c}{} & \multicolumn{1}{|c|}{$\beta$} & \multicolumn{1}{c|}{ $1^{st}$ } & \multicolumn{1}{c|}{ $2^{nd}$ } & \multicolumn{1}{c|}{ $3^{rd}$ } & \multicolumn{1}{c|}{ $4^{th}$ } \\ \hline    
               & 1.90 & 2.224(68) &2.192(75) &2.269(86)&2.209(84) \\ 
$a^{-1}({\rm GeV})$  & 1.95 & 2.416(63) &2.381(73) &2.464(85)&2.400(83) \\ 
               & 2.10 & 3.184(59) &3.137(64) &3.248(75)&3.163(75) \\ \cline{1-6} 
$m_{ud}({\rm GeV})$ &&0.00372(13)&0.00386(17)&0.00365(10)&0.00375(13) \\ \cline{1-6} 
$m_s$({\rm GeV})      &&0.1014(43) &0.1023(39) &0.0992(29)&0.1007(32) \\ \cline{1-6} 
$m_c$({\rm GeV})      && 1.183(34) &1.193(28)  &    1.177(25)&1.219(21) \\ \cline{1-6} 
\multicolumn{1}{||c}{}&\multicolumn{1}{|c|}{1.90} & \multicolumn{4}{c|}{ 0.5290(73) } \\ 
\multicolumn{1}{||c}{$Z_P$}&\multicolumn{1}{|c|}{1.95} & \multicolumn{4}{c|}{ 0.5089(34) } \\ 
\multicolumn{1}{||c}{}&\multicolumn{1}{|c|}{2.10} & \multicolumn{4}{c|}{ 0.5161(27) } \\ \hline    
\end{tabular}
\begin{tabular}{||c|l|c|c|c|c||}
\hline 
\multicolumn{1}{||c}{}&\multicolumn{1}{|c|}{$\beta$} & \multicolumn{1}{c|}{ $5^{th}$ } & \multicolumn{1}{c|}{ $6^{th}$ } & \multicolumn{1}{c|}{ $7^{th}$ } & \multicolumn{1}{c||}{ $8^{th}$ }\\\hline    
               & 1.90 & 2.222(67)&2.195(75)   &2.279(89)&2.219(87) \\ 
$a^{-1}({\rm GeV})$  & 1.95 & 2.414(61)&2.384(73)   &2.475(88)&2.411(86) \\ 
                & 2.10 & 3.181(57)&3.142(64)   &3.262(79)&3.177(78) \\ \cline{1-6} 
$m_{ud}({\rm GeV})$ &&0.00362(12)&0.00377(16)&0.00354(9)&0.00363(12) \\ \cline{1-6} 
$m_s({\rm GeV})$      &&0.0989(44)&0.0995(39) &0.0962(27)&0.0975(30) \\ \cline{1-6} 
$m_c({\rm GeV})$      &&1.150(35) &1.158(27) &1.144(29) &1.182(19) \\ \cline{1-6}    
\multicolumn{1}{||c}{}&\multicolumn{1}{|c|}{1.90} &\multicolumn{4}{c||}{0.5730(42) } \\ 
\multicolumn{1}{||c}{$Z_P$}&\multicolumn{1}{|c|}{1.95} &\multicolumn{4}{c||}{ 0.5440(17) } \\ 
\multicolumn{1}{||c}{}&\multicolumn{1}{|c|}{2.10} &\multicolumn{4}{c||}{ 0.5420(10) } \\ \hline     
\end{tabular} 
\end{center} 
\vspace{-0.25cm}
\caption{\it \small The input parameters for the eight branches of the analysis of Ref.~\cite{Carrasco:2014cwa}. The renormalized quark masses and the RC $Z_P$ are given in the $\overline{\mathrm{MS}}$ scheme at a renormalization scale of 2 GeV. With respect to Ref.~\cite{Carrasco:2014cwa} the table includes an update of the values of the lattice spacing and, consequently, of all the other quantities.}
\label{tab:8branches}
\end{table} 

For each gauge ensemble the PS meson masses are extracted from a single exponential fit (including the proper backward signal) in the range $t_{min} \leq t \leq t_{max}$.
The values chosen for $t_{min}$ and $t_{max}$ at each $\beta$ and lattice volume in the light, strange and charm sectors are collected in Table \ref{tab:timeint}, while the values of the pion, kaon and $D$-meson masses corresponding to pure iso-symmetric QCD, evaluated using the bootstrap samplings of Table \ref{tab:8branches}, are collected in Table \ref{tab:PSmasses}.

\begin{table}[hbt!]
\begin{center}
\begin{tabular}{||c|c||c|c|c||}
\hline
$\beta$ & $T/a$ & $[t_{min}, t_{max}]_{(\ell \ell, \ell s)} / a$ & $[t_{min}, t_{max}]_{(\ell c)}/a$ & $[t_{min}, t_{max}]_{(sc)}/a$ \\
\hline
$1.90$ & $48$ & $[12,23]$ & $[15,21]$ & $[18,23]$ \\
$1.90$ & $64$ & $[12,31]$ & $[15,24]$ & $[18,25]$ \\ \hline
$1.95$ & $48$ & $[13,23]$ & $[16,21]$ & $[19,21]$ \\
$1.95$ & $64$ & $[13,31]$ & $[16,24]$ & $[19,29]$ \\ \hline
$2.10$ & $96$ & $[18,40]$ & $[20,27]$ & $[25,40]$ \\
\hline
\end{tabular}
\end{center}
\caption{\it \small Time intervals $[t_{min}, t_{max}] / a$ adopted for the extraction of the PS meson masses in the light ($\ell$), strange ($s$) and charm ($c$) sectors.} 
\label{tab:timeint}
\end{table}

\begin{table}[htb!]
\begin{center}
\begin{tabular}{||c|c|c||c|c|c||}
\hline
ensemble & $\beta$ & $V / a^4$ & $M_\pi {\rm (MeV)}$ & $M_K {\rm (MeV)}$ & $M_D {\rm (MeV)}$ \\
\hline \hline
$A30.32$ & $1.90$ & $32^{3}\times 64$ & 275 (10) & 568 (22) & 2012 (77) \\
$A40.32$ & & & 316 (12) & 578 (22) & 2008 (77) \\
$A50.32$ & & & 350 (13) & 586 (22) & 2014 (77) \\
\cline{1-1} \cline{3-6} 
$A40.24$ & & $24^{3}\times 48 $ & 322 (13) & 582 (23) & 2017 (77) \\
$A60.24$ & & & 386 (15) & 599 (23) & 2018 (77) \\
$A80.24$ & & & 442 (17) & 618 (24) & 2032 (78) \\
$A100.24$ & & & 495 (19) & 639 (24) & 2044 (78) \\
\cline{1-1} \cline{3-6}
$A40.20$ & & $20^{3}\times 48 $ & 330 (13) & 586 (23) & 2029 (79) \\
\hline \hline
$B25.32$ & $1.95$ & $32^{3}\times 64$ & 259 ~(9) & 546 (19) & 1942 (67) \\
$B35.32$ & & & 302 (10) & 555 (19) & 1945 (67) \\
$B55.32$ & & & 375 (13) & 578 (20) & 1957 (68) \\
$B75.32$ &  & & 436 (15) & 599 (21) & 1970 (68) \\
\cline{1-1} \cline{3-6}
$B85.24$ & & $24^{3}\times 48 $ & 468 (16) & 613 (21) & 1972 (68) \\
\hline \hline
$D15.48$ & $2.10$ & $48^{3}\times 96$ & 223 ~(6) & 529 (14) & 1929 (49) \\ 
$D20.48$ & & & 255 ~(7) & 535 (14) & 1933 (50) \\
$D30.48$ & & & 318 ~(8) & 550 (14) & 1937 (49) \\
 \hline   
\end{tabular}
\end{center}
\vspace{-0.25cm}
\caption{\it \small Values of the pion, kaon and $D$-meson masses evaluated using the bootstrap samplings of Table \ref{tab:8branches} for all the $16$ ETMC gauge ensembles.}
\label{tab:PSmasses}
\end{table}

Following Refs.~\cite{deDivitiis:2013xla,Gasser:2003hk} we impose a specific matching condition between the full QCD+QED and the isospin symmetric QCD theories: in the $\overline{MS}$ scheme at a renormalization scale $\mu = 2$ GeV we require $\widehat{m}_f(\overline{\mathrm{MS}}, 2~\mbox{GeV}) = m_f(\overline{\mathrm{MS}}, 2~\mbox{GeV})$ for $f = (ud), s, c$, where $\widehat{m}$  and $m$ are the renormalized quark masses in the full theory and in isosymmetric QCD. A similar condition is imposed on the strong coupling constants of the two theories (i.e.~the lattice spacing). 
These conditions fix the isosymmetric QCD bare parameters and a unique prescription to define the isosymmetric QCD contribution to each hadronic quantity (see for instance the first term on the r.h.s.~of Eq.~(\ref{eq:MPS})). 
The parameters given in Table \ref{tab:8branches} have been obtained in Ref.~\cite{Carrasco:2014cwa} by using for the isosymmetric QCD contributions to the hadronic inputs the estimates given by FLAG~\cite{FLAG}. 
In this work we provide new results for these inputs that can be used in the future to obtain (slightly) improved determinations of the isosymmetric bare couplings. 
We stress that in the calculation of leading IB observables it is fully legitimate to use the QCD parameters given in Ref.~\cite{Carrasco:2014cwa}  because a change in the prescription that fixes these values has an effect only at higher orders in $\alpha_{em}$ and $(\widehat{m}_d - \widehat{m}_u)$.

\section{Evaluation of the IB corrections}
\label{sec:master}

According to the approach of Ref.~\cite{deDivitiis:2013xla} the e.m.~and strong IB corrections to the mass of a PS meson with charge $Qe$ can be written as 
 \be
      M_{PS^Q} = M_{PS} + \left[ \delta M_{PS^Q} \right]^{QED} + \left[ \delta M_{PS} \right]^{QCD} 
      \label{eq:MPS}
 \ee
 with
  \bea
        \label{eq:MPS_QED}
        \left[ \delta M_{PS^Q} \right]^{QED} & \equiv &  4 \pi \alpha_{em} \left[ \delta M_{PS^Q} \right]^{em} + ... ~ , \\[2mm]
        \label{eq:MPS_QCD}
        \left[ \delta M_{PS} \right]^{QCD} & \equiv & (\widehat{m}_d - \widehat{m}_u) \left[ \delta M_{PS} \right]^{IB} + ... ~ ,
  \eea
where the ellipses stand for higher order terms in $\alpha_{em}$ and $(\widehat{m}_d - \widehat{m}_u)$, while $M_{PS}$ stands for the PS meson mass corresponding to the renormalized quark masses in the isosymmetric QCD theory.
The separation in Eq.~(\ref{eq:MPS}) between the QED and QCD contributions, $\left[ \delta M_{PS^Q} \right]^{QED}$ and $\left[ \delta M_{PS} \right]^{QCD}$, is prescription and renormalization scheme and scale dependent \cite{Gasser:2003hk,Bijnens:1993ae}, as it will be specified in a while. 

\begin{figure}[htb!]
\begin{center}
\includegraphics[scale=1.8]{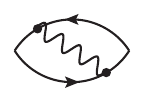}~
\includegraphics[scale=1.8]{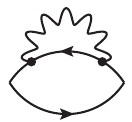}~
\includegraphics[scale=1.8]{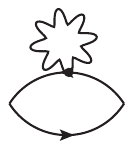}~
\includegraphics[scale=1.8]{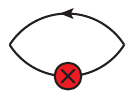}~
\includegraphics[scale=1.8]{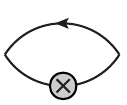}~
\end{center}
\flushleft{\qquad \qquad \qquad \qquad (a) \qquad \qquad \qquad (b) \qquad \qquad \qquad (c) \qquad \qquad \qquad (d) \qquad \qquad  \qquad (e)}
\caption{\it \small Fermionic connected diagrams contributing at $\mathcal{O}(e^2)$ and $\mathcal{O}(m_d - m_u)$ to the IB corrections to meson masses: exchange (a), self energy (b), tadpole (c), pseudoscalar insertion (d) and scalar insertion (e).}
\label{fig:diagrams}
\end{figure}

Throughout this work we adopt the quenched QED approximation, which neglects the sea-quark electric charges and corresponds to consider only (fermionic) connected diagrams.  
Including the contributions coming from the insertions of the e.m.~current and tadpole operators, of the PS and scalar densities (see Refs.~\cite{deDivitiis:2011eh,deDivitiis:2013xla}) the basic diagrams are those depicted schematically in Fig.~\ref{fig:diagrams}.
The insertion of the PS density is related to the the e.m.~shift of the critical mass present in lattice formulations breaking chiral symmetry, as in the case of Wilson and twisted-mass fermions.

In order to evaluate the diagrams (\ref{fig:diagrams}a)-(\ref{fig:diagrams}e) the following correlators are considered:
 \bea
     \label{eq:deltaC_J}
     \delta C^J(t) & = & \sum_{\vec{x}, y_1, y_2} \langle 0| T \left \{ \phi_{PS}^\dagger(\vec{x}, t) ~ J_\mu(y_1) J_\mu(y_2) ~ 
                                   \phi_{PS}(0) \right \} | 0 \rangle ~ , \\
    \label{eq:deltaC_T}
     \delta C^T(t) & = & \sum_{\vec{x}, y} \langle 0| T \left \{ \phi_{PS}^\dagger(\vec{x}, t) ~ T(y) ~ \phi_{PS}(0) \right \} | 0 \rangle ~ , \\
     \label{eq:deltaC_P}
     \delta C^{P_f}(t) & = & \sum_{\vec{x}, y} \langle 0| T \left \{ \phi_{PS}^\dagger(\vec{x}, t) ~ i \overline{\psi}_f(y) \gamma_5  \psi_f(y) ~
                                       \phi_{PS}(0) \right \} | 0 \rangle ~ , \\
     \label{eq:deltaC_S}
     \delta C^{S_f}(t) & = & - \sum_{\vec{x}, y} \langle 0| T \left \{ \phi_{PS}^\dagger(\vec{x}, t) ~  \left[ \overline{\psi}_f(y) \psi_f(y) \right] ~ 
                                          \phi_{PS}(0) \right \} | 0 \rangle ~ ,
 \eea
where $f = \{u, d, s, c\}$, 
 \bea
     J_\mu(y) & = & \sum_f q_f ~ \frac{1}{2} \left[ \bar{\psi}_f(y) (\gamma_\mu - i \tau^3 \gamma_5 ) U_\mu(y) \psi_f(y + a \hat{\mu}) 
                              \right. \nonumber \\
                   & + & \left. \bar{\psi}_f(y + a \hat{\mu}) (\gamma_\mu + i \tau^3 \gamma_5 ) U_\mu^\dagger(y) \psi_f(y) \right]
     \label{eq:Jmu}
 \eea
is the (lattice) conserved e.m.~current, and
 \bea
     T(y) & = & \sum_f q_f^2 ~ \sum_\nu \frac{1}{2} \left[ \bar{\psi}_f(y) (\gamma_\nu - i \tau^3 \gamma_5 ) U_\nu(y) \psi_f(y+ a \hat{\nu}) 
                      \right. \nonumber \\
             & - & \left. \bar{\psi}_f(y + a \hat{\nu}) (\gamma_\nu + i \tau^3 \gamma_5 ) U_\nu^\dagger(y) \psi_f(y) \right] 
     \label{eq:tadpole}
 \eea
is the tadpole operator with $\phi_{PS}(x) = i \overline{\psi}_{f_1}(x) \gamma_5 \psi_{f_2}(x)$ being the interpolating field for a PS meson composed by two valence quarks $f_1$ and $f_2$ with charges $q_1 e$ and $q_2 e$.
In our twisted-mass setup the Wilson parameters of the two valence quarks are chosen to be opposite ($r_1 = - r_2$) in order to guarantee that discretization effects on $M_{PS}$ are of order $\mathcal{O}(a^2 m \Lambda_{QCD})$.

Within the quenched QED approximation the correlator $\delta C^J(t)$ corresponds to the sum of the diagrams (\ref{fig:diagrams}a)-(\ref{fig:diagrams}b), while the correlators $\delta C^T(t)$, $\delta C^{P_f}(t)$ and $\delta C^{S_f}(t)$ represent the contributions of the diagrams (\ref{fig:diagrams}c), (\ref{fig:diagrams}d) and (\ref{fig:diagrams}e), respectively.
The removal of the photon zero-mode is done according to $QED_L$~\cite{Hayakawa:2008an}, i.e.~the photon field $A_\mu$ in momentum space satisfies $A_\mu(k_0, \vec{k} = \vec{0}) \equiv 0$ for all $k_0$.

The statistical accuracy of the meson correlators is based on the use of the so-called ``one-end" stochastic method \cite{McNeile:2006bz}, which includes spatial stochastic sources at a single time slice chosen randomly.
Four stochastic sources (diagonal in the spin variable and dense in the color one) were adopted per each gauge configuration.

A new technique for the lattice evaluation of the photon insertion in the diagrams (a)-(c) of Fig.~\ref{fig:diagrams} and an estimate of the computational cost are presented in the Appendix \ref{sec:appendixA}.

In our analysis the correlators $\delta C^j(t)$ with $j = \{J, T, PS, S\}$ are divided by the tree-level one
 \be
     C(t) \equiv \sum_{\vec{x}} \langle 0| T \left \{ \phi_{PS}^\dagger(\vec{x}, t) \phi_{PS}(0) \right \} | 0 \rangle ~ ,
     \label{eq:tree}
 \ee
obtaining at large time distances, where the PS ground-state is dominant,
 \bea
      \frac{\delta C^j(t)}{C(t)} ~ _{\overrightarrow{t >> a, (T-t) >> a}} ~ \frac{\delta Z_{PS}^j}{Z_{PS} } + 
                                                                                                           \frac{\delta M_{PS}^j}{M_{PS}} f_{PS}(t)
      \label{eq:ratio}
 \eea
where $Z_{PS} \equiv \langle 0 | \phi_{PS}(0) | PS \rangle$ and
 \be
      f_{PS}(t) \equiv  M_{PS} \left( \frac{T}{2} - t \right) \frac{e^{- M_{PS} t} - e^{- M_{PS} (T-t)}}{e^{- M_{PS} t} + e^{- M_{PS} (T-t)}} - 
                               1 - M_{PS} \frac{T}{2}
      \label{eq:fPS}
 \ee
is almost a linear function of the Euclidean time $t$.
Thus, the various e.m.~and strong IB corrections to the PS mass, $\delta M_{PS}^j$ ($j = J, T, P_f, S_f$), can be extracted from the slope of the corresponding ratios $\delta C^j(t) / C(t)$ at large time distances (see Table \ref{tab:timeint} for the chosen fitting intervals).

The difference between the bare quark mass $\widehat{\mu}_f$ in QCD+QED and the bare mass $\mu_f$ in isosymmetric QCD is related to the corresponding difference between the renormalized masses $\widehat{m}_f$ and $m_f$ by
 \be
      \widehat{\mu}_f - \mu_f = \frac{\widehat{m}_f}{\widehat{Z}_{m_f}}  - \frac{m_f}{Z_m} =
                                               \frac{1}{Z_m} \left[ \frac{Z_m}{\widehat{Z}_{m_f}} \widehat{m}_f - m_f\right]
 \ee
where $\widehat{Z}_{m_f}$ ($Z_m$) is the mass renormalization constant in QCD+QED (QCD).
By defining
 \be
       \frac{Z_m}{\widehat{Z}_{m_f}} = 1 + 4 \pi \alpha_{em} \frac{1}{{\cal{Z}}_f}
 \ee
we get
 \be
      \widehat{\mu}_f - \mu_f = \frac{1}{Z_m} \left[ \widehat{m}_f  - m_f \right] + 4 \pi \alpha_{em} \frac{1}{Z_m {\cal{Z}}_f} ~ \widehat{m}_f ~ .
      \label{eq:deltam_bare}
 \ee
For our maximally twisted-mass setup one has $Z_m = 1 / Z_P$, while for $1 / {\cal{Z}}_f$ we use the perturbative result at leading order in $\alpha_{em}$  in the $\overline{\mathrm{MS}}$ scheme at the renormalization scale $\mu$, given by \cite{Aoki:1998ar}
 \be
      \frac{1}{{\cal{Z}}_f}(\overline{\mathrm{MS}}, \mu) = \frac{q_f^2}{16 \pi^2} \left[ 6 \mbox{log}(a \mu) - 22.596 \right] ~ .
      \label{eq:Zud}
 \ee
Once multiplied by the bare quantity $\delta M_{PS}^{S_f}$ related to the insertion of the scalar density, the first term in the r.h.s.~of Eq.~(\ref{eq:deltam_bare}) generates a finite term, which in our prescription~\cite{deDivitiis:2013xla} defines the QCD correction
\be
    \left[ \delta M_{PS} \right]^{QCD}(\overline{\mathrm{MS}}, \mu) = \sum_{f = f_1, f_2} Z_P(\overline{\mathrm{MS}}, \mu) ~
          \left[ \widehat{m}_f(\overline{\mathrm{MS}}, \mu) - m_f(\overline{\mathrm{MS}}, \mu) \right] ~ \delta M_{PS}^{S_f} ~ .
     \label{eq:deltaMPS_QCD}
 \ee
The second term in the r.h.s.~of Eq.~(\ref{eq:deltam_bare}) generates a logarithmic divergent contribution that, when included in the QED correction, compensates the corresponding divergence of the self-energy and tadpole diagrams.
At leading order in $\alpha_{em}$ and $(\widehat{m}_d - \widehat{m}_u)$ one has
 \bea
      \left[ \delta M_{PS^Q} \right]^{QED}(\overline{\mathrm{MS}}, \mu) & = & 4 \pi \alpha_{em} \left \{ \delta M_{PS}^J + \delta M_{PS}^T + 
                              \sum_{f = f_1, f_2} \delta m_f^{crit} ~  \delta M_{PS}^{P_f} \right. \nonumber \\ 
                    & + & \left.  \sum_{f = f_1, f_2} \frac{Z_P(\overline{\mathrm{MS}}, \mu)}{{\cal{Z}}_f(\overline{\mathrm{MS}}, \mu)} ~ 
                              m_f(\overline{\mathrm{MS}}, \mu) ~ \delta M_{PS}^{S_f} \right\} ~ ,
      \label{eq:deltaMPS_QED}
 \eea
where $\delta m_f^{crit}$ is the e.m.~shift of the critical mass for the quark flavor $f$, which will be discussed in details in the next Section.
Note that, since we require $\widehat{m}_f(\overline{\mathrm{MS}}, 2~\mbox{GeV}) = m_f(\overline{\mathrm{MS}}, 2~\mbox{GeV})$ for $f = (ud), s, c$, the r.h.s.~of Eq.~(\ref{eq:deltaMPS_QCD}) at the scale $\mu = 2 $ GeV receives a  non-vanishing contribution only when a valence light quark $u$ or $d$ is present in the PS meson (since $m_d = m_u = m_{ud}$).
In that case $\left[ \delta M_{PS} \right]^{QCD}(\overline{\mathrm{MS}}, 2~\mbox{GeV})$ is proportional to $(\widehat{m}_d - \widehat{m}_u)(\overline{\mathrm{MS}}, 2~\mbox{GeV})$, as anticipated in Eq.~(\ref{eq:MPS_QCD}).
When $PS^Q = \pi^{0,+}$ the contributions coming from the $u$ and $d$ quarks cancel out and $\left[ \delta M_\pi \right]^{QCD}(\overline{\mathrm{MS}}, 2~\mbox{GeV}) = 0$ at leading order in $(\widehat{m}_d - \widehat{m}_u)(\overline{\mathrm{MS}}, 2~\mbox{GeV})$.

\subsection{Determination of $\delta m_f^{crit}$}
\label{sec:mcrit}

In order to extract physical information from Eq.~(\ref{eq:deltaMPS_QED}) it is necessary to determine the e.m.~shift of the critical mass of the quarks.
The strategy chosen in Ref.~\cite{deDivitiis:2013xla} is to use the vector Ward-Takahashi identity, which allows to calculate $\delta m_f^{crit}$ as
 \be
     \delta m_f^{crit}  = - \frac{\nabla_0 \left[ \delta V_f^J(t) + \delta V_f^T(t) \right]}{\nabla_0 ~ \delta V_f^{P_f}(t)}
     \label{eq:deltam_crit}
 \ee
where $\nabla_0$ is the backward time derivative and 
 \bea
     \label{eq:deltaV_J}
     \delta V_f^J(t) & = & \frac{1}{L^6} \sum_{\vec{x}, y_1, y_2} \langle 0| T \left \{ V_{\bar{f} f}^\dagger(\vec{x}, t) ~ J_\mu(y_1) J_\mu(y_2) ~
                                      \phi_{\bar{f} f}(0) \right \} | 0 \rangle ~ , \\
    \label{eq:deltaV_T}
     \delta V_f^T(t) & = & \frac{1}{L^3} \sum_{\vec{x}, y} \langle 0| T \left \{ V_{\bar{f} f}^\dagger(\vec{x}, t) ~ T(y) ~ \phi_{\bar{f} f}(0) \right \} | 0 \rangle ~ , \\
     \label{eq:deltaC_PS}
     \delta V_f^{P_f}(t) & = & \frac{1}{L^3} \sum_{\vec{x}, y} \langle 0| T \left \{ V_{\bar{f} f}^\dagger(\vec{x}, t) ~  i \overline{\psi}_f(y) \gamma_5 \psi_f(y) ~
                                         \phi_{\bar{f} f}(0) \right \} | 0 \rangle ~ ,
 \eea
with $V_{\bar{f} f}(x) \equiv \overline{\psi}_f(x) \gamma_0 \psi_f(x)$. 

\begin{figure}[htb!]
\begin{center}
\includegraphics[scale=0.90]{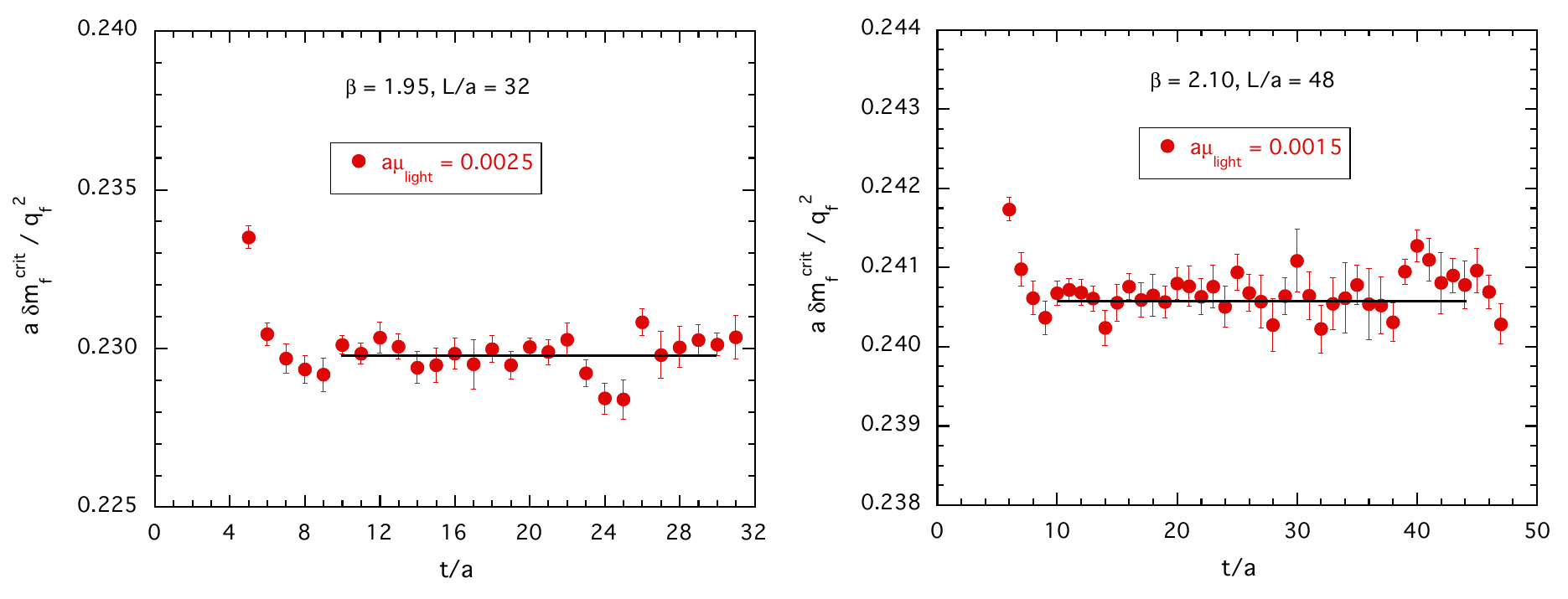}
\end{center}
\vspace{-1.0cm}
\caption{\it \small Results of the r.h.s~of Eq.~(\ref{eq:deltam_crit}) in lattice units calculated for the ETMC gauge ensembles B25.32 (left panel) and D15.48 (right panel). The solid lines represent the value of $\delta m_f^{crit} / q_f^2$ extracted from the corresponding plateau regions.}
\label{fig:mcrit_plateaux}
\end{figure}

Within the quenched QED approximation the shift $\delta m_f^{crit}$ is proportional to $q_f^2$ and can be determined from the plateaux of the r.h.s~of Eq.~(\ref{eq:deltam_crit}), as shown in Fig.~\ref{fig:mcrit_plateaux} for the gauge ensembles B25.32 and D15.48.

\begin{figure}[htb!]
\begin{center}
\includegraphics[scale=0.90]{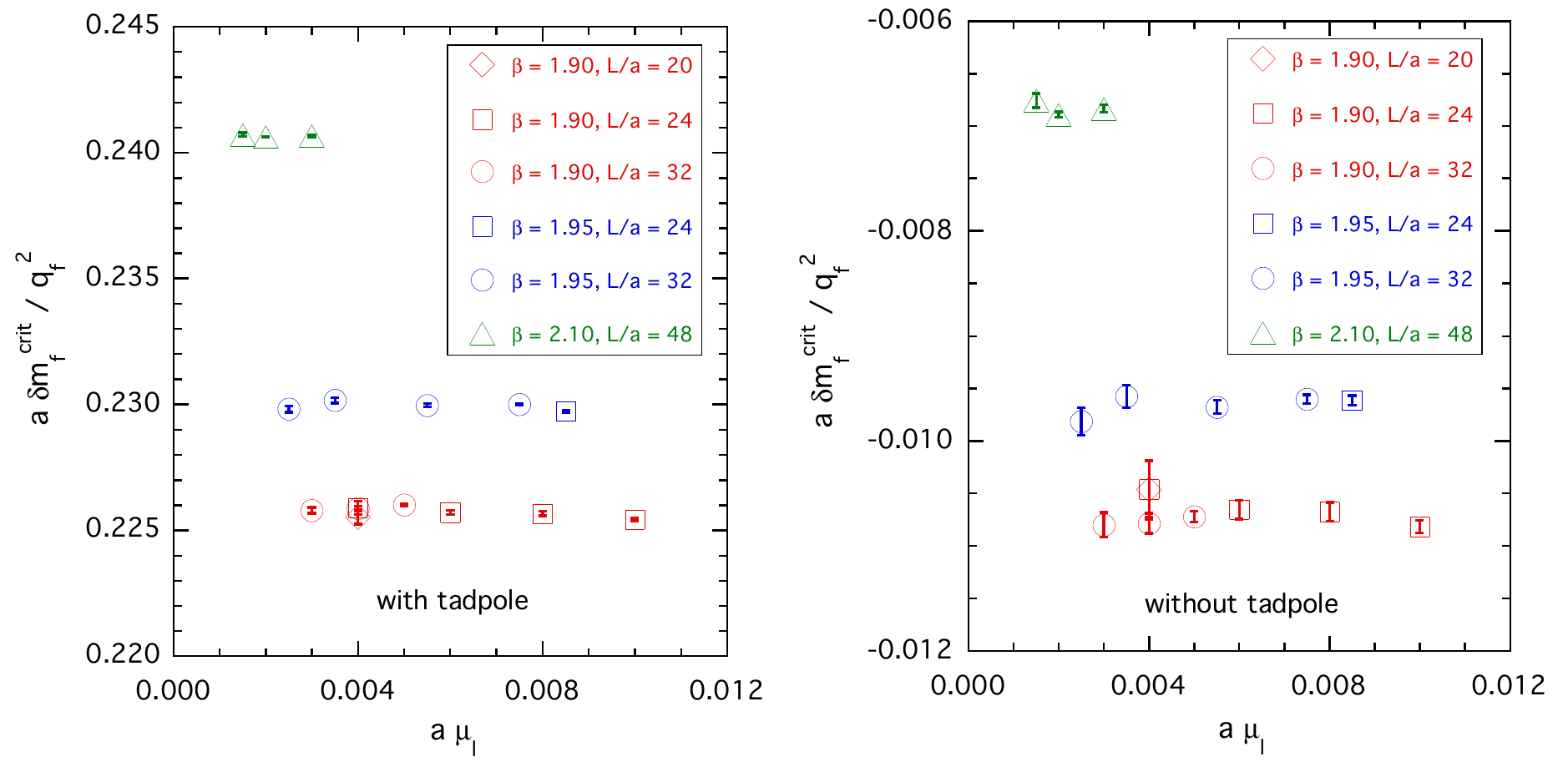}
\end{center}
\vspace{-0.75cm}
\caption{\it \small Values of the e.m.~shift of the critical mass $\delta m_f^{crit} / q_f^2$ versus the bare light-quark mass (in lattice units) calculated for the ETMC gauge ensembles of Table \ref{tab:simudetails}. Left panel: with the tadpole contribution. Right panel: without the tadpole contribution.}
\label{fig:deltam_crit}
\end{figure}

The results of $\delta m_f^{crit} / q_f^2$ for all the ETMC gauge ensembles of Table \ref{tab:simudetails} are collected in Fig.~\ref{fig:deltam_crit}.
It can be seen that: ~ i) the values of $\delta m_f^{crit} / q_f^2$ are determined quite precisely (better than the per mille level), and ~ ii) at each value of the lattice spacing there is a very mild dependence on the value of the light-quark mass.

\subsection{Extraction of the e.m.~and strong IB corrections}
\label{sec:IB}

\begin{figure}[htb!]
\begin{center}
\includegraphics[scale=1.0]{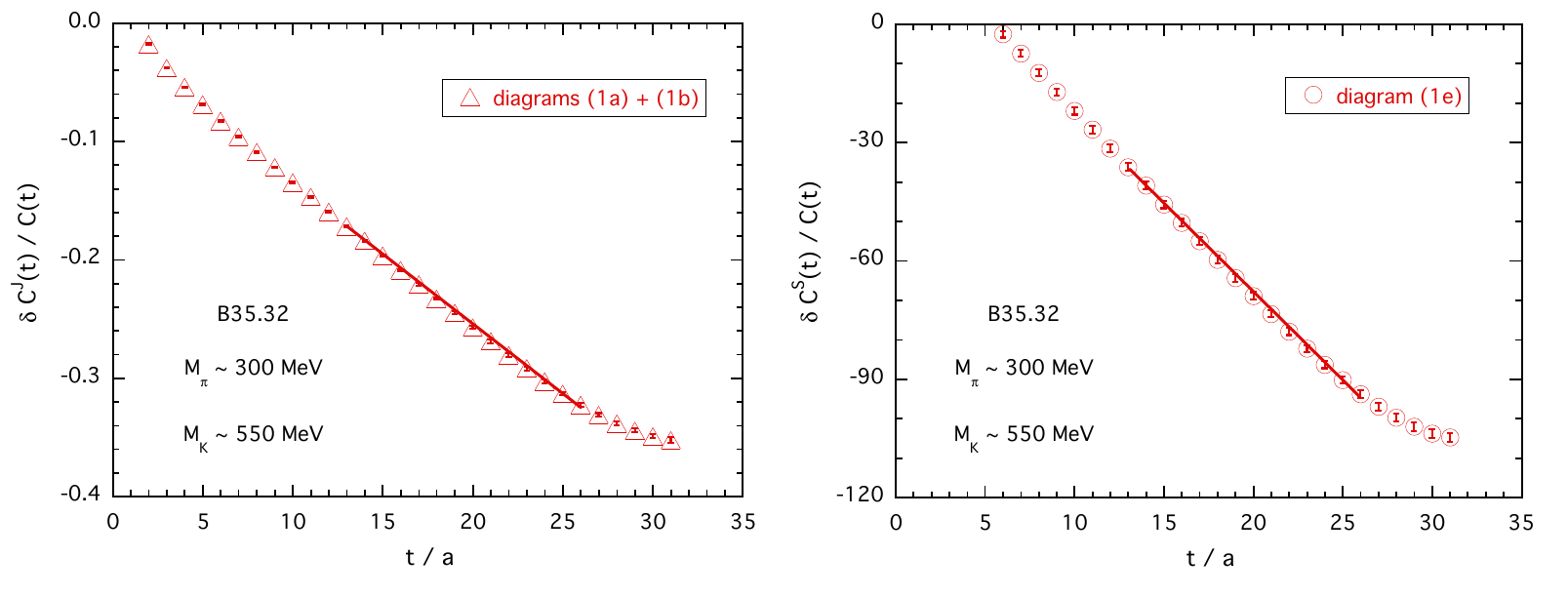}
\end{center}
\vspace{-0.75cm}
\caption{\it \small Ratios $\delta C^J(t) / C(t)$ (left panel) and $\delta C^{S_\ell}(t) / C(t)$ (right panel) in the case of the charged kaon for the gauge ensemble B35.32. The solid lines represent the fit (\protect\ref{eq:ratio}) applied in the time interval where the ground-state is dominant (see Table \ref{tab:timeint}).}
\label{fig:deltaCJS}
\end{figure}

\begin{figure}[htb!]
\begin{center}
\includegraphics[scale=0.75]{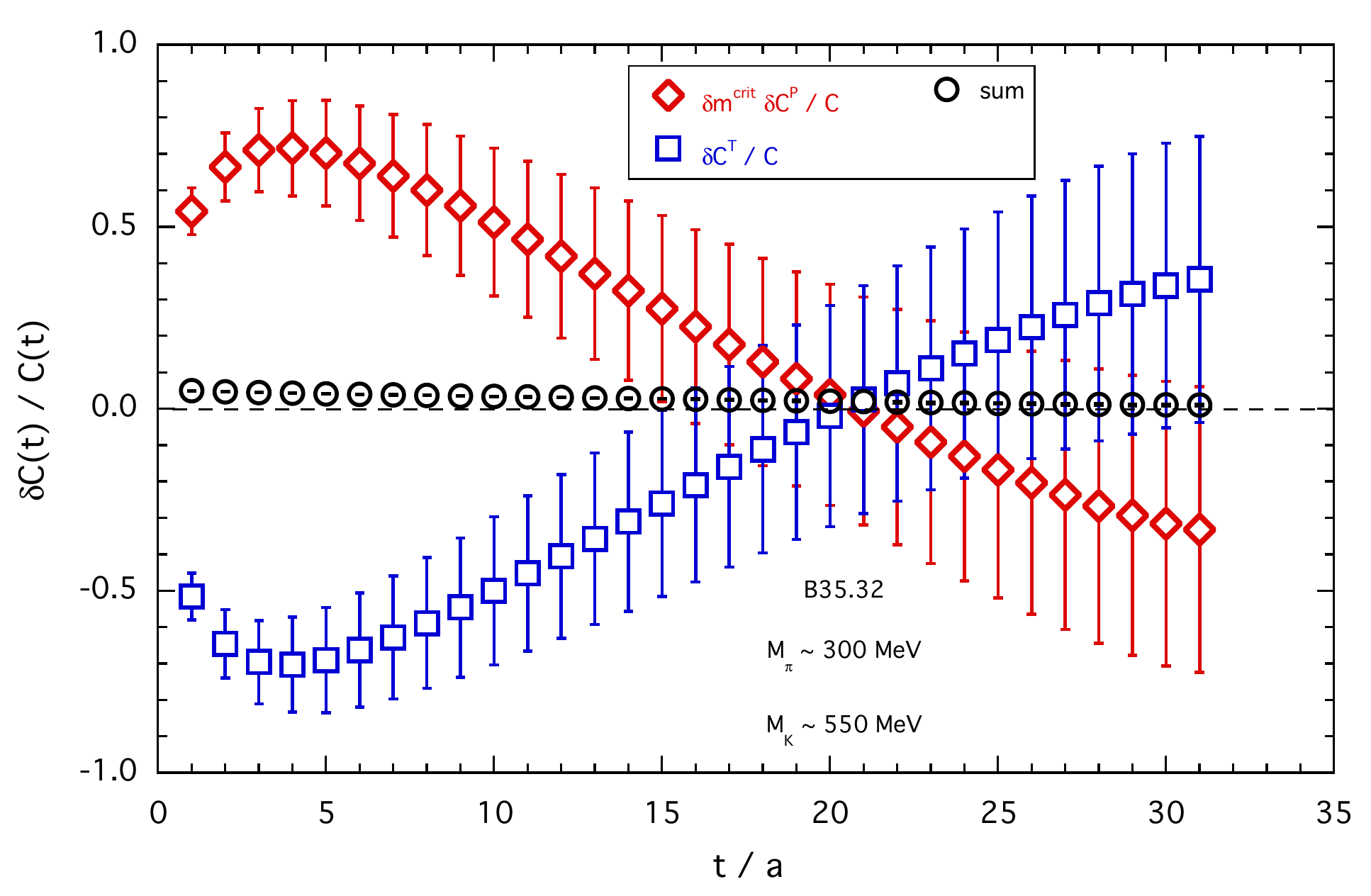}
\end{center}
\vspace{-0.75cm}
\caption{\it \small Ratios $\delta C^T(t) / C(t)$ and $\sum_f \delta m_f^{crit} ~ \delta C^{P_f}(t) / C(t)$ in the case of the charged kaon for the gauge ensemble B35.32. Their sum, shown by the circles, is determined quite precisely.}
\label{fig:deltaCTP}
\end{figure}

In this section we show some plots of the ratios $\delta C^j(t) / C(t)$, used in Eq.~(\ref{eq:ratio}) in order to extract the IB corrections $\delta M^j_{PS}$from the corresponding slopes. 
In Fig.~\ref{fig:deltaCJS} in the case of the kaon for the ensemble B35.32 we show the ratios $\delta C^J(t) / C(t)$ (exchange and self-energy contributions (\ref{fig:diagrams}a)-(\ref{fig:diagrams}b)) and $\delta C^S(t) / C(t)$ (scalar insertion (\ref{fig:diagrams}e)) together with the almost linear fitting curve of Eq.~(\ref{eq:ratio}), performed in the time interval where the ground-state is dominant. 
In Fig.~\ref{fig:deltaCTP} the contributions of the tadpole diagram (\ref{fig:diagrams}c) and of the shift of the critical mass are shown separately. 
It can be seen that the two terms are almost opposite. 
Thanks to the strong correlations due to the dominance of the tadpole contribution in $\delta m^{crit}$ (see Fig.~\ref{fig:deltam_crit}), their sum can be determined with a good precision and turns out to be small compared with the contributions of the self-energy and exchange diagrams.

\section{Analysis of the pion mass splitting $M_{\pi^+} - M_{\pi^0}$}
\label{sec:pion}

According to Ref.~\cite{deDivitiis:2013xla} the charged/neutral pion mass splitting $M_{\pi^+}^2 - M_{\pi^0}^2$ is given by
 \be
    M_{\pi^+}^2 - M_{\pi^0}^2 = 4 \pi \alpha_{em} ~ (q_u - q_d)^2 ~ M_{\pi} ~ \partial_t ~ \frac{\gdllexch - \discgdllexch}{\gdll} ~ ,
    \label{eq:pion_splitting}
 \ee
where, following the notation of Ref.~\cite{deDivitiis:2013xla}, ($-\partial_t$) stands for the operator corresponding to the extraction of the slope $\delta M_{PS}$ from the ratio $\delta C(t) / C(t)$ (see Eq.~(\ref{eq:ratio})).

At first order in the perturbative expansion the pion mass splitting $M_{\pi^+} - M_{\pi^0}$ is a pure e.m.~effect.
Indeed, the strong IB corrections coming from the variation of quark masses do not contribute at leading order to observables that vanish in the isosymmetric theory, like the mass splitting $M_{\pi^+} - M_{\pi^0}$. 
Furthermore all the disconnected diagrams generated by the sea quark charges cancel out in the difference $M_{\pi^+} - M_{\pi^0}$ and therefore Eq.~(\ref{eq:pion_splitting}) holds as well in {\it unquenched} QED.
The only remaining {\it disconnected} diagram in Eq.~(\ref{eq:pion_splitting}) is generated by valence quarks in the neutral pion. 
It vanishes in the $SU(2)$ chiral limit~\cite{deDivitiis:2013xla} and, consequently, it is of order of $O(\alpha_{em} m_{\ell})$.
Thus, at the physical pion mass the disconnected contribution to the pion mass splitting $M_{\pi^+} - M_{\pi^0}$ is expected to be numerically a small correction and has been neglected in the present study.

\begin{figure}[htb!]
\begin{center}
\includegraphics[scale=0.65]{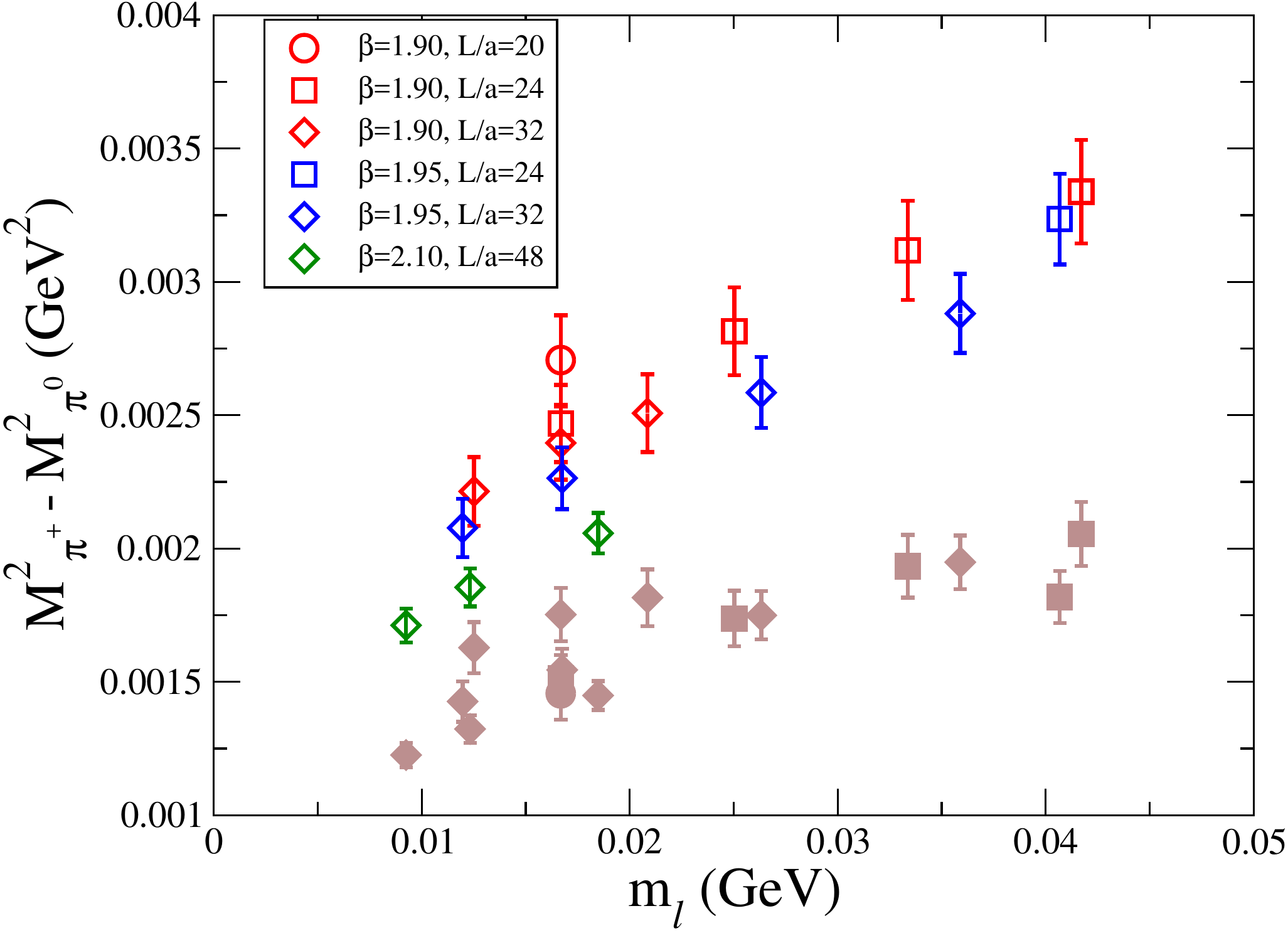}
\end{center}
\vspace{-0.70cm}
\caption{\it \small Results for the pion mass splitting $M_{\pi^+}^2 - M_{\pi^0}^2$ versus the renormalized light-quark mass $m_\ell$, obtained using Eq.~(\ref{eq:pion_splitting}) and neglecting the contribution coming from the {\it disconnected} diagram. Brown full points correspond to the data without any correction for FSEs, while open markers represent the lattice data subtracted by the universal FSEs given by Eq.~(\ref{eq:universal_FSE}).}
\label{fig:dM2Pi_univ}
\end{figure}

Disregarding the disconnected diagram in the r.h.s.~of Eq.~(\ref{eq:pion_splitting}), the results for $M_{\pi^+}^2 - M_{\pi^0}^2$ are shown in Fig.~\ref{fig:dM2Pi_univ} for the ETMC gauge ensembles of Table \ref{tab:simudetails} as a function of the renormalized light-quark mass $m_\ell$.

Putting a massless photon in a finite box yields sizeable finite size effects (FSEs), which have been investigated in Ref.~\cite{Hayakawa:2008an}, using $QED_L$ for the infrared regularization, and for other choices of the zero-mode subtraction in Ref.~\cite{Borsanyi:2014jba}.
The main outcome is that FSEs on hadron masses start at order ${\cal{O}}(1/L)$ and they are {\it universal} up to order ${\cal{O}}(1/L^2)$, i.e.~they depend only on the charge of the hadron and not on its structure.
In the case of $QED_L$ the universal FSEs are given by
 \be
    M_{PS^Q}^2(L) - M_{PS^Q}^2(\infty) = - Q^2 \alpha_{em} \frac{\kappa}{L^2}\left( 1 + 2 M_{PS} L \right)
     \label{eq:universal_FSE}
 \ee
where $\kappa = 2.837297$~\cite{Hayakawa:2008an}.
The universal FSEs are thus present only for the charged pion.
The effect of their subtraction from our lattice data is shown in Fig.~\ref{fig:dM2Pi_univ} by the open markers.
It can be clearly seen that the correction is quite large, approaching $\simeq 40 \%$ at the heaviest light-quark masses.
\begin{figure}[htb!]
\begin{center}
\includegraphics[scale=0.75]{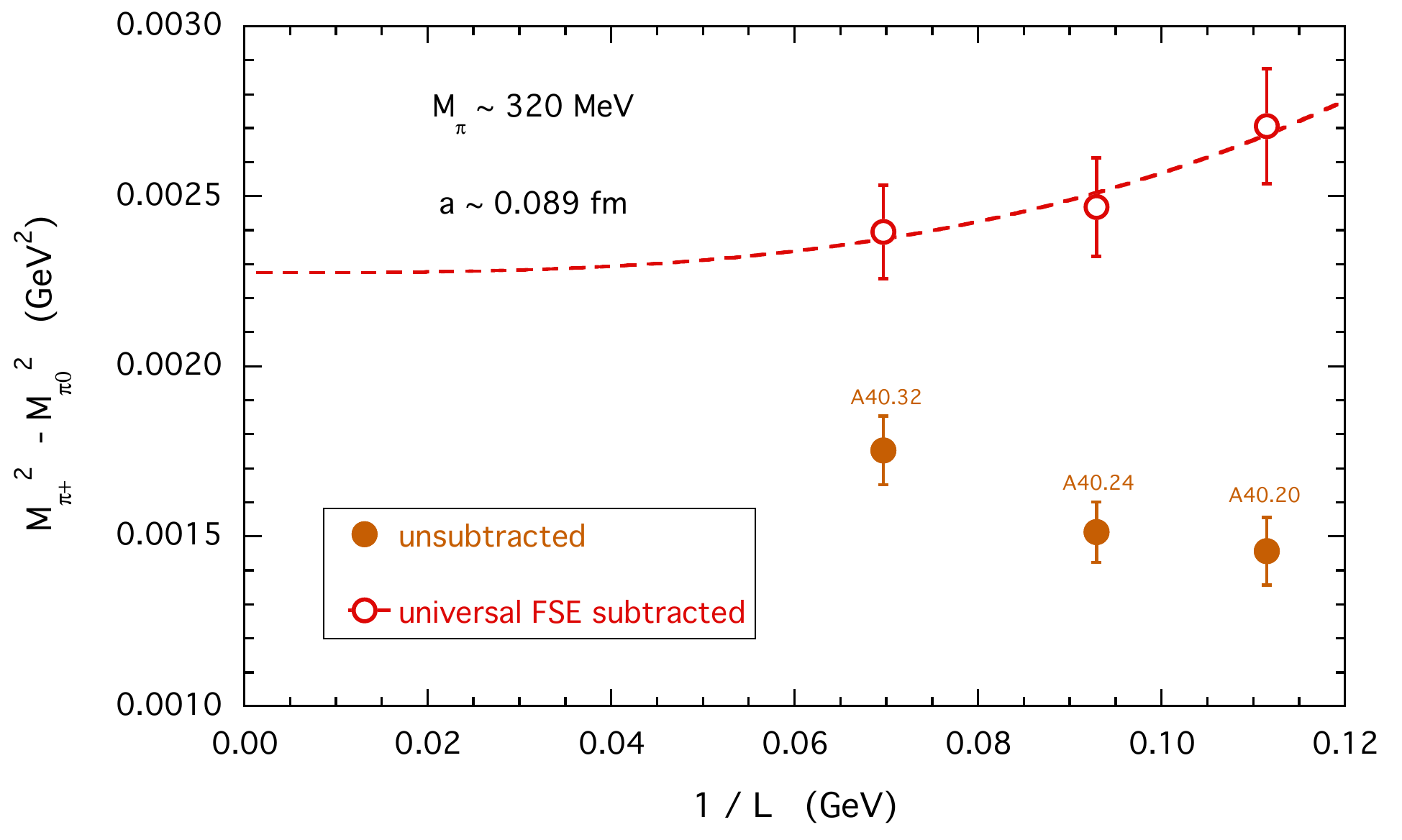}
\vspace{-1.0cm}
\end{center}
\caption{\it \small Results for the pion mass splitting $M_{\pi^+}^2 - M_{\pi^0}^2$ for the gauge ensembles A40.20, A40.24 and A40.32, which share a common value of the pion mass and the lattice spacing, but differ for the lattice size $L$. The brown full points correspond to the data without any correction for FSEs, while the open dots represent the lattice data corrected by the universal FSEs given by Eq.~(\ref{eq:universal_FSE}). The dotted line corresponds to the result of a simple linear fit in $1 / L^3$ (see Eq.~(\ref{eq:pion_savage})).}
\label{fig:FSE_pion}
\end{figure}
In Fig.~\ref{fig:FSE_pion} the data corresponding to the gauge ensembles A40.20, A40.24 and A40.32, which share a common value of the pion mass and the lattice spacing, but differ for the lattice size $L$, are shown.
The presence of residual FSEs after the subtraction of the universal ones is visible, but its impact does not exceed few percent at the largest lattice sizes.
According to the non-relativistic expansion of Ref.~\cite{Davoudi:2014qua}, the structure-dependent (SD) FSEs are expected to be proportional at order $O(1/L^3)$ to the squared pion charge radius $\langle r^2 \rangle_{\pi^+}$, namely
 \be
    \left[ M_{\pi^+}^2(L) - M_{\pi^0}^2(L) \right]^{(SD)} = F \frac{4 \pi \alpha_{em}}{3} \frac{M_\pi}{L^3} \langle r^2 \rangle_{\pi^+} +
                                                                                       {\cal{O}}(\frac{1}{L^4}, \frac{M_\pi}{L^4}),
    \label{eq:pion_savage}
 \ee
where at the physical pion mass $\langle r^2\rangle_{\pi^+} = (0.672 \pm 0.008 ~ \mbox{fm})^2$~\cite{PDG}. 
In Eq.~(\ref{eq:pion_savage}) we have included the multiplicative factor $F$ to account for possible deviations from the theoretical expectation.
The lattice data can be fitted by Eq.~(\ref{eq:pion_savage}) with $F = 2.9 \pm 0.3$, as shown by the dashed line in Fig.~\ref{fig:FSE_pion}.
This highlights a significative deviation of the observed residual SD FSEs from the non-relativistic result.

From now on we always refer to the data for $M_{\pi^+}^2 - M_{\pi^0}^2$ as to the charged/neutral pion mass splitting subtracted by the universal FSEs (\ref{eq:universal_FSE}).

Inspired by the Chiral Perturbation Theory (ChPT) analysis of Ref.~\cite{Hayakawa:2008an}, we perform combined extrapolations to the physical pion mass and to the continuum and infinite volume limits adopting the following fitting function
 \bea
     M_{\pi^+}^2 - M_{\pi^0}^2 & = & 4 \pi \alpha_{em} f_0^2 \left\{ 4 \frac{C}{f_0^4} - \left( 3 + 16 \frac{C}{f_0^4} \right) \frac{\overline{M}^2}{16\pi^2 f_0^2} 
                                                         \log\left(\frac{\overline{M}^2}{16 \pi^2 f_0^2}\right) \right. \nonumber\\
	 				    & + & \left. A_1^\pi \frac{\overline{M}^2}{16 \pi^2 f_0^2} + A_2^\pi \frac{\overline{M}^4}{(4\pi f_0)^4} \right\} + 
					              D^\pi a^2 + D_m^\pi a^2 m_{\ell} \nonumber \\
					    & + & \frac{4 \pi \alpha_{em}}{3} \frac{M_\pi}{L^3} \langle r^2 \rangle_{\pi^+} + F^\pi a^2 \frac{M_\pi}{L^3}
    \label{eq:pion_fit}
 \eea
where $\overline{M}^2 \equiv 2B_0 m_{\ell}$, $B_0$ and $f_0$ are the QCD low-energy constants (LECs) at leading order (LO),
$C$ is the e.m.~LEC at LO, $A_1^\pi$ is a combination of the e.m.~LECs at order $\mathcal{O}(\alpha_{em} m_{\ell})$ (at a ChPT renormalization scale equal to $4 \pi f_0$), $A_2^\pi$ is an effective NNLO LEC, $D^\pi$ and $D_m^\pi$ are fitting parameters that take into account discretization effects.
In Eq.~(\ref{eq:pion_fit}) the SD FSEs are represented by the last two terms in its r.h.s.: the first one is directly given by the non-relativistic result of Ref.~\cite{Davoudi:2014qua}, while the second term, expected from the FSEs related to a heavy intermediate state with mass $\propto 1 / a$~\cite{Tantalo:2016vxk}, is added as a correction with a fitting multiplicative parameter $F^\pi$.

\begin{figure}[htb!]
\begin{center}
\includegraphics[scale=0.65]{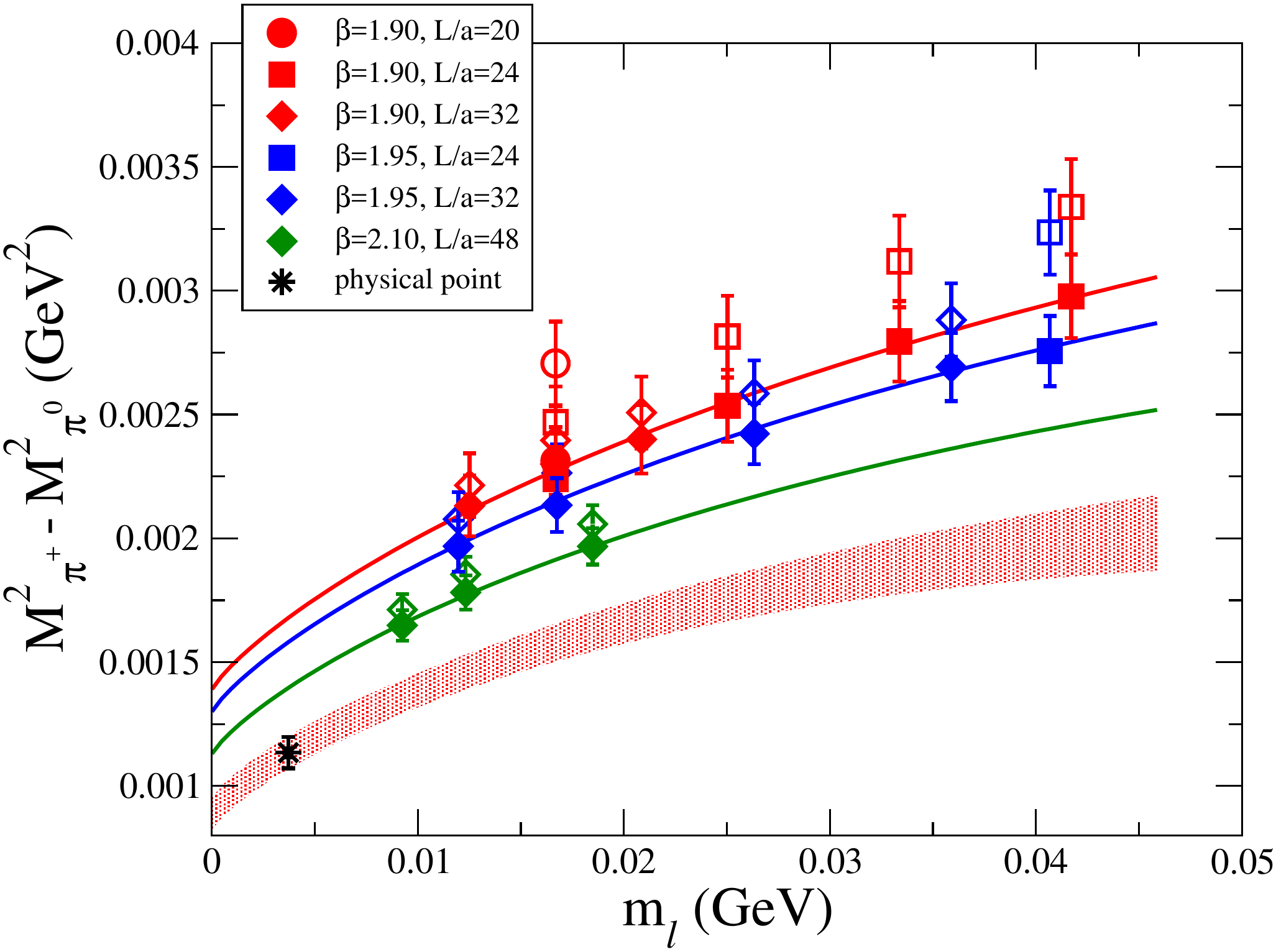}
\end{center}
\vspace{-0.70cm}
\caption{\it \small Results for the pion mass splitting $M_{\pi^+}^2 - M_{\pi^0}^2$ versus the renomalized light-quark mass $m_\ell$. The empty markers correspond to the data after the subtraction of the universal FSEs, while the filled markers represent the lattice data corrected also by the SD FSEs obtained in the fitting procedure (\ref{eq:pion_fit}). The solid lines correspond to the results of the combined fit (\ref{eq:pion_fit}) assuming $A_2^\pi = 0$ obtained in the infinite volume limit at each value of the lattice spacing. The black asterisk represents the pion mass splitting extrapolated at the physical pion mass (corresponding to $m_\ell = m_{ud} = 3.70 (17)~\mbox{MeV}$) and to the continuum limit, while the red area indicates the corresponding uncertainty as a function of $m_\ell$ at the level of one standard deviation.}
\label{fig:dM2Pi_full}
\end{figure}

In Fig.~\ref{fig:dM2Pi_full} the results obtained using the combined fitting function (\ref{eq:pion_fit}) assuming $A_2^\pi = 0$ are shown, i.e.~with $C$, $A_1^\pi$, $D^\pi$, $D_m^\pi$ and $F^\pi$ being free parameters.

As for the lattice spacing $a$ and the renormalization constants $Z_P$, their uncertainties (see Table~\ref{tab:8branches}) are taken into account as follows. 
First, we randomly generate the values $a^i$ and $Z_P^i$ for the bootstrap event $i$ assuming gaussian distributions corresponding to the central values and the standard deviations of Table \ref{tab:8branches}.
Then, we add to the definition of the $\chi^2$ variable the following contribution
 \be
    \sum\nolimits_\beta \frac{\left( \overline{a}^ i  - a^i \right)^2}{\sigma_a^2}  + 
    \sum\nolimits_\beta \frac{\left( \overline{Z}_P^i  - Z_P^i \right)^2}{\sigma_{Z_P}^2} ~ ,
    \label{eq:chi2term}
 \ee
where $\overline{a}^i$ and $\overline{Z}_P^i$ are free parameters of the fitting procedure. 
The use of Eq.~(\ref{eq:chi2term}) allows the quantities $a$ and $Z_P$ to slightly change from their central values (in the given bootstrap event) with a weight in the $\chi^2$ given by their uncertainties.
This procedure corresponds to impose a gaussian prior for $a$ and $Z_P$.

At the physical pion mass and in the continuum and infinite volume limits our result is
\bea
    M_{\pi^+}^2 - M_{\pi^0}^2 & = & 1.137 ~ (63)_{stat+fit} ~ (24)_{disc} ~ (22)_{chir} ~ (10)_{FSE} \cdot 10^{-3} ~ \mbox{GeV}^2 ~ , \nonumber \\
                                             & = & 1.137 ~ (63)_{stat+fit} ~ (34)_{syst} \cdot 10^{-3} ~ \mbox{GeV}^2 ~ , \nonumber \\
                                             & = & 1.137 ~ (72) \cdot 10^{-3} ~ \mbox{GeV}^2 ~ , 
    \label{eq:pion_result}
\eea
where
\begin{itemize}
\item $()_{stat+fit}$ indicates the statistical uncertainty including also the ones induced by the fitting procedure and by the errors of the input parameters of Table \ref{tab:8branches}, namely the values of the average $u/d$ quark mass $m_{ud}$, the lattice spacing and the quark mass RC $1 / Z_P$.
\item $()_{disc}$ is the uncertainty due to discretization effects estimated by comparing the results obtained either including or excluding the $D_m^\pi a^2 m_{\ell}$ term in Eq.~(\ref{eq:pion_fit});
\item $()_{chir}$ is the error coming from including ($A_2^\pi \neq 0$) or excluding ($A_2^\pi = 0$) the term proportional to $m_\ell^2$ in Eq.~(\ref{eq:pion_fit});
\item $()_{FSE}$ is the uncertainty due to FSEs estimated by comparing the results obtained including or excluding the two SD terms in Eq.~(\ref{eq:pion_fit}). In the latter case only the ensembles with $L / a = 32, ~ 48$ have been included in the fit.
\end{itemize}

Our result (\ref{eq:pion_result}) implies
\bea
    M_{\pi^+} - M_{\pi^0} & = & 4.21 ~ (23)_{stat+fit} ~ (13)_{syst} ~ \mbox{MeV} ~ , \nonumber \\
                                      & = & 4.21 ~ (26) ~ \mbox{MeV} ~ , 
\eea
which agrees with the experimental determination
\be
    \left[ M_{\pi^+} - M_{\pi^0} \right]^{exp} = 4.5936 ~ (5) ~ \mbox{MeV}
    \label{eq:pion_exp}
\ee
within $\approx 1.5$ standard deviations.
The difference among the central values, which is equal to $\approx 8 \%$, may be of statistical origin, but it may be due also to the disconnected contribution at order ${\cal{O}}(\alpha_{em} m_{\ell})$ in Eq.~(\ref{eq:pion_splitting}) as well as to possible higher-order effects proportional to $\alpha_{em} (\widehat{m}_d - \widehat{m}_u)$ and to $(\widehat{m}_d - \widehat{m}_u)^2$, which have been neglected.
The latter ones are estimated to be of the order of $\simeq 4 \%$ in Ref.~\cite{FLAG} and therefore the disconnected contribution at order ${\cal{O}}(\alpha_{em} m_{\ell})$ is expected to be of the same size $\approx 4 \%$, which corresponds to $\approx 0.2$ MeV.

\section{Determination of $\epsilon_{\pi^0}$}
\label{sec:pi0}
 
The Dashen's theorem~\cite{Dashen:1969eg} states that in the chiral limit the self-energies of the neutral Nambu-Goldstone bosons vanish. 
Thus, the violation of the Dashen's theorem in the pion sector can be measured through the quantity $\epsilon_{\pi^0}$ defined as~\cite{FLAG}
\be
    \epsilon_{\pi^0}  = \frac{\left[\delta M^2_{\pi^0} \right]^{QED}}{M_{\pi^+}^2 - M_{\pi^0}^2} ~ .
    \label{eq:epsilon_pi0}
\ee
In our analysis the e.m.~contribution $\left[ \delta M^2_{\pi^0} \right]^{QED}$ is computed in the quenched QED approximation and neglecting also the disconnected diagram of Eq.~(\ref{eq:pion_splitting}), namely
\be
     \left[\delta M^2_{\pi^0}\right]^{QED} = 8 \pi \alpha_{em} M_\pi \left[\delta M_{\pi^0}\right]^{em} ~ ,
\ee
where
\bea
    \left[\delta M_{\pi^0}\right]^{em} & = & - \frac{q_u^2+q_d^2}{2}  \partial_t\frac{\gdllexch}{\gdll} -(q_u^2+q_d^2) \partial_t \frac{\gdllself + \gdllphtad}{\gdll}
                                                                \nonumber \\
                                                       & - & (\delta m^{crit}_u + \delta m^{crit}_d) \partial_t \frac{\gdlip}{\gdll} + Z_P \left( \frac{1}{{\cal{Z}}_u} + \frac{1}{{\cal{Z}}_d} \right) 
                                                                m_\ell \partial_t \frac{\gdli}{\gdll} ~ .
    \label{eq:mpi0}
\eea

\begin{figure}[htb!]
\begin{center}
\includegraphics[scale=0.65]{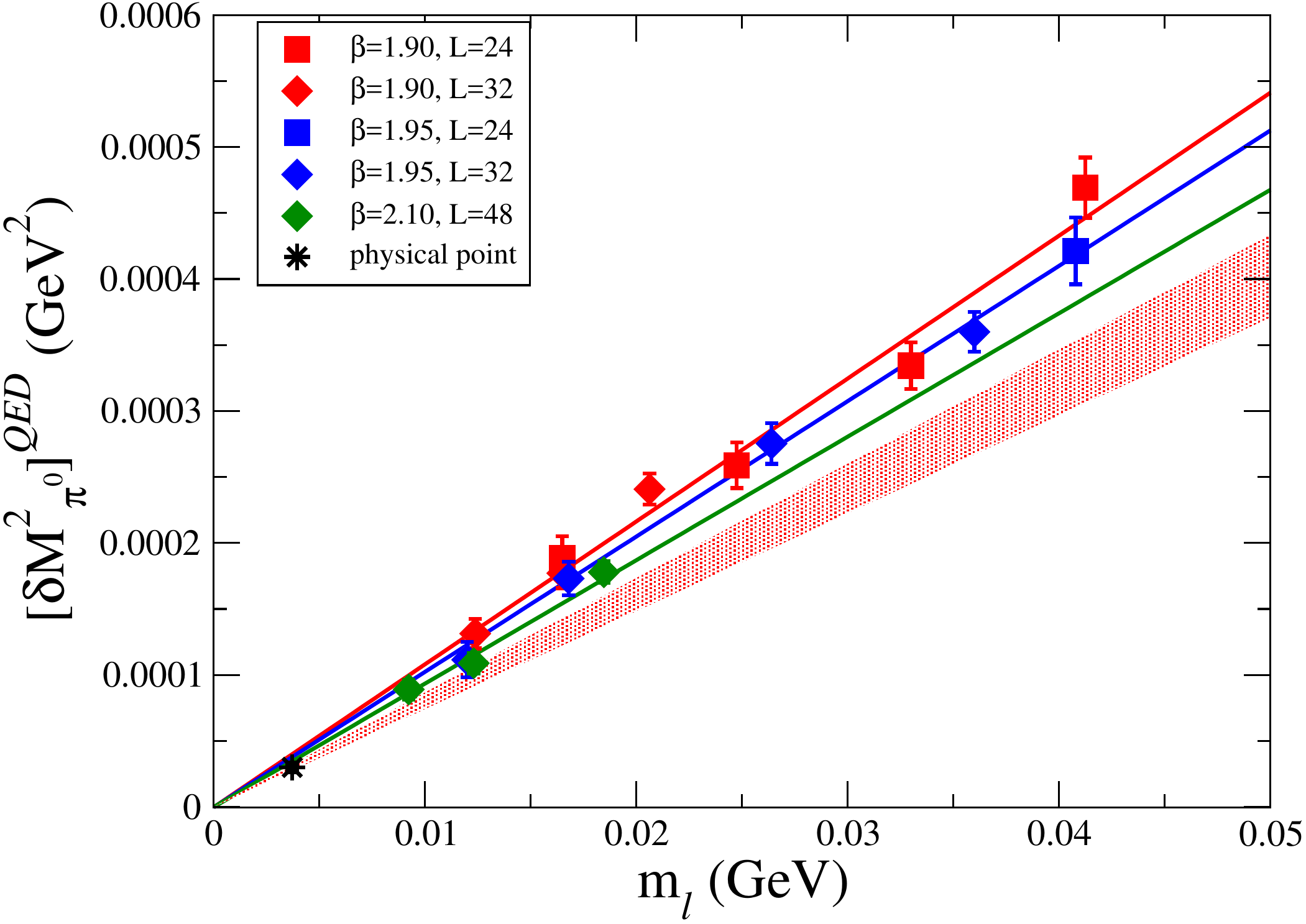}
\end{center}
\vspace{-0.70cm}
\caption{\it \small Results for the quantity $\left[ \delta M^2_{\pi^0} \right]^{QED}$ versus the renomalized light-quark mass $m_\ell$. The filled markers represent the lattice data without FSE corrections. The solid lines correspond to the results of the combined fit (\ref{eq:M2Pi0g_fit}) assuming $\widetilde{A}_2^\pi = 0$ obtained at each value of the lattice spacing. The black asterisk represents the value extrapolated at the physical pion mass $m_\ell = m_{ud} = 3.70 (17)~\mbox{MeV}$ and to the continuum limit, while the red area indicates the corresponding uncertainty as a function of $m_\ell$ at the level of one standard deviation.}
\label{fig:M2Pi0g}
\end{figure}

The lattice data for $\left[ \delta M^2_{\pi^0} \right]^{QED}$ are shown by filled markers in Fig.~\ref{fig:M2Pi0g}.
It can be seen that the data exhibit an almost linear behavior as a function of the light-quark mass $m_\ell$ without any significant FSEs.
Thus for the combined chiral and continuum limit extrapolations we use the following simple Ansatz
\be
    \left[\delta M^2_{\pi^0}\right]^{QED} = \widetilde{A}_1^\pi ~ \frac{\overline{M}^2}{16\pi^2 f_0^2} \left( 1 + 
                                                                \widetilde{A}_2^\pi ~ \frac{\overline{M}^2}{16\pi^2 f_0^2} \right) +
                                                                \widetilde{D}^\pi a^2 + \widetilde{D}_m^\pi a^2m_{\ell},
    \label{eq:M2Pi0g_fit}
\ee
where $\overline{M}^2 \equiv 2 B_0 m_\ell$ and $\widetilde{A}_1^\pi$, $\widetilde{A}_2^\pi$, $\widetilde{D}^\pi$ and $\widetilde{D}_m^\pi$ are free parameters.
The results of the fitting procedure assuming $\widetilde{A}_2^\pi = 0$ are shown in Fig.~\ref{fig:M2Pi0g} by the solid lines at each value of the lattice spacing and by the black asterisk at the physical pion mass and in the continuum limit. 

At the physical pion mass and in the continuum limit we obtain
\bea
      \left[\delta M^2_{\pi^0}\right]^{QED} & = & 0.032 ~ (3)_{stat+fit} ~ (2)_{chir} ~ (2)_{disc} ~ (50)_{qQED} \cdot 10^{-3}~\mbox{GeV}^2 ~ , \nonumber \\
                                                               & = & 0.032 ~ (3)_{stat+fit} ~ (3)_{syst} ~ (50)_{qQED} \cdot 10^{-3}~\mbox{GeV}^2 ~ , \nonumber \\
                                                               & = & 0.032 ~ (50) \cdot 10^{-3}~\mbox{GeV}^2 ~ ,
      \label{eq:M2Pi0g_result}
\eea
where
\begin{itemize}
\item $()_{stat+fit}$ indicates the statistical uncertainty including also the ones induced by the fitting procedure and by the determination of the input parameters of Table \ref{tab:8branches};
\item $()_{chir}$ is the error coming from including ($\widetilde{A}_2^\pi \neq 0$) or excluding ($\widetilde{A}_2^\pi = 0$) the quadratic term;
\item $()_{disc}$ is the uncertainty due to discretization effects estimated by comparing the results obtained including both the $\widetilde{D}^\pi a^2$ and $\widetilde{D}_m^\pi a^2 m_{\ell}$ terms in Eq.~(\ref{eq:M2Pi0g_fit}) or excluding one out of them.
\item $()_{qQED}$ comes from our estimate of the neglect of the neutral pion, disconnected diagram ($0.05 \cdot 10^{-3}$ GeV$^2$), which dominates over all other uncertainties.
\end{itemize}

Using the experimental value $M_{\pi^0} = 134.98$ MeV \cite{PDG} our result (\ref{eq:M2Pi0g_result}) corresponds to a pion mass in pure QCD equal to $M_\pi = 134.9 (2)$ MeV in agreement with the FLAG estimate $M_\pi = 134.8 (3)$ MeV.

Dividing our result (\ref{eq:M2Pi0g_result}) by Eq.~(\ref{eq:pion_result}), we obtain
 \bea
     \epsilon_{\pi^0} & = & 0.028 ~~ (3)_{stat+fit} ~ (2)_{disc} ~ (3)_{chir} ~ (1)_{FSE} ~ (44)_{qQED} ~ , \nonumber \\
                              & = & 0.028 ~~ (3)_{stat+fit} ~ (4)_{syst} ~ (44)_{qQED} ~ , \nonumber \\
                              & = & 0.028 ~ (44) ~ ,
     \label{eq:epsilon_pi0_result}
 \eea
which is consistent with the FLAG estimate $\epsilon_{\pi^0} = 0.07 (7)$~\cite{FLAG}, based on the old determination of Ref.~\cite{Duncan:1996xy} (corrected by FLAG into the value $\epsilon_{\pi^0} = 0.10 (7)$) and on the more recent result $\epsilon_{\pi^0} = 0.03 (2)$ obtained by the QCDSF/UKQCD collaboration~\cite{Horsley:2015vla}.

\section{Analysis of $\epsilon_\gamma$ and determination of $m_d - m_u$}
\label{sec:kaon}

The Dashen's theorem predicts that in the chiral limit the e.m.~corrections to the charged kaon and pion are equal to each other, while the ones for the neutral mesons are vanishing.
Therefore, in the kaon sector the violation of the Dashen's theorem is parameterized in terms of the quantity $\epsilon_\gamma$ defined as~\cite{FLAG}
\be
     \epsilon_\gamma(\overline{\mathrm{MS}}, \mu) = \frac{\left[ M_{K^+}^2 - M_{K^0}^2 \right]^{QED}(\overline{\mathrm{MS}}, \mu)}{M_{\pi^+}^2 - M_{\pi^0}^2} - 1 ~ ,
     \label{eq:epsilon_gamma}
\ee
where $\left[ M_{K^+}^2 - M_{K^0}^2 \right]^{QED}(\overline{\mathrm{MS}}, \mu)$ is the QED contributiparametrisedon to the kaon mass splitting.
Within the quenched QED approximation one has
 \be
    \left[ M_{K^+}^2 -  M_{K^0}^2 \right]^{QED} = 8 \pi \alpha_{em} M_K \left[ M_{K^+} - M_{K^0} \right]^{em} ~ ,
    \label{eq:kaon2_qed}
 \ee
where
\bea
    \left[ M_{K^+} - M_{K^0} \right]^{em} & = & - q_s (q_u - q_d)  \partial_t \frac{\gdslexch}{\gdsl} - (q_u^2 - q_d^2) \partial_t \frac{\gdslselfl + \gdslphtadl}{\gdsl} \nonumber \\
           & - & (\delta m^{crit}_u - \delta m^{crit}_d) \partial_t \frac{\gdsip}{\gdsl} - Z_P  \left( \frac{1}{{\cal{Z}}_d} - \frac{1}{{\cal{Z}}_u} \right) m_\ell \partial_t \frac{\gdsi}{\gdsl} 
    \label{eq:kaon_qed}
\eea
with the red lines representing the strange quark propagator.

\begin{figure}[htb!]
\begin{center}
\includegraphics[scale=0.65]{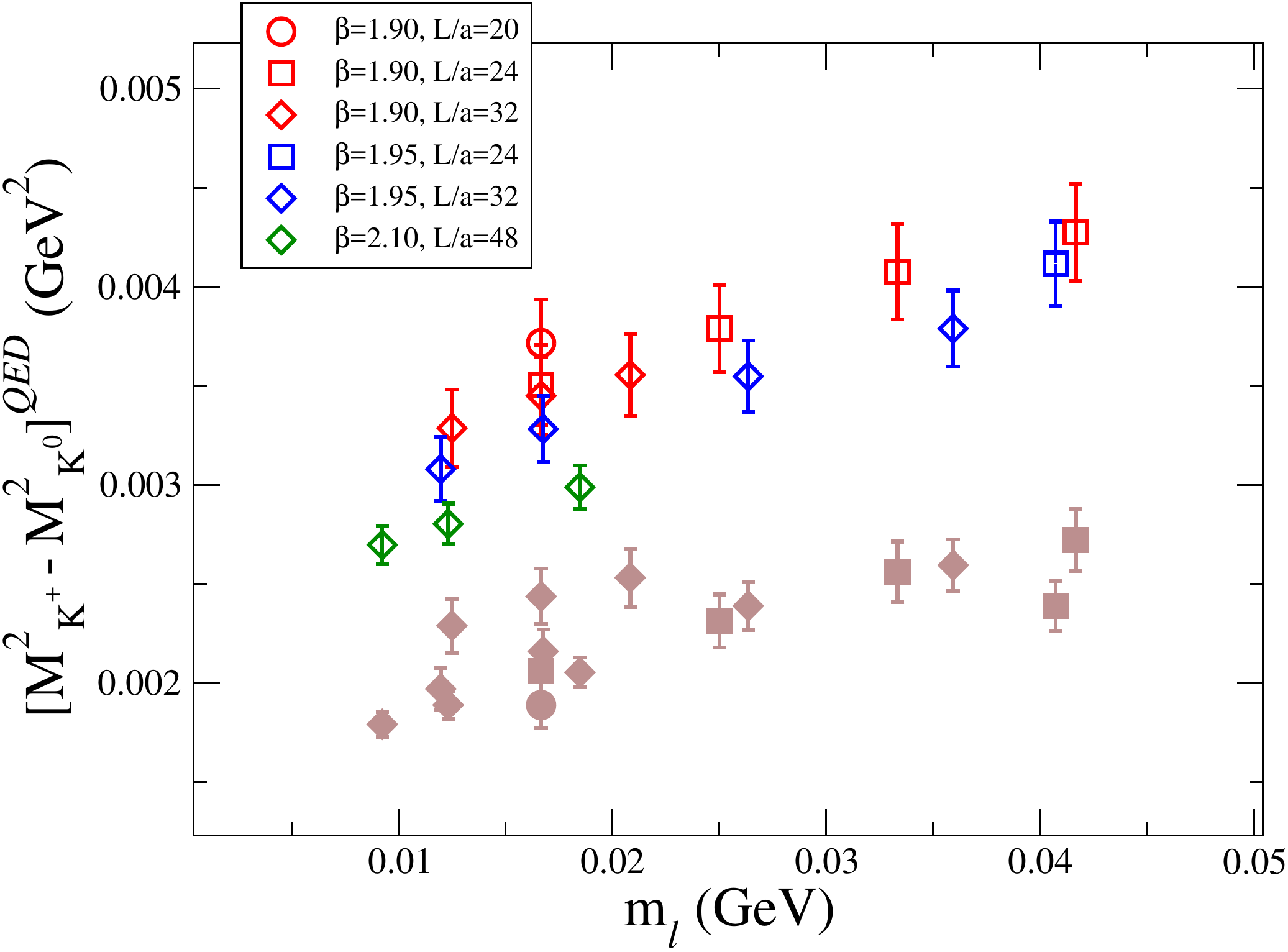}
\end{center}
\vspace{-0.70cm}
\caption{\it \small Results for the kaon mass splitting $\left[ M_{K^+}^2 - M_{K^0}^2 \right]^{QED}$ versus the renormalized light-quark mass $m_\ell$, obtained using Eq.~(\ref{eq:kaon_qed}) in the quenched QED approximation. Brown full points correspond to the data without any correction for FSEs, while open markers represent the lattice data corrected by the universal FSEs given by Eq.~(\ref{eq:universal_FSE}).}
\label{fig:dM2K_QED_univ}
\end{figure}

The results for $\left[ M_{K^+}^2 - M_{K^0}^2 \right]^{QED}$ are shown in Fig.~\ref{fig:dM2K_QED_univ} with and without the subtraction of the universal FSEs, given by Eq.~(\ref{eq:universal_FSE}).
It can be clearly seen that, as in the case of the pion mass splitting, the universal FSE correction is quite large, approaching $\simeq 40 \%$ at the heaviest light-quark masses.

From now on we always refer to the data for $\left[ M_{K^+}^2 - M_{K^0}^2 \right]^{QED}$ as to the QED part of the charged/neutral kaon mass splitting subtracted by the universal FSEs.

\begin{figure}[htb!]
\begin{center}
\includegraphics[scale=0.65]{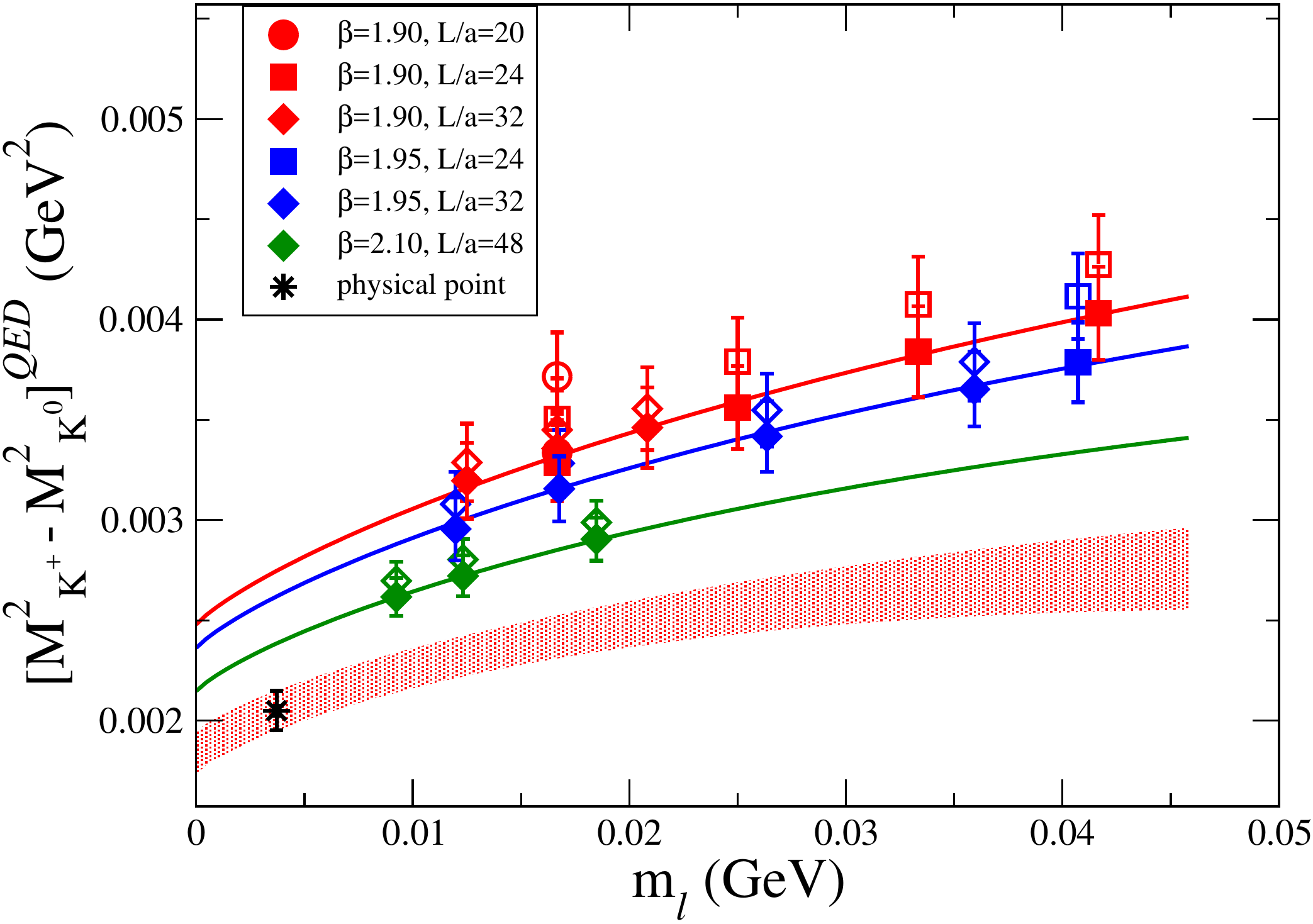}
\end{center}
\vspace{-0.70cm}
\caption{\it \small Results for the kaon mass splitting $\left[ M_{K^+}^2 - M_{K^0}^2 \right]^{QED}$ versus the renormalized light-quark mass $m_\ell$ in the $\overline{MS}$ scheme at a renormalization scale equal to $\mu = 2$ GeV. The empty markers correspond to the data after the subtraction of the universal FSEs, while the filled markers represent the lattice data corrected also by the SD FSEs obtained in the fitting procedure (\ref{eq:kaon_qed_fit}). The solid lines correspond to the results of the combined fit (\ref{eq:kaon_qed_fit}) assuming $A_2^K = 0$ obtained in the infinite volume limit at each value of the lattice spacing. The black asterisk represents the kaon mass splitting extrapolated at the physical pion mass $m_\ell = m_{ud} = 3.70 (17)~\mbox{MeV}$ and to the continuum limit, while the red area indicates the corresponding uncertainty as a function of $m_\ell$ at the level of one standard deviation.}
\label{fig:dM2K_QED_full}
\end{figure}

Inspired by the ChPT analysis of Ref.~\cite{Hayakawa:2008an} we perform combined extrapolations to the physical pion mass and to the continuum and infinite volume limits adopting the following fitting function
\bea
     \left[ M_{K^+}^2 - M_{K^0}^2 \right]^{QED} & = & 16 \pi \alpha_{em} \frac{C}{f_0^2} \left[ A_0^K - \frac{8}{3} \frac{\overline{M}^2}{16\pi^2 f_0^2} 
                      \log\left(\frac{\overline{M}^2}{16 \pi^2 f_0^2}\right) \right. \nonumber\\
            & + & \left. A_1^K \frac{\overline{M}^2}{16\pi^2 f_0^2} + A_2^K \frac{\overline{M}^4}{(4 \pi f_0)^4} \right] + 
                     D^K a^2 + D_m^K a^2 m_{\ell} \nonumber \\
            & + & \frac{4 \pi \alpha_{em}}{3} \frac{M_K}{L^3} \langle r^2 \rangle_{K^+} + F^K a^2 \frac{M_K}{L^3} ~ ,
\label{eq:kaon_qed_fit}
\eea
where the residual SD FSEs are estimated using two terms similar to the ones appearing in Eq.~(\ref{eq:pion_fit}) and with $\langle r^2\rangle_{K^+} = (0.560 \pm 0.031 ~ \mbox{fm})^2$~\cite{PDG}.
The free parameters to be determined by the fitting procedure are $A_0^K$, $A_1^K$, $A_2^K$, $D^K$, $D_m^K$ and $F^K$, while the LEC $C$ is taken from the analysis of the pion mass splitting.
In Fig.~\ref{fig:dM2K_QED_full} we show the results obtained using the combined fitting function (\ref{eq:kaon_qed_fit}) assuming $A_2^K = 0$.
As in the case of the pion mass splitting we obtain a value for the parameter $F^K$ significantly different from zero, which confirms the presence of a deviation from the non-relativistic expansion prediction of Ref.~\cite{Davoudi:2014qua}.

At the physical pion mass and in the continuum and infinite volume limits our result in the $\overline{MS}$ scheme at a renormalization scale equal to $\mu = 2$ GeV is
\bea
    \left[ M_{K^+}^2 - M_{K^0}^2 \right]^{QED} & = & 2.047 ~~ (99)_{stat+fit} ~ (43)_{disc} ~ (23)_{chir} ~ (3)_{FSE} ~ (102)_{qQED} 
                                                                                   \cdot 10^{-3}~\mbox{GeV}^2 ~ , \nonumber \\
                                                                         & = & 2.047 ~~ (99)_{stat+fit} ~ (49)_{syst} ~ (102)_{qQED} \cdot 10^{-3}~\mbox{GeV}^2 ~ , \nonumber \\
                                                                         & = & 2.047 ~ (150) \cdot 10^{-3}~\mbox{GeV}^2 ~ ,
    \label{eq:kaon_qed_result}
\eea
where
\begin{itemize}
\item $()_{stat+fit}$ indicates the statistical uncertainty including also the ones induced by the fitting procedure and by the determination of the input parameters of Table \ref{tab:8branches};
\item $()_{disc}$ is the uncertainty due to discretization effects estimated by comparing the results assuming either $D^K \neq 0$ or $D_m^K = 0$ in Eq.~(\ref{eq:kaon_qed_fit});
\item $()_{chir}$ is the error coming from including ($A_2^K \neq 0$) or excluding ($A_2^K = 0$) the term proportional to $m_\ell^2$;
\item $()_{FSE}$ is the uncertainty due to FSE estimated by comparing the results obtained including or excluding the two phenomenological terms (\ref{eq:kaon_qed_fit}) for the SD FSEs. In the latter case only the ensembles with $L / a = 32, ~ 48$ are considered.
\item $()_{qQED}$ is the estimate of the effects due to the quenched QED approximation ($5 \%)$ taken from Refs.~\cite{Portelli:2012pn,Fodor:2016bgu}.
\end{itemize}
 
Recent results available in the literature for $\left[ M_{K^+}^2 - M_{K^0}^2 \right]^{QED}$ are: $2.075 (395) \cdot 10^{-3}~\mbox{GeV}^2$, obtained using the FLAG inputs~\cite{FLAG}, $2.186 (231) \cdot 10^{-3}~\mbox{GeV}^2$ from the BMW collaboration~\cite{Fodor:2016bgu} at $N_f = 2+1$, and $2.38 (38) \cdot 10^{-3}~\mbox{GeV}^2$ from the latest update of the dispersive analysis of the $\eta \to 3 \pi$ decays~\cite{Colangelo:2016jmc}.
Note that in Ref.~\cite{Fodor:2016bgu} a ``hadronic'' scheme is adopted in which the quark mass difference $(\widehat{m}_d - \widehat{m}_u)$ is replaced by the mass difference of the ``connected'' $\bar{u} u$ and $\bar{d} d$ mesons.
Using our results of Section \ref{sec:pi0} the conversion from the hadronic BMW scheme to the ($\overline{\mathrm{MS}}, 2~\mbox{GeV}$) one amounts to add $0.018(3) \cdot 10^{-3}~\mbox{GeV}^2$ to the result of Ref.~\cite{Fodor:2016bgu}, leading to $\left[ M_{K^+}^2 - M_{K^0}^2 \right]^{QED}(\overline{\mathrm{MS}}, 2~\mbox{GeV}) = 2.204 (231) \cdot 10^{-3}~\mbox{GeV}^2$.
For the other results either the prescription used for evaluating the QED contribution is not clearly defined or the conversion to the ($\overline{\mathrm{MS}}, 2~\mbox{GeV}$) scheme is not known precisely.

Using Eqs.~(\ref{eq:pion_result}) and (\ref{eq:kaon_qed_result}) our estimate for $\epsilon_\gamma$ is
 \bea
    \epsilon_\gamma(\overline{\mathrm{MS}}, 2~\mbox{GeV}) & = & 0.801 ~ (48)_{stat+fit} ~ (8)_{disc} ~ (16)_{chir} ~ (18)_{FSE} ~ (96)_{qQED} , 
                                                                                                           \nonumber \\
                                                                                                 & = & 0.801 ~ (48)_{stat+fit} ~ (25)_{syst} ~ (96)_{qQED} , \nonumber \\
                                                                                                 & = & 0.801 ~ (110) ~ ,
    \label{eq:epsilon_g_result}
 \eea
where now the $()_{qQED}$ error includes also the $4 \%$ effect (added in quadrature) coming from the neglect of the neutral pion, disconnected diagram. 
Our result (\ref{eq:epsilon_g_result}) is consistent with the recent result, converted in the ($\overline{\mathrm{MS}}, 2~\mbox{GeV}$) scheme, $\epsilon_\gamma(\overline{\mathrm{MS}}, 2~\mbox{GeV}) = 0.74 (18)$ from the BMW collaboration~\cite{Fodor:2016bgu} and larger than the recent QCDSF/UKQCD result $\epsilon_\gamma(\overline{\mathrm{MS}}, 2~\mbox{GeV}) = 0.50 (6)$ \cite{Horsley:2015vla} by $\simeq 2.4$ standard deviations.
Note that in Ref.~\cite{Horsley:2015vla} the QED contributions to kaon masses are evaluated in the so-called Dashen scheme, which differs from the ($\overline{\mathrm{MS}}, 2~\mbox{GeV}$) one.
The conversion between the two schemes is taken into account by a perturbative matching performed at leading order in $\alpha_{em}$ in Ref.~\cite{Horsley:2015vla}.

Other results present in the literature are the FLAG estimate $\epsilon_\gamma = 0.7 (3)$~\cite{FLAG} and the two recent findings $\epsilon_\gamma = 0.73 (14)$ from the MILC collaboration~\cite{Basak:2016jnn} and $\epsilon_\gamma = 0.9 (3)$ from the latest update of the dispersive analysis of the $\eta \to 3 \pi$ decays~\cite{Colangelo:2016jmc}.
For the above results either the prescription used for evaluating the QED contribution is not clearly defined or the conversion to the ($\overline{\mathrm{MS}}, 2~\mbox{GeV}$) scheme is not known precisely.

Using the experimental value for the charged/neutral kaon mass splitting, $M_{K^+}^2 - M_{K^0}^2 = -3.903 (3) \cdot 10^{-3}$ GeV$^2$~\cite{PDG}, one gets
 \be
     \left[ M_{K^+}^2 - M_{K^0}^2 \right]^{QCD}(\overline{MS}, 2~\mbox{GeV}) = -5.950 ~ (150) \cdot 10^{-3}~\mbox{GeV}^2 ~ .
     \label{eq:kaon_qcd}
 \ee
In order to estimate the light-quark mass difference $(\widehat{m}_d - \widehat{m}_u)$ from the result (\ref{eq:kaon_qcd}) we need to compute the {\it IB slope} (see Eq.~(\ref{eq:MPS_QCD}))
\be
    \left[ M_{K^+}^2 - M_{K^0}^2 \right]^{IB} \equiv \frac{\left[ M_{K^+}^2 - M_{K^0}^2 \right]^{QCD}}{\widehat{m}_d - \widehat{m}_u} 
                                                                    = - 2 M_K ~ Z_P ~ \partial_t \frac{\gdsi}{\gdsl} ~ .
\ee

The lattice data for $\left[ M_{K^+}^2 - M_{K^0}^2 \right]^{IB}$ have been fitted according to the following Ansatz:
\bea
     \left[ M_{K^+}^2 - M_{K^0}^2 \right]^{IB} & = & \overline{A}_0^K \left[ 1 - \frac{\overline{M}^2}{16\pi^2 f_0^2} \log\left(\frac{\overline{M}^2}{16\pi^2 f_0^2}\right) + 
                                                                               \overline{A}_1^K \frac{\overline{M}^2}{16\pi^2 f_0^2} \right]  ~ \nonumber \\
                                                                     & + & \overline{D}^K a^2 + \overline{F}^K \frac{\overline{M}^2}{16\pi^2 f_0^2} \frac{e^{-\overline{M} L}}{(\overline{M} L)^{3/2}} 
    \label{eq:kaon_qcd_fit}
\eea
where the chiral extrapolation is based on the SU(3) ChPT formulae of Ref.~\cite{Gasser:1984gg} expanded as a power series in terms of the quantity $m_\ell / m_s$, while  
FSEs are described by a phenomenological term inspired by the leading FSE correction in QCD to the pion and kaon masses in the $p$-regime ($\overline{M} L \gg 1$) \cite{Gasser:1986vb}.

\begin{figure}[htb!]
\begin{center}
\includegraphics[scale=0.65]{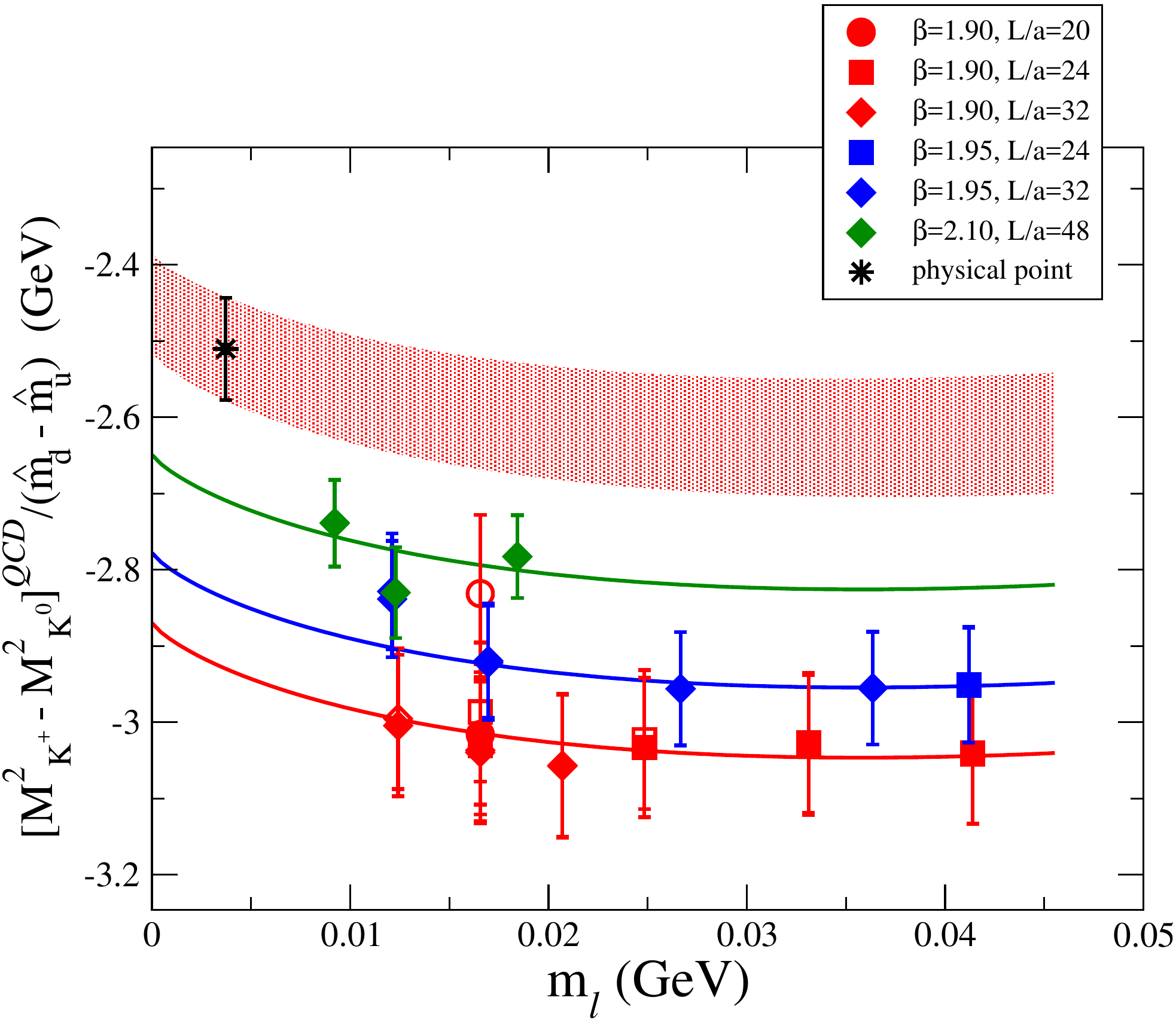}
\end{center}
\vspace{-0.70cm}
\caption{\it \small Results for the IB slope $\left[ M_{K^+}^2 - M_{K^0}^2 \right]^{IB} = \left[ M_{K^+}^2 - M_{K^0}^2 \right]^{QCD} / (\widehat{m}_d - \widehat{m}_u)$ versus the renomalized light-quark mass $m_\ell$. The empty markers correspond to the lattice data, while the filled ones represent the data corrected for the FSEs obtained in the fitting procedure (\ref{eq:kaon_qcd_fit}). The solid lines correspond to the results of the combined fit (\ref{eq:kaon_qcd_fit}) obtained in the infinite volume limit at each value of the lattice spacing. The black asterisk represents the IB slope extrapolated at the physical pion mass $m_\ell = m_{ud} = 3.70 (17)~\mbox{MeV}$ and to the continuum limit, while the red area indicates the corresponding uncertainty as a function of $m_\ell$ at the level of one standard deviation.}
\label{fig:dM2K_QCD}
\end{figure}

The results of the fitting procedure (\ref{eq:kaon_qcd_fit}), using $\overline{A}_0^K$, $\overline{A}_1^K$, $\overline{D}^K$ and $\overline{F}^K$ as free parameters, are shown in Fig.~\ref{fig:dM2K_QCD}.

At the physical pion mass and in the continuum and infinite volume limits we get
\bea
    \left[ M_{K^+}^2 - M_{K^0}^2 \right]^{IB} & = & -2.51 ~ (10)_{stat+fit} ~ (15)_{disc} ~ (1)_{chir} ~ (1)_{FSE} ~ \mbox{GeV} ~ \nonumber \\
            & = & -2.51 ~ (10)_{stat+fit} ~ (15)_{syst} ~ \mbox{GeV} ~ , \nonumber \\
            & = & -2.51 ~ (18) ~ \mbox{GeV} ~ ,
    \label{eq:kaon_qcd_result}
\eea
where
\begin{itemize}
\item $()_{stat+fit}$ indicates the statistical uncertainty including also the ones induced by the fitting procedure and by the determination of the input parameters of Table \ref{tab:8branches};
\item $()_{disc}$ is the uncertainty due to discretization effects estimated by including ($\overline{D}^K \neq 0$) or excluding ($\overline{D}^K = 0$) the discretization term in Eq.~(\ref{eq:kaon_qcd_fit});
\item $()_{chir}$ is the error coming from including the term proportional to the chiral log in Eq.~(\ref{eq:kaon_qcd_fit}) or substituting it with a quadratic term in $m_\ell$ (i.e., $\overline{A}_2^K \overline{M}^4 / (4 \pi f_0)^4$);
\item $()_{FSE}$ is the uncertainty obtained including ($\overline{F}^K \neq 0$) or excluding ($\overline{F}^K = 0$) the FSE term in Eq.~(\ref{eq:kaon_qcd_fit}).
\end{itemize}
Our $N_f = 2+1+1$ result (\ref{eq:kaon_qcd_result}) agrees with the corresponding BMW result, $2.53 (7)$ GeV, obtained at $N_f = 2+1$~\cite{Fodor:2016bgu}.

Putting together the results (\ref{eq:kaon_qcd}) and (\ref{eq:kaon_qcd_result}) with Eq.~(\ref{eq:MPS_QCD}), we get
\bea
    \left[\widehat{m}_d - \widehat{m}_u\right](\overline{MS}, 2~\mbox{GeV}) & = & 2.380 ~~ (87)_{stat+fit} ~ (154)_{disc} ~ (11)_{chir} ~ (11)_{FSE} ~ (41)_{qQED} ~ \mbox{MeV} ~ , 
                                                                                                                                 \nonumber \\
                                                                                                                       & = & 2.380 ~~ (87)_{stat+fit} ~ (155)_{syst} ~ (41)_{qQED} ~ \mbox{MeV} ~ , \nonumber \\
                                                                                                                       & = & 2.380 ~ (182) ~ \mbox{MeV} ~ ,
   \label{eq:deltamud}
\eea
which is consistent with the previous ETMC determination $2.67 (35)$ MeV \cite{Carrasco:2014cwa} at $N_f = 2+1+1$ and with the recent BMW result, converted in the ($\overline{MS}, 2~\mbox{GeV}$) scheme, $2.40 (12)$ MeV~\cite{Fodor:2016bgu} at $N_f = 2+1$.

Combining the result (\ref{eq:deltamud}) with our ETMC determination of the average up/down quark mass $m_{ud}(\overline{MS}, 2~\mbox{GeV}) = 3.70 (17) ~ \mbox{MeV}$ from Ref.~\cite{Carrasco:2014cwa}, we can also compute the $u$- and $d$-quark masses
\bea
      \label{eq:mu}
      \widehat{m}_u(\overline{MS}, 2~\mbox{GeV}) & = & 2.50 ~ (15)_{stat+fit} ~ (8)_{syst} ~ (2)_{qQED} ~ \mbox{MeV} ~ , \nonumber \\
                                                                               & = & 2.50 ~ (17) ~ \mbox{MeV} ~ , \\
      \label{eq:md}
      \widehat{m}_d(\overline{MS}, 2~\mbox{GeV}) & = & 4.88 ~ (18)_{stat+fit} ~ (8)_{syst} ~ (2)_{qQED} ~ \mbox{MeV} ~ , \nonumber \\
                                                                               & = & 4.88 ~ (20) ~ \mbox{MeV}
\eea
and the ratio
\bea
    \frac{\widehat{m}_u}{\widehat{m}_d}(\overline{MS}, 2~\mbox{GeV}) & = & 0.513 ~ (18)_{stat+fit} ~ (24)_{syst} ~ (6)_{qQED} ~ , \nonumber \\
                                                                                                                & = & 0.513 ~ (30) ~ ,
    \label{eq:mumd}
\eea
which are consistent within the uncertainties with the current FLAG estimates~\cite{FLAG} at $N_f = 2+1+1$, based on the ETMC results of Ref.~\cite{Carrasco:2014cwa}, and with the recent BMW results~\cite{Fodor:2016bgu} at $N_f = 2+1$. 

Finally, using the ETMC result $m_s(\overline{MS}, 2~\mbox{GeV}) = 99.6 (4.3) ~ \mbox{MeV}$~\cite{Carrasco:2014cwa} we can obtain a determination of the flavor symmetry breaking parameters $R$ and $Q$, namely
\bea
    \label{eq:R}
    R(\overline{MS}, 2~\mbox{GeV}) \equiv \frac{m_s - m_{ud}}{\widehat{m}_d - \widehat{m}_u} (\overline{MS}, 2~\mbox{GeV}) & = & 40.4 ~ (3.3) ~ , \\[4mm]
    \label{eq:Q}
    Q(\overline{MS}, 2~\mbox{GeV}) \equiv \sqrt{\frac{ m_s^2 - m_{ud}^2}{ \widehat{m}_d^2 - \widehat{m}_u^2}} (\overline{MS}, 2~\mbox{GeV}) & = & 23.8 ~ (1.1) ~ ,
\eea
which are consistent within the errors with the current FLAG estimate $R = 35.6 (5.1)$ and $Q = 22.2 (1.6)$~\cite{FLAG} as well as with the recent BMW results $R = 38.20 (1.95)$ and $Q = 23.40 (64)$~\cite{Fodor:2016bgu}.

Our central value (\ref{eq:Q}) for the parameter $Q$ is $\approx 8 \%$ higher than the recent result of Ref.~\cite{Colangelo:2016jmc}, $Q = 22.0 (7)$, based on the latest update of the dispersive analysis of the $\eta \to 3 \pi$ decay and on the use of the SU(3) ChPT relation 
 \be
      \left[ M_{K^+}^2 - M_{K^0}^2 \right]^{QCD} = \frac{1}{Q^2} \frac{M_K^2}{M_\pi^2} ( M_\pi^2 - M_K^2) \left[ 1 + {\cal{O}}(m_s^2) \right] ~ .
      \label{eq:Q_NNLO}
 \ee
Had we used our result (\ref{eq:kaon_qcd}) in Eq.~(\ref{eq:Q_NNLO}), the value of the parameter $Q$ would have been $Q = 22.6 ~ (3)$, which is $\approx 5 \%$ below the result (\ref{eq:Q}) based on the use of the IB slope (\ref{eq:kaon_qcd_result}) evaluated directly on the lattice.
This seems to suggest that the higher-order corrections to the SU(3) ChPT relation (\ref{eq:Q_NNLO}) may be at the level of $\approx 10 \%$ or equivalently  about one unit for $Q$ (see also Ref.~\cite{Bijnens:2007pr}).

\section{Determination of $\epsilon_{K^0}$}
\label{sec:K0}

The violation of the Dashen's theorem for the neutral kaon mass can be represented by the quantity $\epsilon_{K^0}$ defined as~\cite{FLAG}
\be
    \epsilon_{K^0} = \frac{\left[\delta M^2_{K^0}\right]^{QED}}{M_{\pi^+}^2 - M_{\pi^0}^2} ~ .
    \label{eq:epsilon_k0}
\ee
The e.m.~contribution $\left[\delta M^2_{K^0}\right]^{QED}$ is given within the quenched QED approximation by
\be
    \left[\delta M^2_{K^0}\right]^{QED} = 8\pi \alpha_{em} M_K \left[\delta M_{K^0}\right]^{em},
\ee
where
\bea
    \left[\delta M_{K^0}\right]^{em} & = & q_d q_s  \partial_t \frac{\gdslexch}{\gdsl} - q_d^2 \partial_t \frac{\gdslselfl+\gdslphtadl}{\gdsl} \nonumber \\
                                                     & - & [\delta m^{crit}_d]\partial_t \frac{\gdsip}{\gdsl} + [\delta m^{crit}_s]\partial_t \frac{\gdsipl}{\gdsl} \nonumber \\
                                                     & - & q_s^2 \partial_t \frac{\gdslselfs + \gdslphtads}{\gdsl} + \frac{Z_P}{{\cal{Z}}_s} m_s \partial_t \frac{\gdsil}{\gdsl} \nonumber \\
                                                     & + & \frac{Z_P}{{\cal{Z}}_d} m_\ell \partial_t \frac{\gdsi}{\gdsl} ~ .
    \label{eq:mk0}
\eea

\begin{figure}[htb!]
\begin{center}
\includegraphics[scale=0.65]{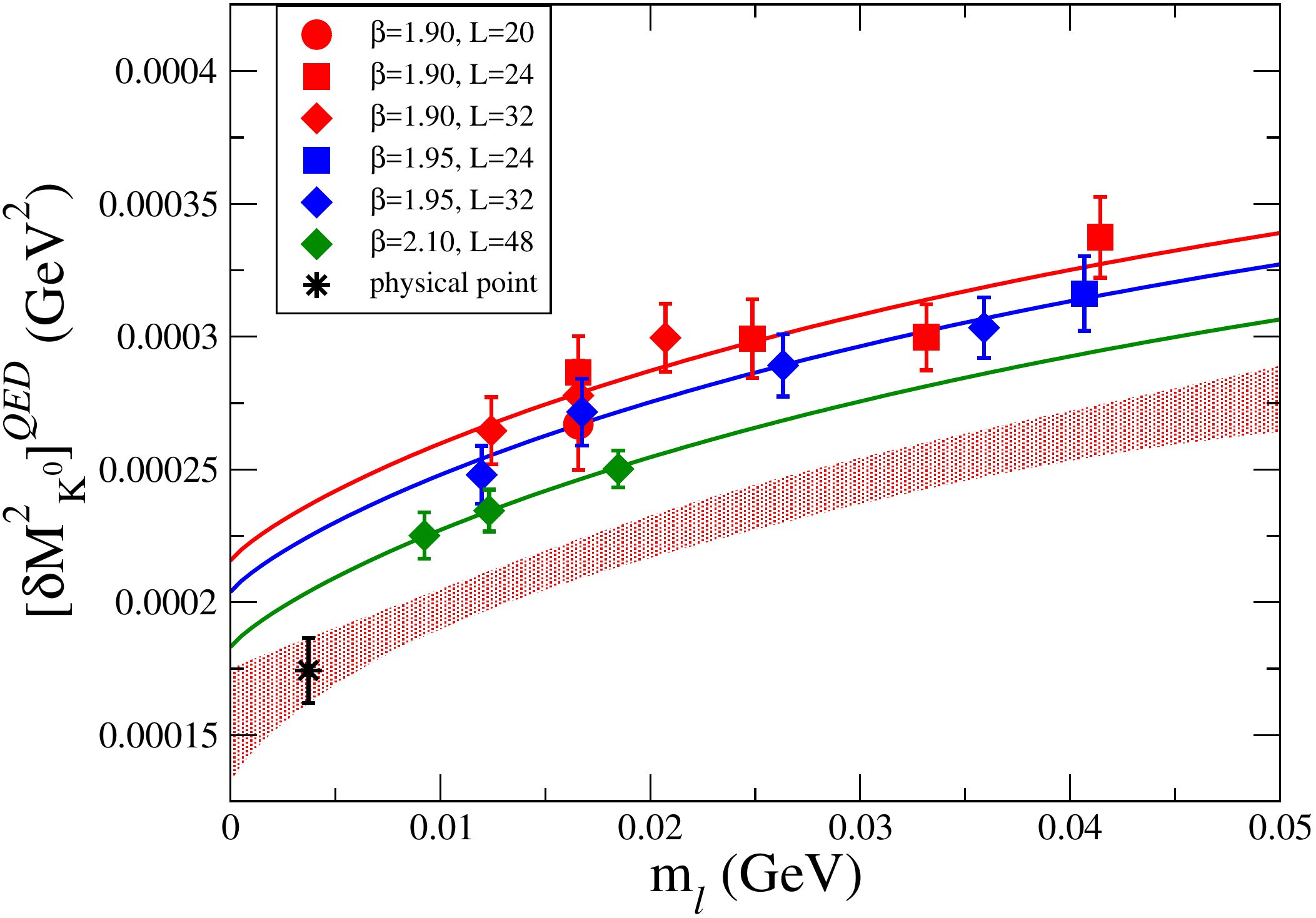}
\end{center}
\vspace{-0.70cm}
\caption{\it \small Results for the quantity $\left[ \delta M^2_{K^0} \right]^{QED}$ versus the renomalized light-quark mass $m_\ell$. The filled markers represent the lattice data without FSE corrections. The solid lines correspond to the results of the combined fit (\ref{eq:M2K0g_fit}) obtained at each value of the lattice spacing. The black asterisk represents the value extrapolated at the physical pion mass $m_\ell = m_{ud} = 3.70 (17)~\mbox{MeV}$ and to the continuum limit, while the red area identifies the corresponding uncertainty at the level of one standard deviation.}
\label{fig:M2K0g}
\end{figure}

The lattice data for $\left[ \delta M^2_{K^0} \right]^{QED}$ are shown by filled markers in Fig.~\ref{fig:M2K0g}.
No significant FSEs are visible and therefore for the combined chiral and continuum limit fitting procedure we use the following simple Ansatz
\be
    \left[\delta M^2_{K^0}\right]^{QED} = \widetilde{A}_0^K \left[1 + \widetilde{A}_L^K \frac{\overline{M}^2}{16\pi^2 f_0^2}\log\left(\frac{\overline{M}^2}{16\pi^2 f_0^2}\right) +
                                                               \widetilde{A}_1^K \frac{\overline{M}^2}{16\pi^2 f_0^2} \right] + \widetilde{D}^K a^2 ~ ,
    \label{eq:M2K0g_fit}
\ee
where $\widetilde{A}_0^K$, $\widetilde{A}_L^K$, $\widetilde{A}_1^K$ and $\widetilde{D}^K$ are free parameters.
The results of the fitting procedure are shown in Fig.~\ref{fig:M2Pi0g} by the solid lines at each value of the lattice spacing and by the black asterisk at the physical pion mass and in the continuum limit.

At the physical pion mass and in the continuum limit we obtain
\bea
    \left[ \delta M_{K^0}^2 \right]^{QED}(\overline{MS}, 2~\mbox{GeV}) & = & 0.174 ~ (12)_{stat+fit} ~ (19)_{disc} ~ (3)_{chir} ~ (9)_{qQED} \cdot 10^{-3} ~ \mbox{GeV}^2 ~ , \nonumber \\
                                                                                                               & = & 0.174 ~ (12)_{stat+fit} ~ (19)_{syst} ~ (9)_{qQED} \cdot 10^{-3} ~ \mbox{GeV}^2 ~  , \nonumber \\
                                                                                                               & = & 0.174 ~ (24) \cdot 10^{-3} ~ \mbox{GeV}^2 ~  ,
    \label{eq:M2K0g_result}
\eea
where
\begin{itemize}
\item $()_{stat+fit}$ indicates the statistical uncertainty including also the ones induced by the fitting procedure and by the determination of the input parameters of Table \ref{tab:8branches};
\item $()_{disc}$ is the uncertainty due to discretization effects estimated by comparing the results obtained including ($\widetilde{D}^K \neq 0$) or excluding ($\widetilde{D}^K = 0$) the discretization term  in Eq.~(\ref{eq:M2K0g_fit});
\item $()_{chir}$ is the error coming from including the term proportional to the chiral log in Eq.~(\ref{eq:M2K0g_fit}) or substituting it with a quadratic term in $m_\ell$ (i.e., $\widetilde{A}_2^K \overline{M}^4 / (4 \pi f_0)^4$);
\item $()_{qQED}$ is the $5 \%$ estimate of the effects due to the quenched QED approximation taken from Refs.~\cite{Portelli:2012pn,Fodor:2016bgu}.
\end{itemize}

Using the experimental value $M_{K^0}  = 497.611 (13)$ MeV \cite{PDG} our results (\ref{eq:M2K0g_result}) and (\ref{eq:kaon_qcd}) correspond to a kaon mass in pure QCD equal to $M_K = 494.4 (1)$ MeV in agreement with the FLAG estimate $M_K = 494.2 (3)$ MeV.

Dividing our result (\ref{eq:M2K0g_fit}) by Eq.~(\ref{eq:pion_result}), we obtain
\bea
     \epsilon_{K^0}(\overline{MS}, 2~\mbox{GeV}) & = & 0.154 ~ (14)_{stat+fit} ~ (20)_{disc} ~ (1)_{chir} ~ (1)_{FSE} ~ (10)_{qQED} ~ , \nonumber \\
                                                                              & = & 0.154 ~ (14)_{stat+fit} ~ (20)_{syst} ~ (10)_{qQED} ~ , \nonumber \\
                                                                              & = & 0.154 ~ (26) ~ , 
    \label{eq:epsilon_K0_result}
\eea
where now the $()_{qQED}$ error includes also the $4 \%$ effect coming from the disconnected diagram neglected in the pion mass splitting analysis. 
Our result (\ref{eq:epsilon_K0_result}) is in agreement with (and more precise than) both the estimate quoted by FLAG, namely $\epsilon_{K^0} = 0.3 (3)$~\cite{FLAG}, and the recent QCDSF/UKQCD result $\epsilon_{K^0}(\overline{MS}, 2~\mbox{GeV}) = 0.2 (1)$~\cite{Horsley:2015vla}.

\section{QED and strong IB corrections in charmed mesons}
\label{sec:charm}

In this section using the RM123 approach we address the evaluation of the leading-order e.m.~and strong IB corrections to the $D$-meson mass splitting ($M_{D^+} - M_{D^0}$), and the determination of the leading-order e.m.~corrections to the $D$-meson mass combination ($M_{D^+} + M_{D^0}$) and to the $D_s$-meson mass $M_{D_s^+}$.
In the case of $D$-meson mass splitting  we make use of the determination (\ref{eq:deltamud}) of the $u$- and $d$-quark mass difference done in the kaon sector (see Section \ref{sec:kaon}) to evaluate the strong IB correction and therefore to predict for the first time the physical mass splitting ($M_{D^+} - M_{D^0}$) on the lattice.

\subsection{Electromagnetic and strong IB corrections to $M_{D^+} - M_{D^0}$}
\label{sec:Dmeson}

Within the quenched QED approximation and the RM123 prescription described in Section \ref{sec:master}, the QED contribution to the $D$-meson mass splitting is given by
 \be
    \left[ M_{D^+}^2 - M_{D^0}^2 \right]^{QED} = 8 \pi \alpha_{em} M_D \left[ M_{D^+} - M_{D^0} \right]^{em} ~ ,
    \label{eq:Dmeson2_qed}
 \ee
where
\bea
    \left[ M_{D^+} - M_{D^0} \right]^{em} & = & (q_u - q_d) q_c  \partial_t \frac{\gdclexch}{\gdcl} - (q_d^2 - q_u^2) \partial_t \frac{\gdclselfl + \gdclphtadl}{\gdcl} \nonumber \\
           & - & (\delta m^{crit}_d - \delta m^{crit}_u) \partial_t \frac{\gdlipc}{\gdcl} + Z_P \left( \frac{1}{{\cal{Z}}_d} - \frac{1}{{\cal{Z}}_u} \right) m_\ell \partial_t \frac{\gdlic}{\gdcl} ~ 
    \label{eq:Dmeson_qed}
\eea
with the green lines representing the charm quark propagator.

\begin{figure}[htb!]
\begin{center}
\includegraphics[scale=0.65]{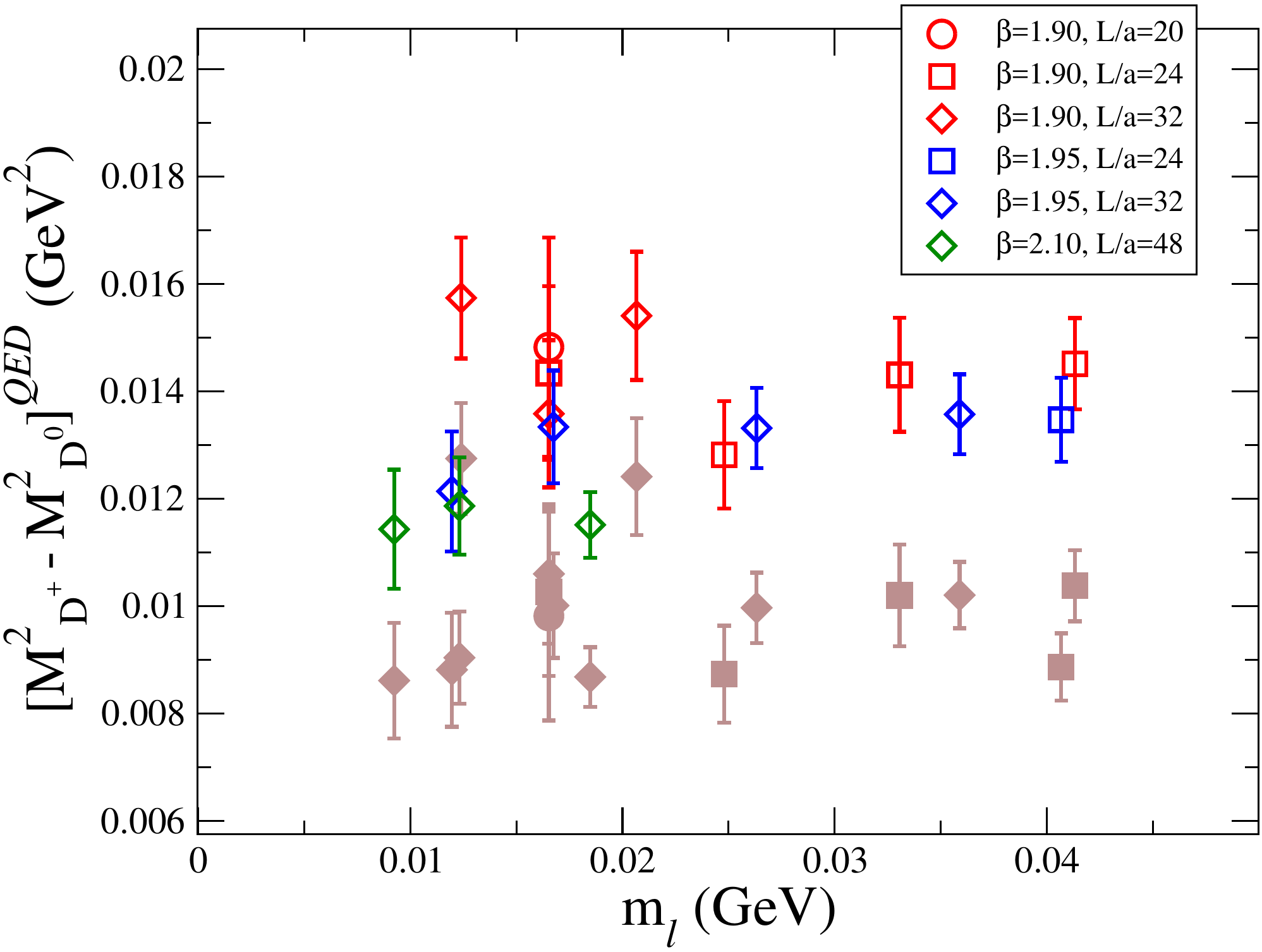}
\end{center}
\vspace{-0.70cm}
\caption{\it \small Results for the $D$-meson mass splitting $\left[ M_{D^+}^2 -M_{D^0}^2 \right]^{QED}$ versus the renormalized light-quark mass $m_\ell$, obtained using Eq.~(\ref{eq:Dmeson_qed}) in the quenched QED approximation. Brown full points correspond to the data without any correction for FSEs, while open markers represent the lattice data corrected by the universal FSEs given by Eq.~(\ref{eq:universal_FSE}).}
\label{fig:dM2D_QED_univ}
\end{figure}

In Fig.~\ref{fig:dM2D_QED_univ} the data for $\left[ M_{D^+}^2 - M_{D^0}^2 \right]^{QED}$ are shown before and after the subtraction of the universal FSEs, given by Eq.~(\ref{eq:universal_FSE}).
It can be clearly seen that, as in the case of the pion and kaon mass splittings, the universal FSE correction is quite large, approaching $\simeq 30 \%$ at the heaviest light-quark masses.

From now on we always refer to the data for $\left[ M_{D^+}^2 - M_{D^0}^2 \right]^{QED}$ as to the QED part of the charged/neutral $D$-meson mass splitting already subtracted by the universal FSEs.

\begin{figure}[htb!]
\begin{center}
\includegraphics[scale=0.65]{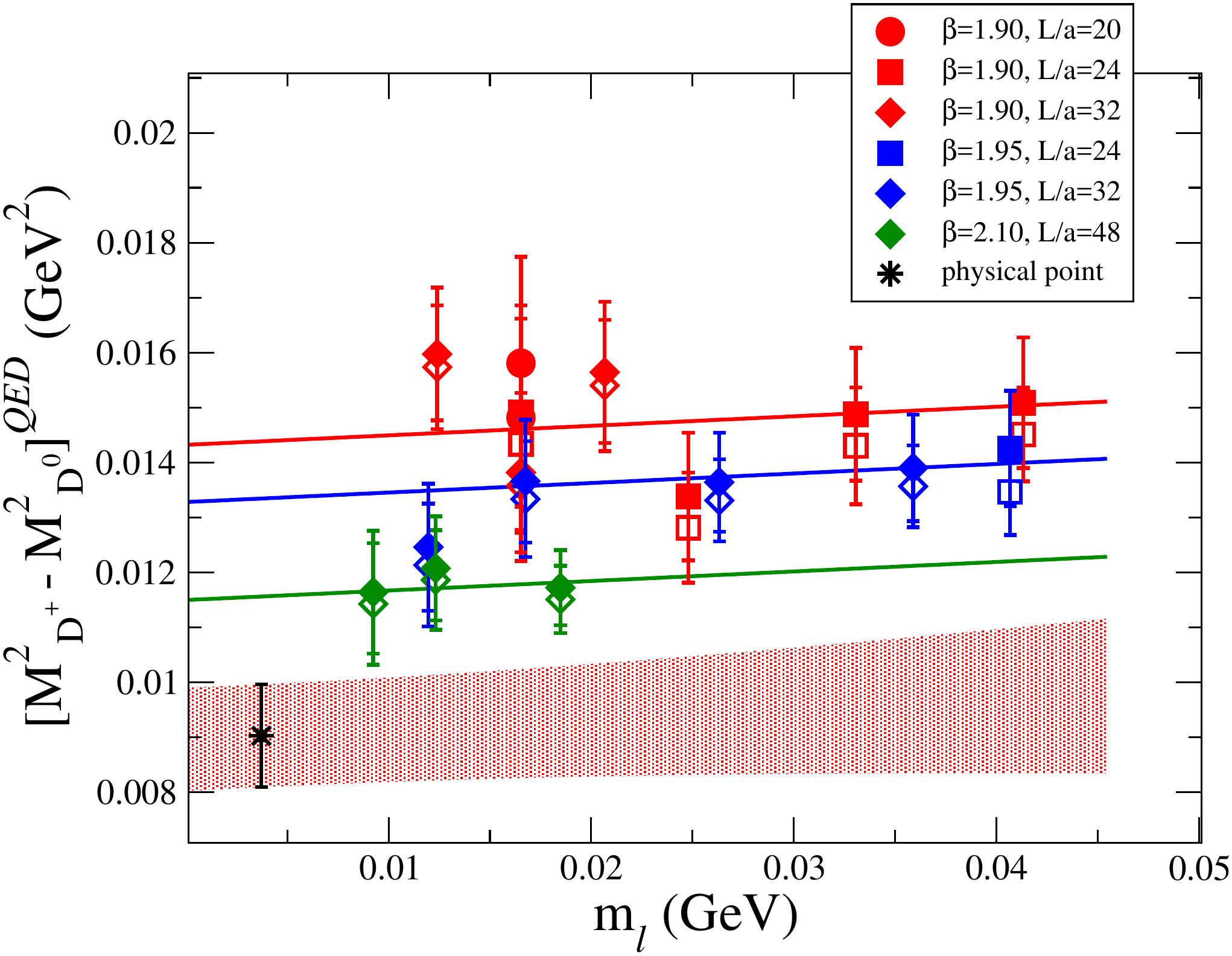}
\end{center}
\vspace{-0.70cm}
\caption{\it \small Results for the $D$-meson mass splitting $\left[ M_{D^+}^2 - M_{D^0}^2 \right]^{QED}$ versus the renormalized light-quark mass $m_\ell$. The empty markers correspond to the data after the subtraction of the universal FSEs, while the filled markers represent the lattice data corrected also by the SD FSEs obtained in the fitting procedure. The solid lines correspond to the results of the combined fit (\ref{eq:Dmeson_qed_fit}) obtained in the infinite volume limit at each value of the lattice spacing. The black asterisk represents the D-meson mass splitting extrapolated at the physical pion mass $m_\ell = m_{ud} = 3.70 (17)~\mbox{MeV}$ and to the continuum limit, while the red area identifies the corresponding uncertainty at the level of one standard deviation.}
\label{fig:dM2D_QED_full}
\end{figure}

We have performed combined chiral, continuum and infinite volume extrapolations adopting the following fitting function
\be
     \left[ M_{D^+}^2 - M_{D^0}^2 \right]^{QED} = 4 \pi \alpha_{em} \left[ A_0^D + A_1^D m_\ell + D^D a^2 + F^D \frac{M_D}{L^3} \right] ~ ,
     \label{eq:Dmeson_qed_fit}
\ee
where $A_0^D$, $A_1^D$, $D^D$ and $F^D$ are free parameters.
In Fig.~\ref{fig:dM2D_QED_full} we show the results obtained using the combined fitting function (\ref{eq:Dmeson_qed_fit}).

At the physical pion mass and in the continuum and infinite volume limits our result in the $\overline{MS}$ scheme at a renormalization scale equal to $\mu = 2$ GeV is
\bea
    \left[ M_{D^+}^2 - M_{D^0}^2 \right]^{QED} & = & 9.03 ~ (0.84)_{stat+fit} ~ (1.65)_{disc} ~ (0.12)_{chir} ~ (0.07)_{FSE} ~ (0.45)_{qQED} \cdot 10^{-3}~\mbox{GeV}^2 ~ , \nonumber \\
                                                                         & = & 9.03 ~ (0.84)_{stat+fit} ~ (1.65)_{syst} ~ (0.45)_{qQED} \cdot 10^{-3}~\mbox{GeV}^2 ~ , \nonumber \\
                                                                         & = & 9.03 ~ (1.90) \cdot 10^{-3}~\mbox{GeV}^2 ~ ,
    \label{eq:Dmeson_qed_result}
\eea
where
\begin{itemize}
\item $()_{stat+fit}$ indicates the statistical uncertainty including also the ones induced by the fitting procedure and by the determination of the input parameters of Table \ref{tab:8branches};
\item $()_{disc}$ is the uncertainty due to discretization effects estimated by comparing the results assuming either $D^D \neq 0$ or $D^D = 0$ in Eq.~(\ref{eq:Dmeson_qed_fit});
\item $()_{chir}$ is the error coming from including ($A_1^D \neq 0$) or excluding ($A_1^D = 0$) the linear term in the light-quark mass;
\item $()_{FSE}$ is the uncertainty due to FSE estimated by comparing the results obtained including ($F^D \neq 0$) or excluding ($F^D= 0$) the phenomenological term for the SD FSEs;
\item $()_{qQED}$ is the estimate of the effects due to the quenched QED approximation ($5 \%)$ taken from Refs.~\cite{Portelli:2012pn,Fodor:2016bgu} and extended to the case of charmed mesons.
\end{itemize}

\begin{figure}[htb!]
\begin{center}
\includegraphics[scale=0.65]{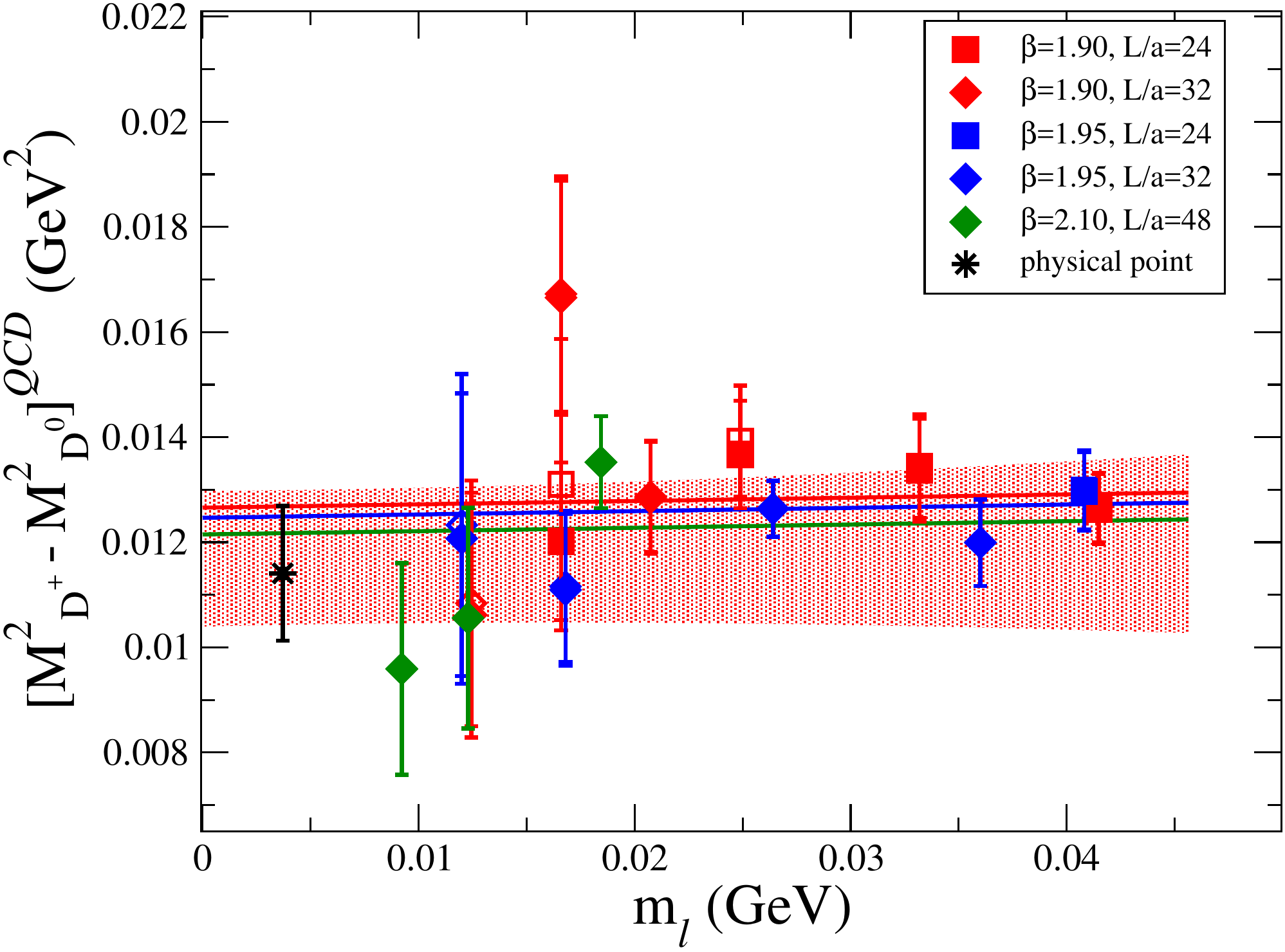}
\end{center}
\vspace{-0.70cm}
\caption{\it \small Results for the QCD contribution $\left[ M_{D^+}^2 - M_{D^0}^2 \right]^{QCD}$ versus the renomalized light-quark mass $m_\ell$. The empty markers correspond to the lattice data, while the filled ones represent the data corrected for the FSEs obtained in the fitting procedure (\ref{eq:Dmeson_QCD_fit}). The solid lines correspond to the results of the combined fit (\ref{eq:Dmeson_QCD_fit}) obtained in the infinite volume limit at each value of the lattice spacing. The black asterisk represents the IB slope extrapolated at the physical pion mass $m_\ell = m_{ud} = 3.70 (17)~\mbox{MeV}$ and to the continuum limit, while the red area identifies the corresponding uncertainty at the level of one standard deviation.}
\label{fig:dM2D_QCD}
\end{figure}

We need now to compute the QCD contribution
\be
    \left[ M_{D^+}^2 - M_{D^0}^2 \right]^{QCD} = 2 M_D ~ Z_P ~ (\widehat{m}_d - \widehat{m}_u) ~ \partial_t \frac{\gdci}{\gdcl} ~ ,
\ee
where for $(\widehat{m}_d - \widehat{m}_u)$ we make use of the result (\ref{eq:deltamud}).
The lattice data for $\left[ M_{D^+}^2 - M_{D^0}^2 \right]^{QCD}$ are shown in Fig.~\ref{fig:dM2D_QCD} and FSEs are visible.
Thus the data have been fitted according to the following simple Ansatz:
\be
     \left[ M_{D^+}^2 - M_{D^0}^2 \right]^{QCD} = \overline{A}_0^D + \overline{A}_1^D m_\ell + \overline{D}^D a^2 + \overline{F}^D \frac{\overline{M}^2}{16\pi^2 f_0^2} 
                                                                             \frac{e^{-\overline{M} L}}{(\overline{M} L)^{3/2}} ~ ,
    \label{eq:Dmeson_QCD_fit}
\ee
where we recall $\overline{M}^2 = 2B_0 m_{\ell}$.
The results of the linear fit (\ref{eq:Dmeson_QCD_fit}) with the four free parameters $\overline{A}_0^D$, $\overline{A}_1^D$, $\overline{D}^D$ and $\overline{F}^D$, are shown in Fig.~\ref{fig:dM2D_QCD}.

At the physical pion mass and in the continuum and infinite volume limits we get
\bea
    \left[ M_{D^+}^2 - M_{D^0}^2 \right]^{QCD}(\overline{\mathrm{MS}}, 2~\mbox{GeV}) & = & 11.41 ~ (99)_{stat+fit} ~ (21)_{disc} ~ (13)_{chir} ~ (9)_{FSE} 
                      \cdot 10^{-3} ~ \mbox{GeV}^2 ~ \nonumber \\
            & = & 11.41 ~ (99)_{stat+fit} ~ (26)_{syst} \cdot 10^{-3} ~ \mbox{GeV}^2 ~ , \nonumber \\
            & = & 11.41 ~ (1.02) \cdot 10^{-3} ~ \mbox{GeV}^2 ~ .
    \label{eq:Dmeson_qcd_result}
\eea
where
\begin{itemize}
\item $()_{stat+fit}$ indicates the statistical uncertainty including also the ones induced by the fitting procedure and by the determination of the input parameters of Table \ref{tab:8branches};
\item $()_{disc}$ is the uncertainty due to discretization effects estimated by including ($\overline{D}^D \neq 0$) or excluding ($\overline{D}^D = 0$) the discretization term in Eq.~(\ref{eq:Dmeson_QCD_fit});
\item $()_{chir}$ is the error coming from including ($\overline{A}_1^D \neq 0$) or excluding ($\overline{A}_1^D = 0$) the linear term in the light-quark mass.
\item $()_{FSE}$ is the uncertainty obtained including ($\overline{F}^D \neq 0$) or excluding ($\overline{F}^D = 0$) the FSE term in Eq.~(\ref{eq:Dmeson_QCD_fit}).
\end{itemize}

Thus, putting together the results (\ref{eq:Dmeson_qed_result}) and (\ref{eq:Dmeson_qcd_result}) we get the prediction
\bea
       M_{D^+} - M_{D^0} & = & 5.47 ~ (30)_{stat+fit} ~ (40)_{disc} ~ (6)_{chir} ~ (3)_{FSE} ~ (12)_{qQED} ~ \mbox{MeV} ~ , \nonumber \\
                                       & = & 5.47 ~ (30)_{stat+fit} ~ (42)_{syst} ~ (12)_{qQED} ~ \mbox{MeV} ~ , \nonumber \\
                                       & = & 5.47 ~ (53) ~ \mbox{MeV} ~ ,
   \label{eq:deltaMD}
\eea
which is consistent with the experimental value $M_{D^+} - M_{D^0} = 4.75 (8)$ MeV \cite{PDG} within $\simeq 1.4$ standard deviations.

\subsection{Electromagnetic corrections to $M_{D^+} + M_{D^0}$}
\label{sec:Daverage}

The $D$-meson mass combination ($M_{D^+} + M_{D^0}$), being isospin symmetric, does not receive any strong IB correction at leading order ${\cal{O}}(\widehat{m}_d - \widehat{m}_u)$.
Within the quenched QED approximation one has
\bea
    \delta M_{D^+} + \delta M_{D^0} & = & 4 \pi \alpha_{em} \left\{ - (q_u + q_d) q_c  \partial_t \frac{\gdclexch}{\gdcl} - (q_d^2 + q_u^2) \partial_t \frac{\gdclselfl + \gdclphtadl}{\gdcl} 
                                                                   \right. \nonumber \\
                                                        & - & 2 \left. q_c^2 \partial_t \frac{\gdclselfc + \gdclphtadc}{\gdcl} + 2 \delta m^{crit}_c \partial_t \frac{\gdcipl}{\gdcl} \right. \nonumber \\
                                                        & - & \left. (\delta m^{crit}_d + \delta m^{crit}_u) \partial_t \frac{\gdlipc}{\gdcl} + 2 \frac{Z_P}{{\cal{Z}}_c} m_c \partial_t \frac{\gdcil}{\gdcl} \right. \nonumber \\
                                                        & + & \left. Z_P \left( \frac{1}{{\cal{Z}}_u} + \frac{1}{{\cal{Z}}_d} \right) m_\ell \partial_t \frac{\gdci}{\gdcl} \right\}~  .
    \label{eq:Dmeson_plus}
\eea

\begin{figure}[htb!]
\begin{center}
\includegraphics[scale=0.65]{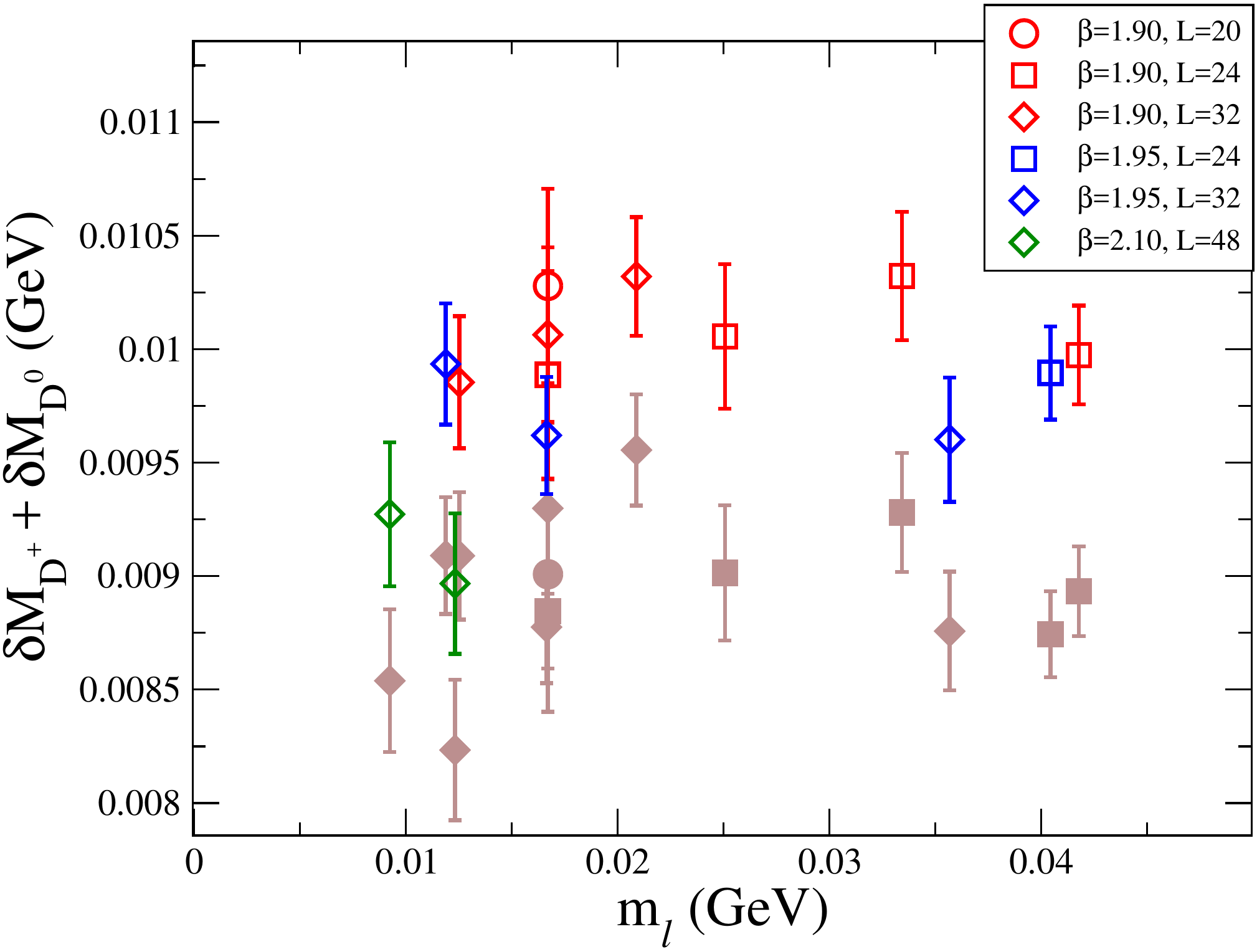}
\end{center}
\vspace{-0.70cm}
\caption{\it \small Results for the e.m.~correction to the charge-averaged $D$-meson mass $\delta M_{D^+} + \delta M_{D^0}$ versus the renormalized light-quark mass $m_\ell$, obtained using Eq.~(\ref{eq:Dmeson_plus}) in the quenched QED approximation. Brown full points correspond to the data without any correction for FSEs, while open markers represent the lattice data corrected by the universal FSEs given by Eq.~(\ref{eq:universal_FSE}).}
\label{fig:sMD_univ}
\end{figure}

The data for $\delta M_{D^+} + \delta M_{D^0}$ after the subtraction of the universal FSEs are shown in Fig.~\ref{fig:sMD_univ}.

\begin{figure}[htb!]
\begin{center}
\includegraphics[scale=0.65]{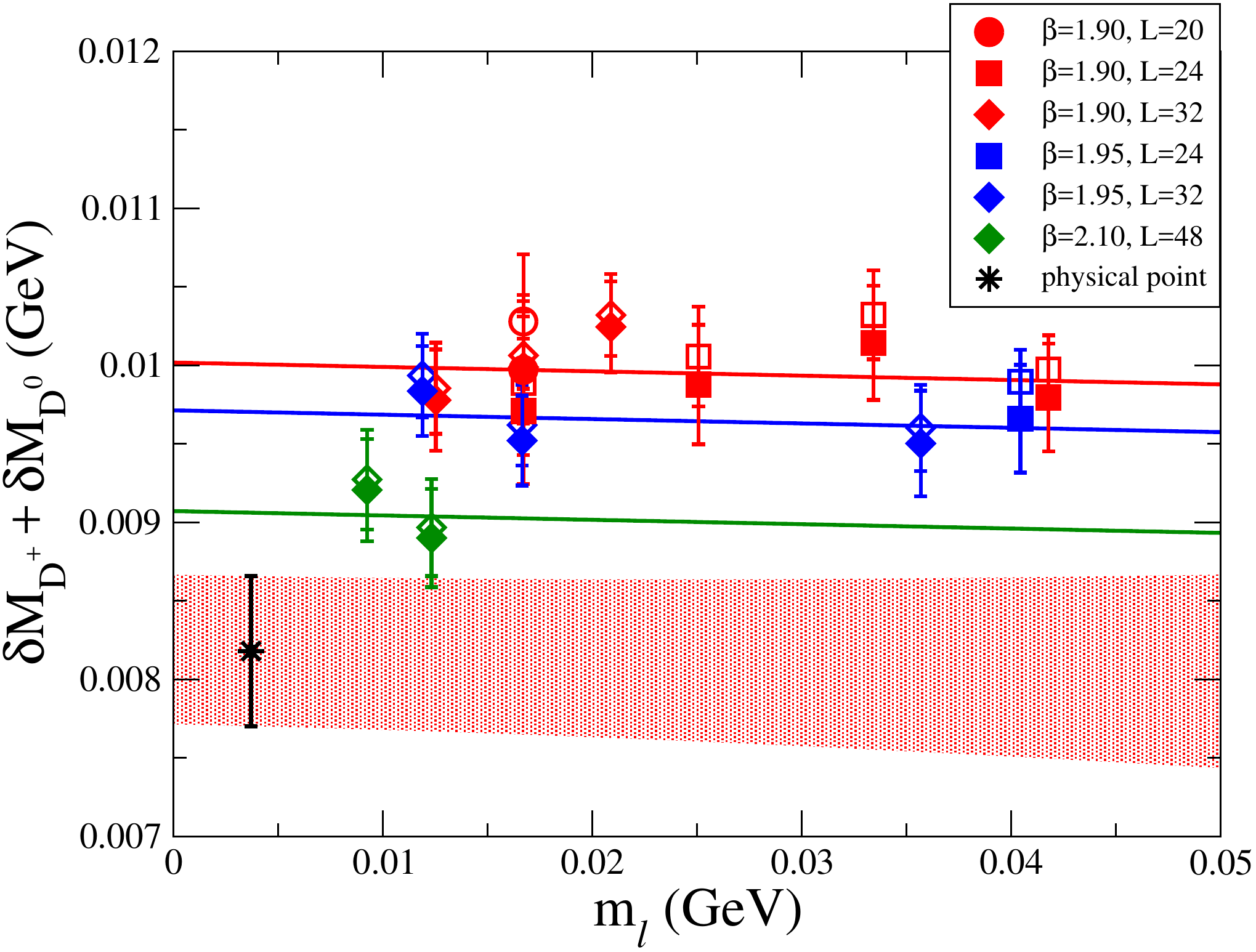}
\end{center}
\vspace{-0.70cm}
\caption{\it \small Results for the e.m.~correction to the charge-averaged $D$-meson mass $\delta M_{D^+} + \delta M_{D^0}$ versus the renormalized light-quark mass $m_\ell$. The empty markers correspond to the data after the subtraction of the universal FSEs, while the filled markers represent the lattice data corrected also by the SD FSEs obtained in the fitting procedure. The solid lines correspond to the results of the combined fit (\ref{eq:Dmeson_plus_fit}) obtained in the infinite volume limit at each value of the lattice spacing. The black asterisk represents the value extrapolated at the physical pion mass $m_\ell = m_{ud} = 3.70 (17)~\mbox{MeV}$ and to the continuum limit, while the red area identifies the corresponding uncertainty at the level of one standard deviation.}
\label{fig:sMD_full}
\end{figure}

We have performed combined chiral, continuum and infinite volume extrapolations adopting the following fitting function
\be
     \delta M_{D^+} + \delta M_{D^0} = \widetilde{A}_0^D + \widetilde{A}_1^D m_\ell + \widetilde{D}^D a^2 + \widetilde{F}^D \frac{M_D}{L^3} ~ ,
     \label{eq:Dmeson_plus_fit}
\ee
where $\widetilde{A}_0^D$, $\widetilde{A}_1^D$, $\widetilde{D}^D$ and $\widetilde{F}^D$ are free parameters.
In Fig.~\ref{fig:sMD_full} we show the results obtained using the combined fitting function (\ref{eq:Dmeson_plus_fit}).

At the physical pion mass and in the continuum and infinite volume limits our result is
\bea
    \delta M_{D^+} + \delta M_{D^0} & = & 8.18 ~ (37)_{stat+fit} ~ (66)_{disc} ~ (2)_{chir} ~ (4)_{FSE} ~ (41)_{qQED} ~ \mbox{MeV} ~ , \nonumber \\
                                                        & = & 8.18 ~ (37)_{stat+fit} ~ (66)_{syst} ~ (41)_{qQED} ~ \mbox{MeV} ~ , \nonumber \\
                                                        & = & 8.18 ~ (86) ~  \mbox{MeV} ~ ,
    \label{eq:Dmeson_plus_result}
\eea
where
\begin{itemize}
\item $()_{stat+fit}$ indicates the statistical uncertainty including also the ones induced by the fitting procedure and by the determination of the input parameters of Table \ref{tab:8branches};
\item $()_{disc}$ is the uncertainty due to discretization effects estimated by comparing the results assuming either $\widetilde{D}^D \neq 0$ or $\widetilde{D}^D = 0$ in Eq.~(\ref{eq:Dmeson_plus_fit});
\item $()_{chir}$ is the error coming from including ($\widetilde{A}_1^D \neq 0$) or excluding ($\widetilde{A}_1^D = 0$) the linear term in the light-quark mass;
\item $()_{FSE}$ is the uncertainty due to FSE estimated by comparing the results obtained including ($\widetilde{F}^D \neq 0$) or excluding ($\widetilde{F}^D = 0$) the phenomenological term for the SD FSEs;
\item $()_{qQED}$ is the estimate of the effects due to the quenched QED approximation ($5 \%)$ taken from Refs.~\cite{Portelli:2012pn,Fodor:2016bgu} and extended to the case of charmed mesons.
\end{itemize}

Using the experimental value $(M_{D^+} + M_{D^0}) / 2 = 1867.2 (4)$ MeV \cite{PDG} our result (\ref{eq:Dmeson_plus_result}) corresponds to a $D$-meson mass in pure QCD equal to $1863.1 (6)$ MeV.

\subsection{Electromagnetic corrections to the $D_s^+$-meson mass}
\label{sec:Dsplus}

Finally we have computed also the e.m.~corrections to the mass of the $D_s^+$-meson, that, within the quenched QED approximation, are given by
\bea
    \delta M_{D_s^+} & = & 4 \pi \alpha_{em} \left\{ - q_c q_s \partial_t \frac{\gdcsexch}{\gdcs} - q_s^2 \partial_t \frac{\gdcsselfs + \gdcsphtads}{\gdcs} \right. \nonumber \\
                                 & - & \left. q_c^2 \partial_t \frac{\gdcsselfc + \gdcsphtadc}{\gdcs} - \delta m^{crit}_s \partial_t \frac{\gdsipc}{\gdcs} +  \delta m^{crit}_c \partial_t \frac{\gdcips}{\gdcs} 
                                           \right. \nonumber \\
                                 & + & \left. \frac{Z_P}{{\cal{Z}}_s} m_s \partial_t \frac{\gdsic}{\gdcs} + \frac{Z_P}{{\cal{Z}}_c} m_c \partial_t \frac{\gdcis}{\gdcs} \right\} ~  .
    \label{eq:Dsmeson_plus}
\eea

\begin{figure}[htb!]
\begin{center}
\includegraphics[scale=0.65]{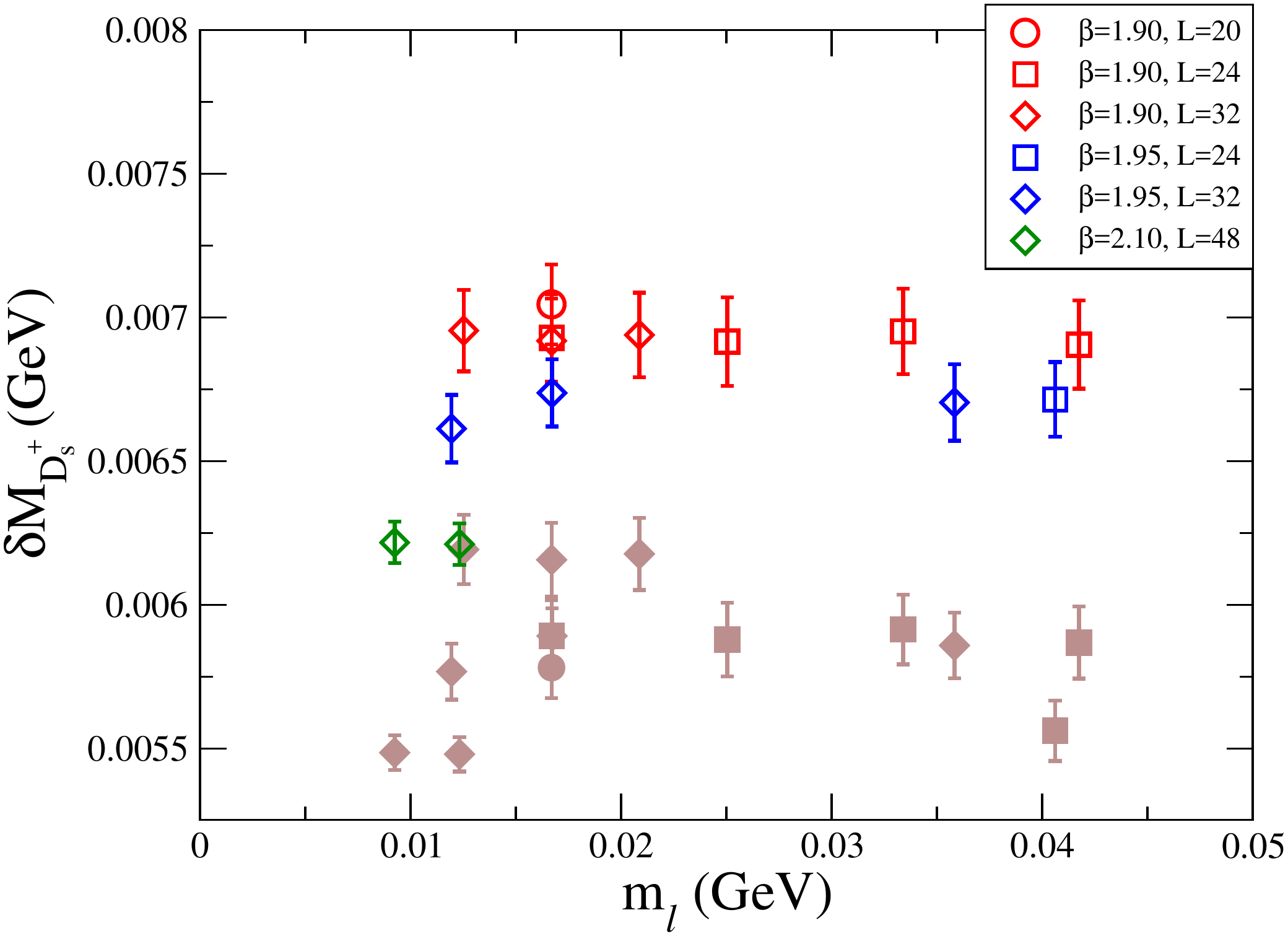}
\end{center}
\vspace{-0.70cm}
\caption{\it \small Results for the e.m.~correction $\delta M_{D_s^+}$ versus the renormalized light-quark mass $m_\ell$, obtained using Eq.~(\ref{eq:Dsmeson_plus}) in the quenched QED approximation. Brown full points correspond to the data without any correction for FSEs, while open markers represent the lattice data corrected by the universal FSEs given by Eq.~(\ref{eq:universal_FSE}).}
\label{fig:MDs_univ}
\end{figure}

The data for $\delta M_{D_s^+}$ after the subtraction of the universal FSEs are shown in Fig.~\ref{fig:MDs_univ}.

\begin{figure}[htb!]
\begin{center}
\includegraphics[scale=0.65]{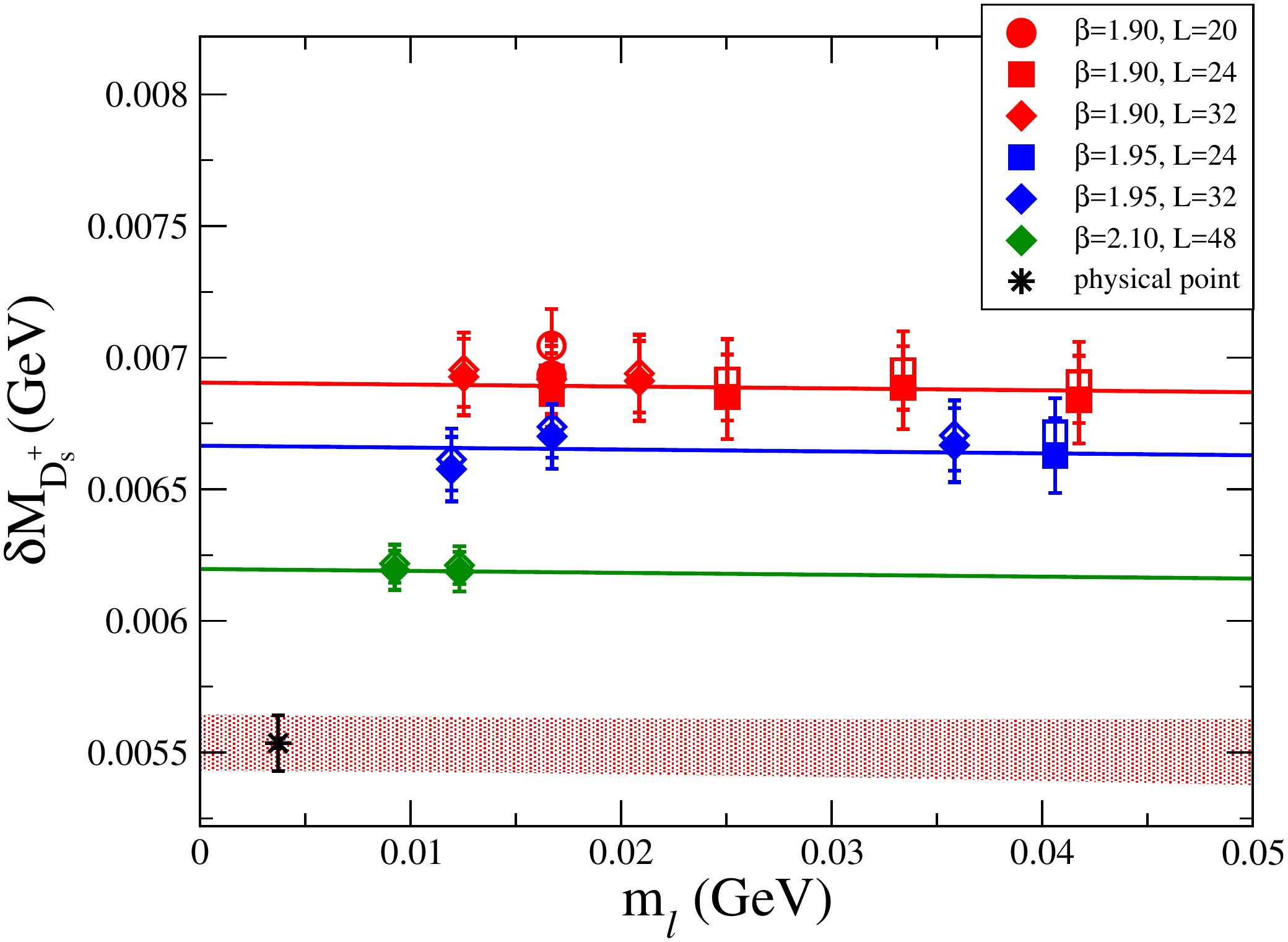}
\end{center}
\vspace{-0.70cm}
\caption{\it \small Results for the e.m.~correction $\delta M_{D_s^+}$ versus the renormalized light-quark mass $m_\ell$. The empty markers correspond to the data after the subtraction of the universal FSEs, while the filled markers represent the lattice data corrected also by the SD FSEs obtained in the fitting procedure. The solid lines correspond to the results of the combined fit (\ref{eq:Dsmeson_fit}) obtained in the infinite volume limit at each value of the lattice spacing. The black asterisk represents the value extrapolated at the physical pion mass $m_\ell = m_{ud} = 3.70 (17)~\mbox{MeV}$ and to the continuum limit, while the red area identifies the corresponding uncertainty at the level of one standard deviation.}
\label{fig:MDs_full}
\end{figure}

We have performed combined chiral, continuum and infinite volume extrapolations adopting the following fitting function
\be
     \delta M_{D_s^+} = A_0^{D_s} + A_1^{D_s} m_\ell + D^{D_s} a^2 + F^{D_s} \frac{M_{D_s}}{L^3} ~ ,
     \label{eq:Dsmeson_fit}
\ee
where $A_0^{D_s}$, $A_1^{D_s}$, $D^{D_s}$ and $F^{D_s}$ are free parameters.
In Fig.~\ref{fig:MDs_full} we show the results obtained using the combined fitting function (\ref{eq:Dsmeson_fit}).

At the physical pion mass and in the continuum and infinite volume limits our result is
\bea
    \delta M_{D_s^+} & = & 5.54 ~ (11)_{stat+fit} ~ (46)_{disc} ~ (1)_{chir} ~ (2)_{FSE} ~ (28)_{qQED} ~ \mbox{MeV} ~ , \nonumber \\
                                & = & 5.54 ~ (11)_{stat+fit} ~ (46)_{syst} ~ (28)_{qQED} ~ \mbox{MeV} ~ , \nonumber \\
                                & = & 5.54 ~ (55) ~  \mbox{MeV} ~ ,
    \label{eq:Dsmeson_result}
\eea
where
\begin{itemize}
\item $()_{stat+fit}$ indicates the statistical uncertainty including also the ones induced by the fitting procedure and by the determination of the input parameters of Table \ref{tab:8branches};
\item $()_{disc}$ is the uncertainty due to discretization effects estimated by comparing the results assuming either $D^{D_s} \neq 0$ or $D^{D_s} = 0$ in Eq.~(\ref{eq:Dsmeson_fit});
\item $()_{chir}$ is the error coming from including ($A_1^{D_s} \neq 0$) or excluding ($A_1^{D_s} = 0$) the linear term in the light-quark mass;
\item $()_{FSE}$ is the uncertainty due to FSE estimated by comparing the results obtained including ($F^{D_s} \neq 0$) or excluding ($F^{D_s} = 0$) the phenomenological term for the SD FSEs;
\item $()_{qQED}$ is the estimate of the effects due to the quenched QED approximation ($5 \%)$ taken from Refs.~\cite{Portelli:2012pn,Fodor:2016bgu} and extended to the case of charmed mesons.
\end{itemize}

Using the experimental value $M_{D_s^+} = 1969.0 (1.4)$ MeV \cite{PDG} our result (\ref{eq:Dsmeson_result}) corresponds to a $D_s$-meson mass in pure QCD equal to $1963.5 (1.5)$ MeV.

\section{Conclusions}
\label{sec:conclusions}

We have presented a lattice computation of the isospin-breaking corrections to pseudoscalar meson masses using the gauge configurations produced by the European Twisted Mass collaboration with $N_f = 2 + 1 + 1$ dynamical quarks at three values of the lattice spacing ($a \simeq 0.062, 0.082, 0.089$ fm) with pion masses in the range $M_\pi \simeq 210 - 450$ MeV. 
The strange and charm quark masses are tuned at their physical values. 

We have adopted the RM123 method of Ref.~\cite{deDivitiis:2013xla}, which is based on the combined expansion of the path integral in powers of the $d$- and $u$-quark mass difference ($\widehat{m}_d - \widehat{m}_u$) and of the electromagnetic coupling $\alpha_{em}$.
All the calculations are performed assuming the quenched QED approximation, which neglects the effects of the sea-quark electric charges.

After extrapolation to the physical pion mass and to the continuum and infinite volume limits we have obtained results for several quantities, as the pion, kaon and (for the first time) charmed-meson mass splittings, the prescription-dependent parameters $\epsilon_\gamma(\overline{MS}, 2~\mbox{GeV})$, $\epsilon_{\pi^0}$, $\epsilon_{K^0}(\overline{MS}, 2~\mbox{GeV})$, related to the violations of the Dashen's theorem, and the light quark mass difference $(\widehat{m}_d - \widehat{m}_u)(\overline{MS}, 2~\mbox{GeV})$.
Using the latter result we make the first lattice determination of the physical D-meson mass splitting $M_{D^+} - M_{D^0}$.
Our results are collected in Eqs.~(\ref{eq:pion}-\ref{eq:Dsplus}).

We have also estimated the pion, kaon, $D$- and $D_s$-meson masses in isospin-symmetric QCD obtaining the values given in Eqs.~(\ref{eq:pion_QCD}-\ref{eq:Dsmeson_QCD}).

A complete evaluation of the isospin-breaking corrections for the meson masses considered in this work requires the removal of the quenched QED approximation.
The development of the appropriate lattice regularization for the full unquenched QED+QCD action using maximally twisted-mass fermions \cite{Frezzotti:2016lwv} as well as the numerical determination of (fermionic) disconnected diagrams related to the sea-quark charges are currently underway and will be the subject of our future investigations.

\section*{Acknowledgements}
We warmly thank R.~Frezzotti and G.C.~Rossi for fruitful discussions and the ETMC members for having generated the gauge configurations used for this study. 
We thank G.~Colangelo for useful comments.
We gratefully acknowledge the CPU time provided by PRACE under the project Pra10-2693 {\em ``QED corrections to meson decay rates in Lattice QCD''} and by CINECA under the specific initiative INFN-LQCD123 on the BG/Q system Fermi at CINECA (Italy).
V.L., G.M., S.S., C.T.~thank MIUR (Italy) for partial support under the contract PRIN 2015. 
G.M.~also acknowledges partial support from ERC Ideas Advanced Grant n. 267985 ``DaMeSyFla''.

\appendix

\section{Numerical evaluation of the diagrams \ref{fig:diagrams}(a) - \ref{fig:diagrams}(e)}
\label{sec:appendixA}

In this paper we consider QED at $\mathcal{O}\left(\alpha_{em}\right)$, evaluating explicitly the fermionic connected diagrams contributing to meson masses.

For the diagrams \ref{fig:diagrams}(a) and \ref{fig:diagrams}(b) the numerical cost scales badly with the volume. 
Therefore, stochastic approaches are needed to avoid computing explicitly the integrals over the beginning and end of the photon propagator, the cost of which would be exceedingly too large for realistic volumes. 
Here we adopt a variation of the technique used in Ref.~\cite{deDivitiis:2013xla}. 

Let us first recall the technique used in the past. \\

For the sake of simplicity let us discuss the case of the ``exchange'' diagram \ref{fig:diagrams}(a) for a bilinear $\bar{\psi}\Gamma\psi$:
\[
\delta C^{exch}\left(t\right)\equiv\sum_{\vec{x},y_{1},y_{2}}\left\langle S\left(0;y_{1}\right)V_{\mu}\left(y_{1}\right)S\left(y_{1};\vec{x},t\right)\Gamma S\left(\vec{x},t;y_{2}\right)V_{\nu}\left(y_{2}\right)S\left(y_{2};0\right)\Gamma\right\rangle G_{\mu\nu}\left(y_{1},y_{2}\right).
\]
The nested summation over $y_1$ and $y_2$ is prohibitively costly and scales like $V^2$. 
We can split them into two separate summations, each scaling as $V$, by introducing a set of real stochastic fields $\eta_{\mu}\left(x\right)=\pm1\ \forall\mu,x\,$. 
The expectation value of the product of two fields is given by: 
\begin{equation}
\lim_{n\to\infty}\frac{1}{n}\sum_{i=1}^{n}\eta_{\mu}^{i}\left(x\right)\eta_{\nu}^{i}\left(y\right)=\delta_{\mu\nu}\delta\left(x,y\right),\label{eq:stoch-exp}
\end{equation}
from which we can write the photon propagator as: 
\[
G_{\mu\nu}\left(y_{1},y_{2}\right)=\lim_{n\to\infty}\frac{1}{n}\sum_{i=1}^{n}\phi_{\mu}^{i}\left(y_{1}\right)\eta_{\nu}^{i}\left(y_{2}\right),
\]
where $\phi_{\mu}^{i}\left(y_{1}\right)=G_{\mu\rho}\left(y_{1},y_{3}\right)\eta_{\rho}^{i}\left(y_{3}\right)$.
Taking advantage of the $\gamma_5$ hermiticity of the quark propagator, the correlation function can be obtained in the Feynman gauge by evaluating:
\begin{equation}
\delta C^{exch}\left(t\right)\equiv\lim_{n\to\infty}\frac{1}{n}\sum_{i=1}^{n}\sum_{\mu,\vec{x}}\left\langle \left.S^{V_{\mu}\phi_{\mu}^{i}}\right.^{\dagger}\left(\vec{x},t;0\right)\gamma_{5}\Gamma S^{V_{\mu}\eta_{\mu}^{i}}\left(\vec{x},t;0\right)\Gamma\gamma_{5}\right\rangle ,\label{eq:diluted}
\end{equation}
where 
\[
S^{V_{\mu}\varphi_{\mu}}\left(\vec{x},t;0\right)\equiv S\left(\vec{x},t;y\right)V_{\mu}\left(y\right)\varphi_{\mu}\left(y\right)S\left(y;0\right)
\]
is a \emph{sequential propagator}, in which the component $\mu$ of the (conserved) vector current coupled to the external field $\varphi$ has been inserted over all possible points of the quark line. 
For the case of interest, in which $\varphi$ can be either $\eta$ or $\phi$, this can computed by solving an appropriate Dirac equation, with a numerical cost similar to that of computing $S\left(z;0\right)$.
It is actually possible to obtain the same correlation function by considering
\begin{equation}
\delta C^{exch}\left(t\right)\equiv\lim_{n\to\infty}\frac{1}{n}\sum_{i=1}^{n}\sum_{\vec{x}}\left\langle \left.S^{V\phi^{i}}\right.^{\dagger}\left(\vec{x},t;0\right)\gamma_{5}\Gamma S^{V\eta^{i}}\left(\vec{x},t;0\right)\Gamma\gamma_{5}\right\rangle ,\label{eq:undiluted}
\end{equation}
where the sum over the Lorentz index $\mu$ has been absorbed inside a single sequential propagator:
\[
S^{V\varphi}\left(\vec{x},t;0\right)\equiv S\left(\vec{x},t;y\right)\left[\sum_{\mu}V_{\mu}\left(y\right)\varphi_{\mu}\left(y\right)\right]S\left(y;0\right).
\]

The difference between Eq.~(\ref{eq:diluted}) and Eq.~(\ref{eq:undiluted}) corresponds to the terms 
\[
\left\langle \left.S^{V\phi_{\mu}^{i}}\right.^{\dagger}\left(\vec{x},t;0\right)\gamma_{5}\Gamma S^{V\eta_{\nu}^{i}}\left(\vec{x},t;0\right)\Gamma\gamma_{5}\right\rangle \quad\mu\neq\nu,
\]
which average to zero in the Feynman gauge. 
We checked that in the PS channel, this terms are of negligible entity, so that Eq.~(\ref{eq:undiluted}) is four time more efficient than Eq.~(\ref{eq:diluted}).
In short, the calculation of $\delta C^{exch}$ with this framework requires to compute three propagators, and average over several (ideally infinite) stochastic sources $\eta$. 
This is the method adopted in Ref.~\cite{deDivitiis:2013xla}.\\

In this work we have adopted a slightly different approach. 
Instead of using Eq.~(\ref{eq:stoch-exp}), we define the photon propagator in terms of expectation value of the time-orderd product of photon fields:
\[
G_{\mu\nu}\left(y_{1},y_{2}\right)=\left\langle A_{\mu}\left(y_{1}\right)A_{\nu}\left(y_{2}\right)\right\rangle ,
\]
where the photon field $A_{\mu}\left(y\right)$ must be generated from the distribution of probability:
\[
P\left(A\right)dA\propto\exp\left[-A_{\mu}\left(y_{1}\right)G_{\mu\nu}^{-1}\left(y_{1},y_{2}\right)A_{\nu}\left(y_{2}\right)\right].
\]

This can be readily obtained drawing each mode of the photon field in momentum space in which the probability distribution is local in $k$, as was first noted in Ref.~\cite{Duncan:1996xy}:
\[
P\left(\tilde{A}\right)d\tilde{A}\propto\exp\left[-\tilde{A}_{\mu}\left(k\right)\tilde{G}_{\mu\nu}^{-1}\left(k\right)\tilde{A}_{\nu}\left(k\right)\right].
\]

After the local change of variable $\tilde{B}_{\rho}\left(k\right)=\sqrt{G_{\rho\nu}^{-1}\left(k\right)}\tilde{A}_{\nu}\left(k\right)$ each component of $\tilde{B}$ can be drawn independently: 
\[
P\left(\tilde{B}\right)d\tilde{B}\propto\exp\left[-\tilde{B}_{\mu}^{2}\left(k\right)\right],
\]
and the value of $\tilde{A}_{\mu}\left(k\right)$ can be constructed via 
\[
\tilde{A}_{\nu}\left(k\right)=\sqrt{\tilde{G}_{\rho\nu}\left(k\right)}\tilde{B}_{\rho}\left(k\right).
\]

The matrix $\sqrt{\tilde{G}_{\rho\nu}\left(k\right)}$ can be easily computed, and for the Wilson action in the Feynman gauge it amounts simply to
\[
\sqrt{\tilde{G}_{\rho\nu}\left(k\right)}=\delta_{\rho\nu}\sqrt{\frac{1}{\hat{k}^{2}}}.
\]

In this way the correlation function can be computed as:
\[
\delta C^{exch}\left(t\right)\equiv\lim_{n\to\infty}\frac{1}{n}\sum_{i=1}^{n}\sum_{\mu,\vec{x}}\left\langle \left.S^{VA_{\mu}^{i}}\left(\vec{x},t;0\right)\right.^{\dagger}\gamma_5\Gamma S^{VA_{\mu}^i}\left(\vec{x},t;0\right)\Gamma\gamma_5\right\rangle ,
\]
or through a single sequential propagator $S^{A^{i}}$, in a way similar to Eq.~(\ref{eq:undiluted}). 
This has a clear benefit: only two quark inversions are required to compute the exchange diagram. 
The case of the PS channel is of special interest, since in this case the correlation function is obtained by computing
\[
\delta C_{PP}^{exch}\left(t\right)\equiv\lim_{n\to\infty}\frac{1}{n}\sum_{i=1}^{n}\sum_{\vec{x}}\left\langle \left|S^{A_{i}}\left(\vec{x},t;0\right)\right|^2\right\rangle ~ .
\]
The result is a factor $50\%$ more precise than the corresponding one computed with $\eta - \phi$ representation of the propagator.\\

A similar reasoning suggests that the diagram \ref{fig:diagrams}(b) for the ``self-energy'' can be obtained by computing 
\[
\delta C^{self}\left(t\right)\equiv\lim_{n\to\infty}\frac{1}{n}\sum_{i=1}^{n}\sum_{\vec{x}}\left\langle \left.S^{VA^{i}\,VA^{i}}\left(\vec{x},t;0\right)\right.^{\dagger}\gamma_5\Gamma S\left(\vec{x},t;0\right)\Gamma\gamma_5\right\rangle ,
\]
with the sequential propagator defined recursively as 
\[
S^{VA^{i}\,VA^{i}}\left(\vec{x},t;0\right)\equiv S\left(\vec{x},t;y\right)\left[\sum_{\mu}V_{\mu}\left(y\right)A_{\mu}^{i}\left(y\right)\right]S^{VA^{i}}\left(y;0\right).
\]
\\

The ``tadpole'' diagram \ref{fig:diagrams}(c) instead can be obtained immediately at the cost of a single sequential propagator, without introducing any additional stochastic noise at all, by noting that the relation
\[
\delta C^T\left(t\right)\equiv\sum_{\mu,\vec{x}}\left\langle \left.S^{T_{\mu}}\left(\vec{x},t;0\right)\right.^{\dagger}\gamma_5\Gamma S\left(\vec{x},t;0\right)\Gamma\gamma_5\right\rangle =\sum_{\vec{x}}\left\langle \left.S^T\left(\vec{x},t;0\right)\right.^{\dagger}\gamma_5\Gamma S\left(\vec{x},t;0\right)\Gamma\gamma_5\right\rangle ,
\]
holds exactly, i.e.~without relying on gauge symmetry.\\

In summary, the QED corrections to meson masses can be computed through four inversions, namely those required to obtain the propagators $S$, $S^{VA^{i}}$, $S^{VA^{i}\,VA^{i}}$ and $S^{T}$. 
An additional propagator $S^{PS}$, corresponding to the PS insertion, is needed to compute the correction due to the shift of the critical mass, diagram \ref{fig:diagrams}(d), which arises specifically in our Twisted-Mass setup. 
Moreover, in order to take into account the mass difference between $u$ and $d$ quarks, an additional inversion is needed to compute the sequential propagator $S^S$ in which the scalar density is inserted, as depicted in diagram \ref{fig:diagrams}(e).

We note that working in the isosymmetric theory, there is no need to compute this diagrams for $u$ or $d$ quark separately\footnote{More specifically in the twisted-mass regularization and for the correlators analyzed in this work, we can obtain the $d$-quark propagator (regularized with an $r$-parameter having opposite sign to the one of the $u$-quark) by employing the r-$\gamma_5$ symmetry of the propagator: $S_u = \gamma_5 S_d^{\dagger} \gamma_5$.}.

Therefore, the number of light inversions, which dominates the numerical cost, is given by $\#_{INV}=4_{QED}+1_{TM}+1_{MASS}=6$. 
Finally, we remark that in order to improve the quality of the signal, we employed sixteen different time source positions, using a different realization of the photon field $A_\mu$ per source position. 
For the stochastic source for the quark interpolator we used $Z_2$ noise, diluted in spin but not in color.
Hence a total number of $4_{spin} \times 6_{prop} \times 16_{time} = 384$ Dirac equations has been solved for each gauge configuration.

\end{document}